\documentclass[preprint]{aastex}
\usepackage{natbib}

\shorttitle{Bondi-Hoyle Accretion in Star Clusters}
\shortauthors{Throop \& Bally}

\newcommand{\rbh}	{\ensuremath{R_{\rm{B}}}}
\newcommand{\rtidal}	{\ensuremath{R_{\rm{T}}}}
\newcommand{\racc}	{\ensuremath{R_{\rm{Acc}}}}
\newcommand{\dmdt}	{\ensuremath{\dot{M}}}
\newcommand{\dmdtbh}	{\ensuremath{\dot{M}_{\rm{B}}}}
\newcommand{\mdotbh}	{\dmdtbh}
\newcommand{\dmbh}	{\ensuremath{{\Delta M}_{\rm{B}}}}

\newcommand{\kms}	{\ensuremath{\rm{km\ s^{-1}}}}
\newcommand{\msol}	{\ensuremath{M_{\odot}}}
\newcommand{\msolyr}	{\ensuremath{M_{\odot}\ \rm{yr^{-1}}}}
\newcommand{\mdot}	{\ensuremath{\dot{M}}}
\newcommand{\eg}	{\textit{e.g.}}
\newcommand{\ie}	{\textit{i.e.}}
\newcommand{\hii}	{\ion{H}{2}}

\newcommand{\degrees}   {\ensuremath{{}^\circ}}

\newcommand{\promille}{ 
  \relax\ifmmode\promillezeichen
        \else\leavevmode\(\mathsurround=0pt\promillezeichen\)\fi}
\newcommand{\promillezeichen}{%
  \kern-.05em%
  \raise.5ex\hbox{\the\scriptfont0 0}%
  \kern-.15em/\kern-.15em%
  \lower.25ex\hbox{\the\scriptfont0 00}}

\begin{document} 

\bibliographystyle{apj}
\title{`Tail-end' Bondi-Hoyle Accretion in Young Star Clusters: Implications for Disks, Planets, and Stars\\
\ \\
Submitted to \textsl{AJ}, 7-Nov-2007 \\
Revised, 1-Apr-2008\\
Accepted, 2-Apr-2008}

\author{Henry B. Throop}
\affil{Southwest Research Institute}
\affil{Department of Space Studies}
\affil{1050 Walnut St, Ste 300, Boulder, CO  80302}
\email{throop@boulder.swri.edu}

\author{John Bally}
\affil{Center for Astrophysics and Space Astronomy}
\affil{University of Colorado}
\affil{UCB 389, Boulder, CO  80309-0389}

\begin{abstract}

Young stars orbiting in the gravitational potential well of forming star clusters pass through the cluster's dense
molecular gas and can experience Bondi-Hoyle accretion from reservoirs outside their individual protostellar cloud
cores.  Accretion can occur for several million years after the stars form, but before the cluster disperses.  This
accretion is predominantly onto the disk and not the star.

N-body simulations of stars orbiting in three young model clusters containing 30, 300, and 3000 stars are presented.
The simulations include the gravitational potential of the molecular gas which smoothly disperses over time.  The
clusters have a star formation efficiency of 33\% and a radius of 0.22~pc.  We find that the disks surrounding
solar-mass stars in the N=30 cluster accretes $\sim0.01\ \msol$ (1 minimum-mass solar nebula, MMSN) per Myr, with a
$1\sigma$ width of 50 times due to variations in initial stellar positions and velocities within the cluster.  The
accretion rate scales as $M^{2.1\pm0.1}$ for stars of mass $M$.  The accretion rate is $\sim$ 5 times lower for N=3000
cluster, due to its higher stellar velocities and higher temperature.  The Bondi-Hoyle accretion rates onto the disks
are several times lower than accretion rates observed directly onto young stars \citep[\eg,][]{mlb05}: these two
accretion rates follow the same $M^2$ behavior and may be related.  The accreted disk mass is large enough that it may
have a substantial and unappreciated effect on disk structure and the formation of planetary systems.  We discuss a
variety of implications of this process, including its effect on metallicity differences between cluster stars,
compositional differences between a star and its disk, the formation of terrestrial and gas-giant planets, and isotopic
anomalies observed in our Solar System.

\end{abstract}

\section{Introduction} 
\singlespace

Stars form from the collapse of dense cores in giant molecular clouds (GMCs).  While some stars form in relative
isolation or in small groups, the majority of stars in the Solar neighborhood appear to be born in transient clusters
containing hundreds or thousands of members \cite[\eg,][]{apf06,ll03}.  Dense cores have masses ranging from under $1
\msol$ to over $10^4 \msol$ for the most massive cluster-forming environments.  Self-gravity and efficient cooling by
dust and molecular line radiation leads to collapse and fragmentation.  The timescale for the birth of an isolated star
can be estimated by considering the accretion rate onto a protostar from an isothermal sphere:  $\tau_{acc} \sim M  /
\dot M \approx GM / c_s^3$, where $M$ is the final stellar mass and $c_s$ is the effective sound speed in the core.
Assuming a 1 $\msol$ star in a cloud with $c_s$ = 0.35~\kms, we find that $\tau_{acc} \approx 10^5\ \rm{yr}$.  In
contrast, the observed age spread of young stellar objects (YSOs) in clusters indicates that the formation time-scale
for an entire cluster is a few million years.  Thus, only about 3 to 10\% of the final population of YSOs in a cluster
are expected to be in their main accretion phase (Class 0 / Class I) at any time.  Star formation then can be
characterized by a hierarchy of at least three timescales: the $10^5\ \rm{yr}$ scale to form an individual star; the
$10^6\ \rm{yr}$ to form a cluster; and the $10^7\ \rm{yr}$ timescale of an OB association.  Planet formation operates on
roughly comparable timescales: $10^5\ \rm{yr}$ to form meter- to km-sized bodies; $10^6\ \rm{yr}$ to form planetesimals
and cores (for the terrestrial and giant planets, respectively); and $10^7\ \rm{yr}$ to complete the planets
\citep{ls07,ntk07,ll03,tbe01,bal01,wei97}.

As the cluster gives birth to more stars, the oldest stars will evolve into Class II and III YSOs surrounded by
protoplanetary disks.  For at least a few million years, as these YSOs move through the cluster, they have a chance to
accrete additional material from reservoirs of dense gas remaining in the region.  Bondi-Hoyle (BH) accretion
describes the rate at which material will be added to the star \citep{bon52,bh44}.  The accretion rate is generally much
less than that experienced by stars in the collapse and accretion phases of their formation: $10^{-8}\msolyr$ vs.
$10^{-5}\msolyr$ \citep{bb05}.  The total mass accreted is small compared to the star, but it can be large compared to
the mass of the protoplanetary disk.  Moreover, the disk intercepts material falling toward the star, so mass is
deposited onto the disk and not the star.  Thus, this `tail-end accretion' can have profound consequences for the
evolution of very young planetary and pre-planetary systems.  Because accretion scales as $\dot{M} \propto M^2$, it is
likely to be most important for the larger members of a forming cluster that spend the longest time moving through dense
gas.

Typical star-formation efficiency (SFE) in a cluster-forming cloud core is $10\% - 30\%$ \citep[\eg,][]{jjk07,ll03}.
The majority of the gas is not consumed by stellar birth, but remains in the cloud for millions of years, even in the
presence of ongoing star formation.  For instance, dense molecular cores located as close as 0.1~pc to the ONC cluster
core continue to produce stars; the OMC-1S core 90\arcsec\ southwest of the Trapezium stars contains several dozen 
currently accreting protostars \citep{zhr06}.  Likewise, the moderate mass cluster IC~348 in the Perseus Molecular cloud
contains more than 400 young stars with an age spread of least 2.5--5 Myr \citep{mll07} -- yet, this cluster contains at
least 20 class 0/I protostars, many of which drive outflows at it southern periphery \citep{wbh06}.  \citet{mll07}
propose that this cluster was built over several crossing-times and that the current generation of class 0/I YSOs will
eventually fall into the cluster core and merge with it.

Stellar winds, supernovae, and other processes eventually remove the gas; clusters with gas remaining beyond 10~Mya are
rare \citep{ll03}.  When only low- to intermediate-mass stars are formed, protostellar winds and jets are the most
likely agents responsible for the removal of remaining gas \citep[\eg,][]{wbr05}.  In massive cloud cores, outflows may
not inject enough energy and momentum to disperse the gas, enabling star formation to continue.  Furthermore, winds may
increase the turbulent motions and internal pressures of the remaining gas, thereby increasing the tendency to form more
massive stars \citep{tm04}.  When massive O and B stars form, their intense UV radiation fields terminate star formation
by heating and removing the remaining gas.  UV heating and ionization will bring star formation to a halt in a crossing
time which can be estimated from the radius of the region, $L$, and the sound speed in photo-ionized plasma, $c_s \approx
11\ \kms$.  Typically, this time-scale ranges from $10^5$ to a few times $10^6$ years.  In some cases, UV radiation may
also trigger star birth: as expanding ionization fronts overrun and surround cores, the rocket effect of photo-ablating
plasma combined with the high pressure of the \hii\ region can result in radiation-driven implosion, gravitational
collapse, and triggered star birth.  


Star formation is observed over a broad range of size scales.  The Orion nebula is the prototypical example of a nearby
massive region that has spawned over 2000 stars in the last few Myr \citep{hc00,hil97}.  The dynamics of the gas is now
dominated by UV radiation and winds emerging from the massive Trapezium stars at its core, but for several Myr, the
region was cool and dark as low-mass stars formed and orbited through the cluster.  The Orion nebula's solar-mass stars
are observed to have ages of up to 1--2~Myr, far older than the $10^4$--$10^5$ year-old Trapezium stars
\citep{bsd98}.  The NGC 1333 region in the Perseus molecular cloud \citep[\eg,][]{jjk07,lal96} provides a good example
of a medium-sized cluster that has formed about 150 low- to intermediate-mass stars within the last 1--2 Myr.
Protostellar winds appear to dominate the gas dynamics and may regulate the rate of star formation.  At smaller size
scales, star formation in Taurus is divided into numerous small, distinct clumps of $\sim 15$ stars \citep{ghk93}.
These small clusters have far lower spatial density than Orion, but with similar SFE.  

Across the range of stellar number, the timescales and spatial scales of star formation are remarkably uniform: all
three have ages (and lifetimes) of several Myr, and diameters $0.5 - 1\ \rm{pc}$.  Large regions are characterized by
higher gas and stellar densities, but only slight increases in spatial scale: for instance, the width of the Orion
nebula core is $\sim$0.1--0.2~pc, comparable to the low-mass cores in Taurus with their 100-times lower density
\citep{jb99,omk96}.


This paper reports on numerical simulations of Bondi-Hoyle (BH) accretion onto stars and disks during the several Myr
after stars have formed, but before clusters have dispersed.  This is assumed to be a distinct, long-lived phase after
the initial mass function (IMF) has largely been established.  Accretion onto the star/disk continues, but at a rate far
lower than any accretion during stellar formation.  This `tail-end' phase of accretion has received little attention,
but may prove to have substantial consequences, because it is during this 1--10~Myr period when the disk exists that
planetesimals and young planets complete all but the final stages of their formation.


BH accretion is a well-studied process with several applications to stellar formation and evolution.  It has been
studied for its role in determining the stellar IMF.  The origin of the IMF is not well understood, with debate
primarily between two models.  In the first, stars fragment from the cloud with only a small seed mass, and accrete the
rest of their material via BH `competitive accretion' from the cluster over a few $10^5\ \rm{yr}$ \citep[cf.][and
references therein]{blz07}.  This model is supported by extensive numerical simulations; for instance, the work of
\citet{bb05} used SPH calculations to model the collapse of a cloud with radius of 0.19~pc and mass 50~\msol, for
$\sim300\ \rm{kyr}$, during which the IMF was almost completely determined.  The second model proposes that the IMF is
set at the time of cloud fragmentation.  A molecular cloud breaks into a spectrum of pre-stellar cores, and these cores
condense into stars without accreting substantially more mass \citep[cf.][and references therein]{kmk05b}.  This model
is supported by direct millimeter observations of young cores, which show pre-stellar condensations in a distribution
that closely matches a Kroupa-type IMF over the range 0.05 -- 5\ \msol \citep{abm07,man98}.  

\label{sect:bh_imf}

Recently, \citet{pkn05} invoked BH accretion to explain the observed accretion signatures onto young stars, which we
address later in this paper.  On longer timescales of $10^{10}\ \rm{yr}$, BH accretion within globular clusters has been
used to explain the `self-pollution' of high-metallicity veneers onto stars \citep{tjg02,pjm00,jmn99}.  As a historical
note, an early paper by \citet{lyt61} proposed that the entire Solar System was accreted by a BH process as the Sun
traveled through thousands of cold molecular clouds, but such a scenario was shown implausible by \citet{aw73}.  

In this paper, we use N-body simulations to model stellar dynamics in three clusters containing 30, 300, and 3000 stars,
and containing a large reservoir of gas which disperses on timescales of several Myr.  This paper focuses on the roughly
solar-mass stars since they are the ones of greatest interest to studies of planet formation.  The accretion of gas onto
the stars and their disks is calculated using the Bondi-Hoyle equations, with no new hydrodynamic simulations.  In
\S~\ref{sect:background} we give background and motivation.  Numerical simulations are detailed in
\S~\ref{sect:simulations}, and \S~\ref{sect:results} discusses the results.  We compare to observations in
\S~\ref{sect:observations}, and we discuss implications for the formation of stars, disks, and planets in
\S~\ref{sect:discussion}.


\section{Background and Motivation}
\label{sect:background}

\subsection{Bondi-Hoyle accretion}
\label{sect:bh}

The Bondi-Hoyle accretion (BHA) rate can be written as \citep[\eg,][]{edg04}:
\begin{equation}
\mdotbh = {{\pi\, n\, \mu\, m_H\, v\, \racc^2}},
\label{eq:dmdtbh}
\end{equation}
where $n$ is the gas number density, $\mu\, m_H$ is the mean molecular weight,
$v$ is the stellar speed relative to the gas, and \racc\ is the accretion radius, which is the distance outward to
which the star's influence on the gas can be felt.  For an isolated star, \racc\ is the BH radius \rbh\ given by

\begin{equation}
\rbh = {{2\, G\, M} \over {(v^2 + c_s^2)}} ,
\label{eq:rbh}
\end{equation}
where $G$ is the gravitational constant, $M$ is the stellar mass, and
$c_s$ is the sound speed.  For stars in a dense cluster, accretion can happen only out to the tidal radius \rtidal,
where 

\begin{equation}
\rtidal = {a\over 2} \left({M \over M_{enc}}\right)^{1/3}
\end{equation}
for a star at distance $a$ from the cluster core, enclosing a cluster mass $M_{\rm{enc}}$ within its orbital distance.
The star accretes material out to \racc, which is the lesser of the gravitational and tidal radii \citep{bbc01}:

\begin{equation}
\racc = {\rm min}(\rbh, \rtidal) \ \ .
\end{equation}
For the parameter space explored here, $\rbh \ll \rtidal$ in almost all cases.  In our simulations, the two become
comparable only for short periods at the apoapses of highly elongated orbits. 

For a typical cool molecular cloud of 25~K and molecular weight $2.3\, m_H$, the sound speed is $c_s \approx 0.3\
\rm{km\ s^{-1}}$.  Using $n = 10^4\ \rm{cm^{-3}}$ and $v=2\ \rm{km\ s^{-1}}$, we find $\rbh \approx 500\ \rm{AU}$, and the
accretion rate is $\mdotbh \approx 10^{-8}\ \msol\ \rm{yr}^{-1}$.  For most YSOs in a cluster, \rbh\ will be larger than
typical disk sizes of 10--200~AU, so the \textit{gravitational} cross-section for accretion is many times higher than
the YSO's \textit{physical} cross-section.   

The velocity $v$ is the stellar speed relative to the gas.  $v$ is difficult to calculate directly because the gas is
turbulent and characterized by many velocities on different scales.  A simplification could be made that the gas is at
rest, but this would result in a velocity estimate that is usually too low, causing an overestimate of the
accretion radius and the accretion rate.  We use two different methods to estimate $v$.  Both are based on the
assumption that the gas can be thought of as being composed of small clumps that travel primarily on gravitational
orbits through the cluster region, similar to the stellar motion.  This assumption is supported by observations of the
motions of pre-stellar condensations in young clusters, moving at roughly virial velocities \citep{abm07}.  Therefore,
in our first approach, we estimate $v$ based on the stellar velocity $v_*$ and virial velocity $v_v$ by

\begin{equation}
v = (v_*^2 + v_v^2)^{1/2}
\label{eq:v_v}
\end{equation}
where

\begin{equation}
v_v = \left[2\ (KE + PE) / M\right]^{1/2}.
\end{equation}
KE and PE are the total kinetic and potential energy of the system, and M is the total mass.  The virial velocity
evolves with time, and generally gets smaller as the system evolves.

The second method also assumes that the gas clumps behave dynamically like stars, and thus calculates $v$ by recording
the actual velocity $v_n$ between a star and its nearest stellar neighbor, at each timestep, and then taking

\begin{equation}
v = v_n \ .
\label{eq:v_n}
\end{equation}
In this way the distribution of velocities between stars and gas clumps orbiting within the cluster's potential well can
be estimated.  A byproduct of this method is that transitions between velocities are not as smooth as they would be in
reality; that is, when a star's nearest-neighbor changes, its $v_n$ also immediately changes.  For conceptual simplicity
some of our plots therefore use eq.~[\ref{eq:v_v}], but all of our final results use eq.~[\ref{eq:v_n}].


Bondi-Hoyle accretion is a simple and robust process, but it can be affected by several environmental factors.  Stellar
winds from Class II YSOs or T Tauri stars have $\dot{M}_{\rm{w}} \sim 10^{-8}\ \msol\ \rm{yr^{-1}}$ and $v_w \sim 200\
\kms$ \citep{rdb07}.  The outward ram pressure from such a wind ($P = \dot M v/R^2$) could be enough to prevent
accretion from a competing inward BH flow, because the mass rates are comparable, and the wind velocities greatly exceed
the BH velocities.  However, these winds and jets tend to be quite narrow, and can only stop accretion for material that
passes directly through their paths.  HST observations by \citet{hep04} of several young T Tauri stars found the jets to
have opening angles of $\sim 20\degrees$ at distances 20--50~AU.  At greater distances, the opening angles are smaller
and the jets become essentially collimated.  Thus, near the stars the winds cover solid angle $\Omega \lesssim 1\
\rm{sr}$, and this decreases substantially at \rbh.  Stellar winds may therefore be responsible for reducing BH
accretion by a fraction $\Omega/(4\pi) \approx 0.1$.  Conceivably, if the flow passes through the wind cone several
times this fraction could be increased; however, we argue below that the BH flow accretes onto the disk rapidly, and
once on the disk it is certainly immune from polar outflows.  The detailed nature of the outflow-inflow interaction
could be studied with future simulations.


\citet{ec04} showed that radiation pressure from low-mass stars is too weak to affect the accretion rate, but that above
$10\msol$, radiation pressure becomes strong enough to inhibit BH accretion entirely.  The transition between these two
regimes is sharp.  Because we focus on solar-mass stars and our models have very few high-mass stars, this effect has no
consequence on our results.

\begin{figure}
\centerline{              {\includegraphics[width=6in]{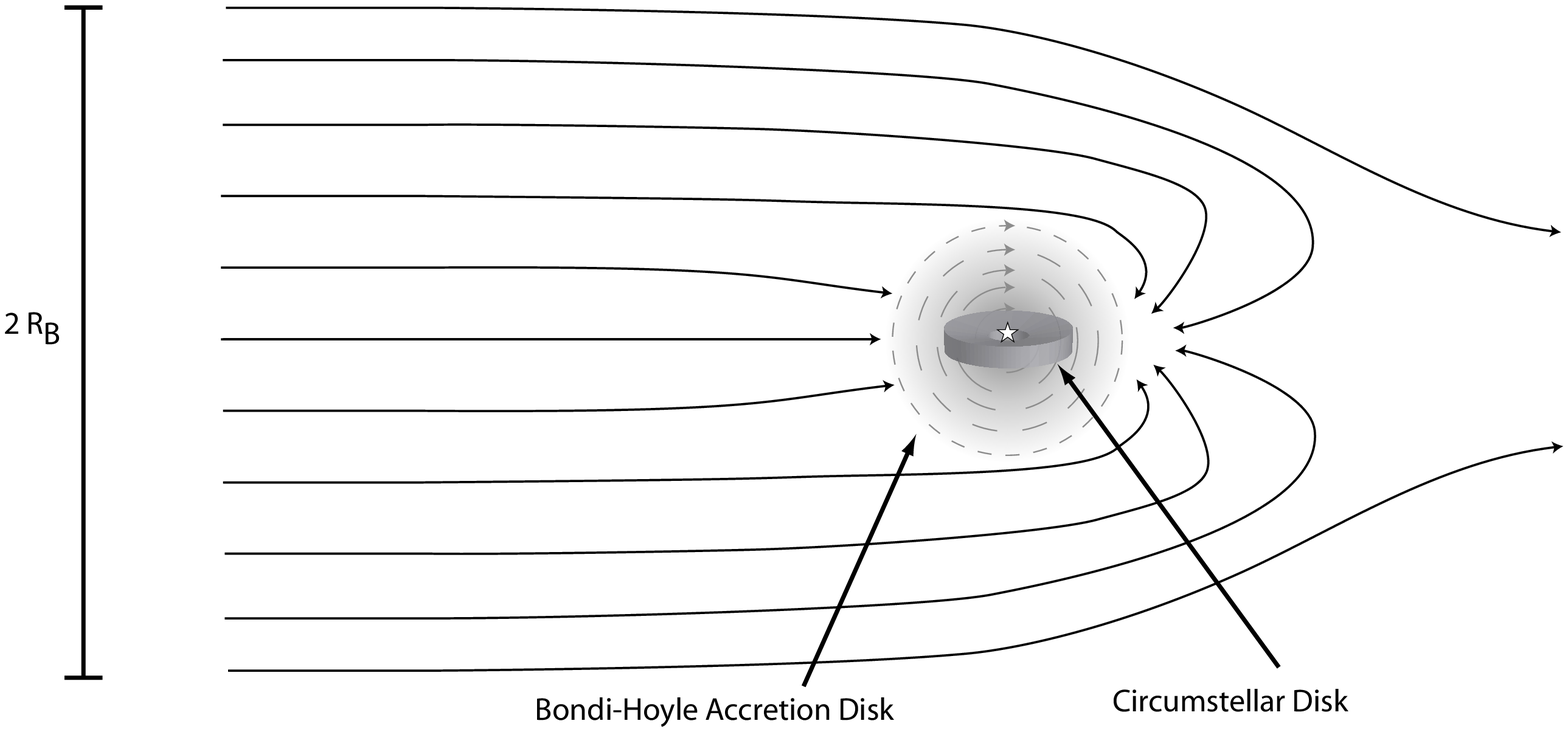}}}
\vskip -5 in
\caption{Cartoon of Bondi-Hoyle accretion.  The star is moving relative to the ISM gas.  Material is focused toward the
star.  Most material accretes \textit{behind} the star, where opposing streamlines meet and cancel their velocities.
Accretion is onto a large, temporary accretion disk of size $\sim 0.2\ \rbh$, which deposits material onto the star's
circumstellar disk of size $\lesssim 0.1\ \rbh$.} \label{fig:bh_cartoon} \end{figure}

BHA is normally modeled as accretion onto a point source or star.  The flow is gravitationally focused from the initial
width of \rbh\ to a much smaller size as it nears the star (Figure~\ref{fig:bh_cartoon}).  The flow does not accrete
onto the star's front hemisphere, but continues beyond the star, where it loses much of its energy when it collides with
opposing streamlines.  The flow behavior can be modeled analytically for subsonic flows, and has been the subject of
numerous 2D and 3D simulations for higher-speed, turbulent flows at Mach 3--10
\citep[\eg,][]{kmk05,edg04,ruf99,fr97,ruf97,blt97}.  These simulations find that as the accreted material collects
behind the star, it sets up a transient, tenuous `accretion disk' of size $\sim 0.2$~\rbh.  The disk oscillates unstably
between prograde and retrograde rotation (the so-called `flip-flop' instability), and eventually accretes onto the star
on Keplerian timescales.  The accretion flow has rate variations of a factor of 2--5, and the tenuous disk has mass
$\sim 10^{-3}$ MMSN, for conditions in our model cluster.  Most of the mass contained inside \rbh\ is ultimately
accreted onto the star.  Angular momentum is also accreted onto the disk, but less efficiently: a gradient of a few
percent in the background gas velocity can inhibit most of the angular momentum accretion onto the star \citep{ruf99}.

These simulations have been performed using stars as accretors, but to our knowledge there are no published numerical
simulations of BHA onto stars with pre-existing disks.  The existing work shows that the accretion flow is not directly
onto the star, but passes through the plane of the disk many times before being accreted onto the star.  However,
upwards of 80\% of young stars in clusters are observed to have disks \citep{sbs05}, and for the MMSN case, the disk
density is on the order of $10^3$ times higher than the accretion flow density.  Infalling material will be rapidly
intercepted and absorbed by the disk.  Only in the relatively unlikely circumstance that the star moves through the
medium with its disk aligned precisely edge-on, or with no disk at all, can accretion directly onto the star be
significant.

In some ways the accretion flow onto the disk can be compared to the proto-Jovian nebula `starved disk' model of
\citet{cw02}, where gas accretes from the solar nebula disk onto the circum-Jovian disk orbiting within it.  The two disks 
are co-planar and move at constant velocity to each other, making mass transfer between them a smooth, linear process.
The molecular cloud case is much more complex, because the orientation and velocity of the disk relative to the ISM
is constantly changing.  The accretion flow will almost always have a different angular momentum vector than the
star's disk.  In the short term, accretion onto the disk will always result in mass gain.  However, if the incoming
angular momentum is opposite that of the disk, one can imagine that orbital decay and mass loss onto the star will soon
follow as the disk is perturbed.  There are thus two distinct accretion rates: one from the ISM onto the disk, and one
from the disk onto the star.  The two are related, but the disk acts as a buffer between them, on the viscous timescale
of $10^5-10^7\ \rm{yr}$.  Our current understanding of the interaction between the accretion flow and the disk is very
limited; we discuss the relationship between the two accretion rates further in \S~\ref{sect:observed_rates}.

The numerical simulations presented here include large-scale density gradients in the background gas field, but do not
include gas turbulence.  Work by \citet{kmk06} generalized BHA to a supersonically turbulent medium.  At high
turbulence, the accretion rate is limited by large-scale vorticity, which can suppress accretion.  In general, they
found vorticity to be important in the cases $\rbh /l \gtrsim 1$, where $l$ is the largest size scale of the turbulence,
comparable to the distance between protostellar cores.  In the core of our densest simulations, $n_*=10^6\
\rm{pc}^{-3}$, so $\rbh / l \approx 1/4$.  This can cause a moderate reduction in accretion; however, such dense
regions occur only in our largest simulation, only at its core, and only at the beginning.  In the vast majority of our
parameter space, $\rbh / l \ll 1$, and the BH equations remain valid.  \citet{kmk06} argued for some inhibited accretion
in molecular clouds, but they assumed far larger $\rbh$ than used here (20,000~AU vs.  500~AU), due to their initial
conditions of colder gas and lower stellar velocities.  

 
\section{Numerical simulations} \label{sect:simulations}

We use N-body simulations of three model clusters to trace the dynamical and environmental histories of individual stars.
Each cluster contains a background distribution of gas from which these stars can accrete.  The code NBODY6, a
sophisticated Hermite-method integrator designed for accuracy in handling of close approaches between stars, is used to
simulate the evolution of each star cluster \citep{aar99}.  NBODY6 is fast and proven with a long heritage.  It allows
for flexible control over initial input conditions, and enables the inclusion of a background molecular cloud whose
gravity may dominate the motion of stars.  The cloud is modeled as a gravitational Plummer potential of a given width
and mass, which decays smoothly after an initial delay.  NBODY6 treats the gravitational effect of the gas on the stars,
but the stars have no direct feedback on the gas: the cloud shape, density structure, and evolution are directly
controlled only by input parameters.

Our calculations are performed in two discrete steps.  First, N-body simulations of a cluster generate the positions
and velocities of each star, outputting at fixed intervals.  Second, these quantities are used to calculate \dmdtbh\ at
each step for each star, using eqs.~[\ref{eq:dmdtbh}--\ref{eq:v_n}].  The output timestep used is $2\times10^4\
\rm{yr}$, which is sufficient for convergence.  

Several simplifying assumptions have been made.  First, the simulations are not entirely self-consistent in that mass
accreted onto stars is removed only as a general global process, and not locally in response to accretion.  However,
mass lost to accretion by the end of the simulations is less than $1\%$ of the original cloud mass, so ignoring this
has only a minor effect on stellar motion.  \citet{ruf96} shows that the drag on an accreting body is approximately
$\dot{M}\, v$, so the stars' velocity change is also insignificantly small.  Because the sound speed is comparable to
the stellar velocities, any `tunnels' created in the cloud by BH accretion will be rapidly filled, making a local
treatment unnecessary.  Second, the increase in mass of individual stars due to BH accretion is ignored during the
N-body simulations; it is only calculated afterward.  The increase in stellar mass is usually $< 1\%$, making only small
changes to a star's motion.  This assumption may underestimate growth for some high-mass stars, by allowing them to
retain their full velocity even as their mass increases.  However, our emphasis is on the consequences of BHA for
solar-mass stars, and ignoring this small effect sets a conservative lower-bound.

For simplicity, it is assumed that all stars are formed at $t=0$.  Observations show that star formation is an on-going
process that begins as the cloud collapses and operates continually until the cluster disperses
\citep[\eg,]{sw07,amg07,tkm06}.  Therefore, our model overestimates the total mass accreted by the latest-formed stars,
by assuming that all stars form at the start.  However, the \textit{rate} of accretion onto a star is not affected by
this simplification.  We implicitly assume that the star formation timescale is much shorter than the cluster lifetime
(\citealt{elm00,har01}, but also see \citealt{tkm06,tm04b}).  Stars are formed with a pre-set
IMF; any mass evolution of the IMF due to ongoing accretion naturally shows up in our results.  

\begin{figure}
\centerline{              {\includegraphics[width=6in]{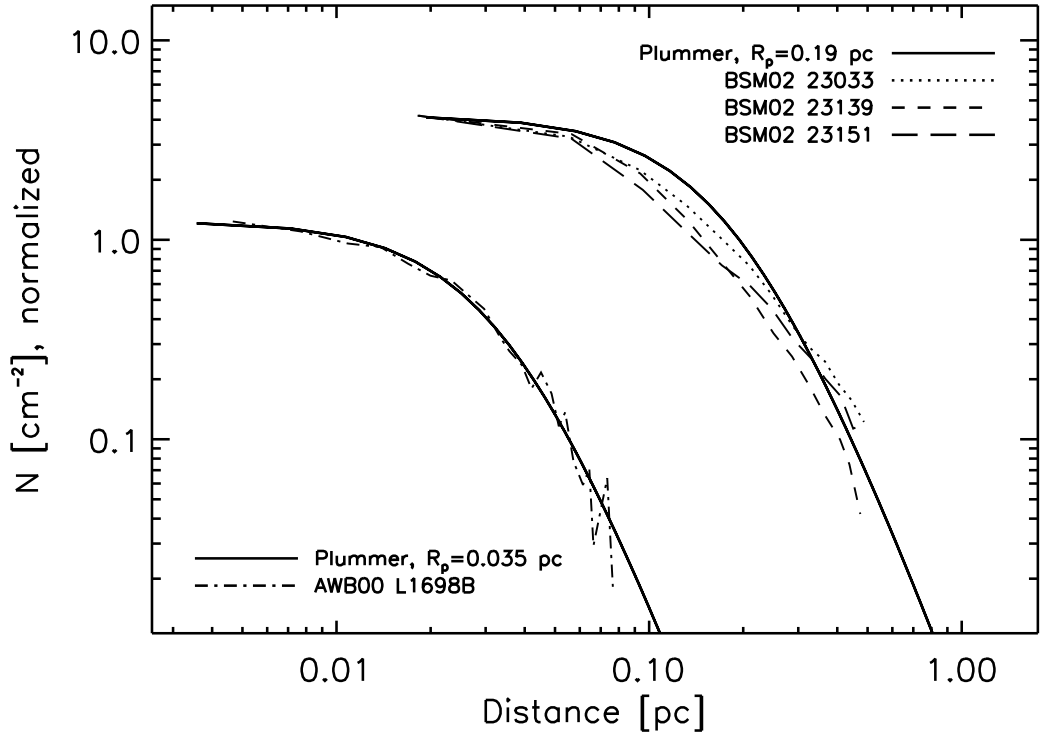}}}
\caption{Comparison of observed cores to Plummer spheres.  The lower curve plots the column density of a
pre-stellar condensation in L1689 \citep{awb00}, while the upper curves show dense cores in Cassiopeia
\citep{bsm02}.  All are fit well by the column density of a Plummer sphere.}
\label{fig:plummer} \end{figure}

We take the initial stellar and gas distributions to be three-dimensional Plummer spheres, with a profile given by

\begin{equation}
 \rho_s(r)	\propto  {[1+(r/r_0)^2]}^{-5/2} ,
 \label{eq:plummer}
\end{equation}
where $r$ is the distance from the cluster core, and $r_0$ is a measure of the cluster's size.  40\% of the mass is
contained within this distance.  This symmetrical distribution idealizes the shape and radial distribution of real
cloud cores.  Although some regions (such as Orion) are highly asymmetrical, observations by \citet{sey03} found a
majority of dense cloud cores ($10^2$--$10^4$~\msol) to be symmetric, with aspect ratios $< 1.3$.  For cores such as
these, Plummer spheres provide a remarkably good fit to the radial distribution over a variety of size ranges.
Figure~\ref{fig:plummer} shows the line-of-sight density measured in four optically thin condensations measured by
\citet{bsm02} and \citet{awb00}.  The largest three of these are young high-mass star-forming regions in Cassiopeia,
with masses 1200-2300~\msol, while the smallest is an individual pre-stellar core in L1689.  All four have flat inner
regions with steep outer edges, and all are readily fit by Plummer profiles (eq.~\ref{eq:plummer}, integrated
line-of-sight).  Because of the broad appropriateness of the Plummer distribution, and its easy analytic form, we have
adopted it for our initial distributions.  An improved future model could explore the effects of an asymmetrical cloud
core shape.


The Plummer sphere is a smooth function that does not have the local density variations of a realistic molecular cloud.
Its density changes only with radial distance.  However, because the accretion rate for an individual star depends
linearly on density (eq.~\ref{eq:dmdtbh}), local variations in density will in general average themselves out over an
orbital period.

Our simulations do not consider the effects of binaries.  Long-lived binaries are almost always the result of
primordial binaries that have been hardened due to subsequent three-body interactions \citep[\eg,][]{gkg07}.  We
include no primordial binaries, because we are mostly interested in the evolution of single stars around which
planetary formation is best understood.  For binary stars separated by much less than $\rbh\ \sim 500\ \rm{AU}$,
accretion will depend only on the total mass of the system, regardless of whether it is a single or multiple star.


As with previous simulations, we ignore magnetic fields.  The effect on accretion is most likely small \citep{kmk06},
but requires further study.

\subsection{Initial parameters}

The parameters for our three simulations are listed in Table~\ref{table:params}.  The {\sc small} cluster model is a
small, dark cloud, similar to one in Taurus, with $N=30$ stars and a gas mass of $M_g = 30~\msol$.  Masses are
distributed using the IMF of \citet{ktg93}.  The minimum stellar mass is 0.1~\msol, and the mean is 0.5~\msol, giving a
total stellar mass of $M_s = 15~\msol$.  The maximum is determined by the IMF, and is 1.2~\msol.  Star formation
efficiency (SFE) is the ratio $M_s / (M_s + M_g)$ and is 33\%.  The gas is retained entirely until $t=2\ \rm{Myr}$,
after which it is smoothly lost to simulate the dispersal by stellar winds and other processes.  The simulation ends at
$t=4\ \rm{Myr}$, which is picked to correspond to typical estimates for the lifetime of star formation
\citep{bps07,amg07}.  

The {\sc medium} and {\sc large} models assume the same SFE, gas radius, cluster radius, IMF, and gas timescales as the
{\sc small} cluster.  Parameters that change are the number of stars, the maximum stellar mass, and the total system
mass.  The {\sc medium} cluster models a region like IC348, with $N=500$ stars, while the {\sc large} cluster models the
environment in a young version of the Orion Nebula Cluster (ONC) with $N=3000$, before O/B stars have heated the gas and
stopped BH accretion.  


The Plummer radius in all cases is taken to be $r_0 = 0.22\ \rm{pc}$, which implies a FWHM in column density of 0.3~pc.
This size results in properly normalized values for the masses and peak densities of clusters: \eg, with this $r_0$, our
{\sc large} case has a peak density of $N = 10^6\ \rm{pc}^{-3}$ and $n = 10^6\ \rm{cm}^{-3}$, which fit observations of
Orion.  Peak densities are also correct for the two other models.

We take initial stellar velocities to be isotropic, drawn from a distribution specified by $g(q) = q^2 (1-q^2)^{7/2}$,
where $q = v / v_e$ is the ratio of stellar to escape velocities \citep{aar03}.  Stars are initially virialized.  Work
by \citet{apf06} explored the effect of initial stellar velocities and viriality on cluster evolution.  They found that
stellar velocities in general became more radial with time, especially as gas is removed.  We show in
\S\ref{sect:results} that most of the BH activity occurs during brief passages through the cluster core by stars on
radial orbits.  Our assumption of isotropic initial conditions thus represents a conservative case: choosing subvirial
(\ie, radial) velocities would increase the number of close passages through the cluster core, increasing BH accretion.

The initial radial profile of the gas envelope is the same as for the stars.  The gas mass then decays over time, given
by \citep{kah01}

\begin{equation}
M_g(t) = \left\{ \begin{array}{r@{\quad:\quad}l} t < t_d & M_g(0) \\ 
                                                 t \ge t_d & M_g(0) / { [1 + (t-t_d)/D]} \end{array} \right. .
\end{equation}
Here $t$ is the time, $t_d$ is the delay time before loss begins, and $D$ is the timescale for loss to half the initial
mass.  $D$ is set so that the gas mass at 4~Myr is 5\% of its value at 2~Myr.

We assume two different gas temperature models.  In the first, we take T=5~K, T=15~K, and T=40~K for the {\sc small},
{\sc medium}, and {\sc large} clusters, respectively.   In the second, we use T=25K for all cases.  Except where otherwise
mentioned, all of our results are reported using the first model; the second model is used only as a test case to
`unwrap' the effects of cluster size and temperature on accretion rate.  In all cases, the sound speed $c_s$ is given by

\begin{equation}
c_s = \left({{k_B T} \over {\mu m_H}}\right)^{1/2} .
\end{equation}


No new stars are produced during the runs; a few stars are removed as they are ejected beyond $\sim 5\ \rm{pc}$ from the
core.  All runs are stopped at 4~Myr; by this time the gas has mostly dispersed, and the accretion rate is
insignificant.  

We performed multiple runs for each set of initial conditions, varying only by the random seed, and combined these
together in order to have equal statistical sampling of each cluster.  Our findings reflect the results from 6000 stars
for each case: 200 runs for N=30, 20 runs for N=300, and 2 runs for N=3000.


\section{Results}
\label{sect:results}

\subsection{Cluster morphology and validation}

\begin{figure}   
\vskip -2 in
\centerline{     \scalebox{0.6}{\includegraphics*[0, 70][450, 800]{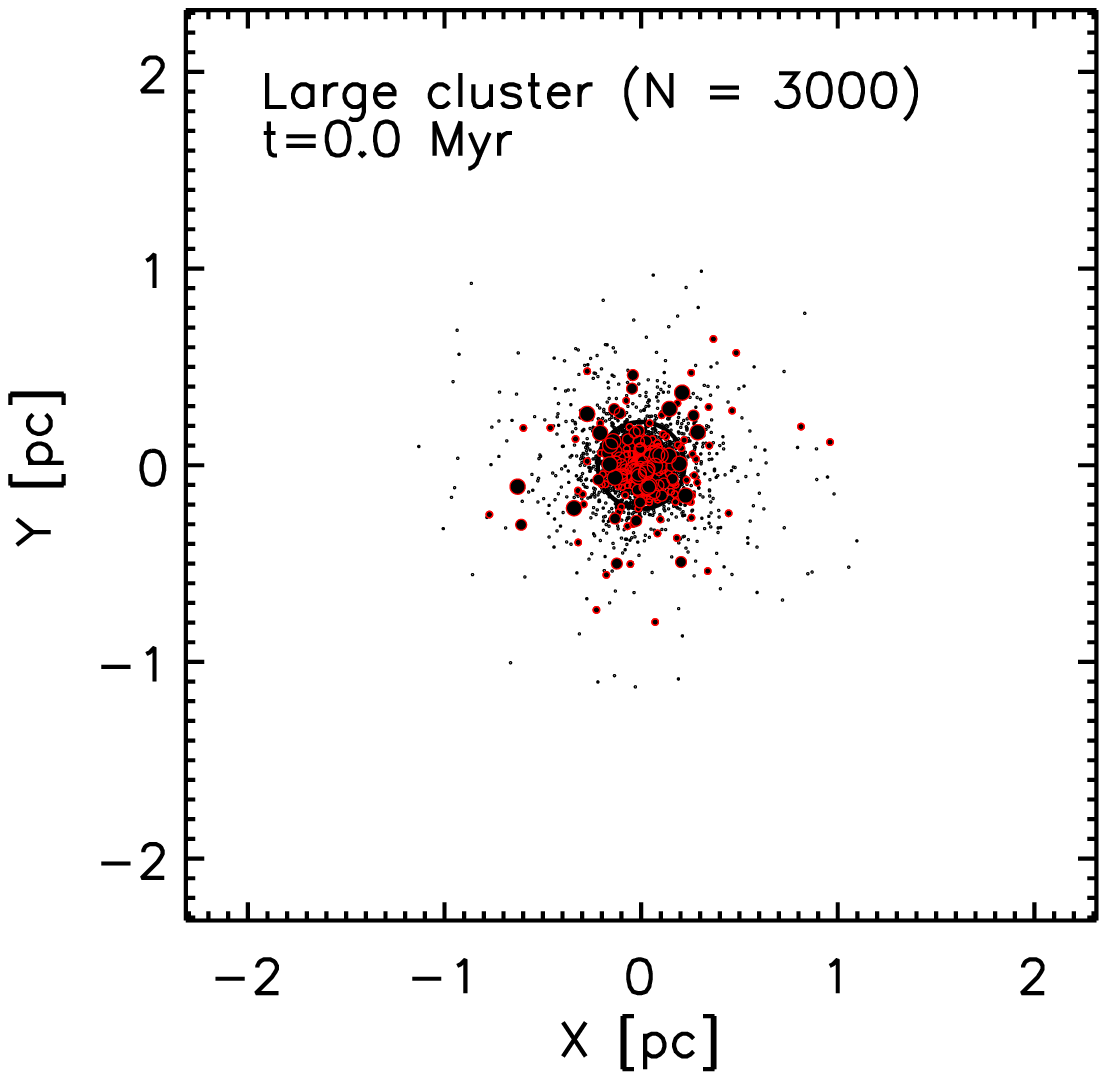}}
\hskip -0.8in    \scalebox{0.6}{\includegraphics*[60,70][450, 800]{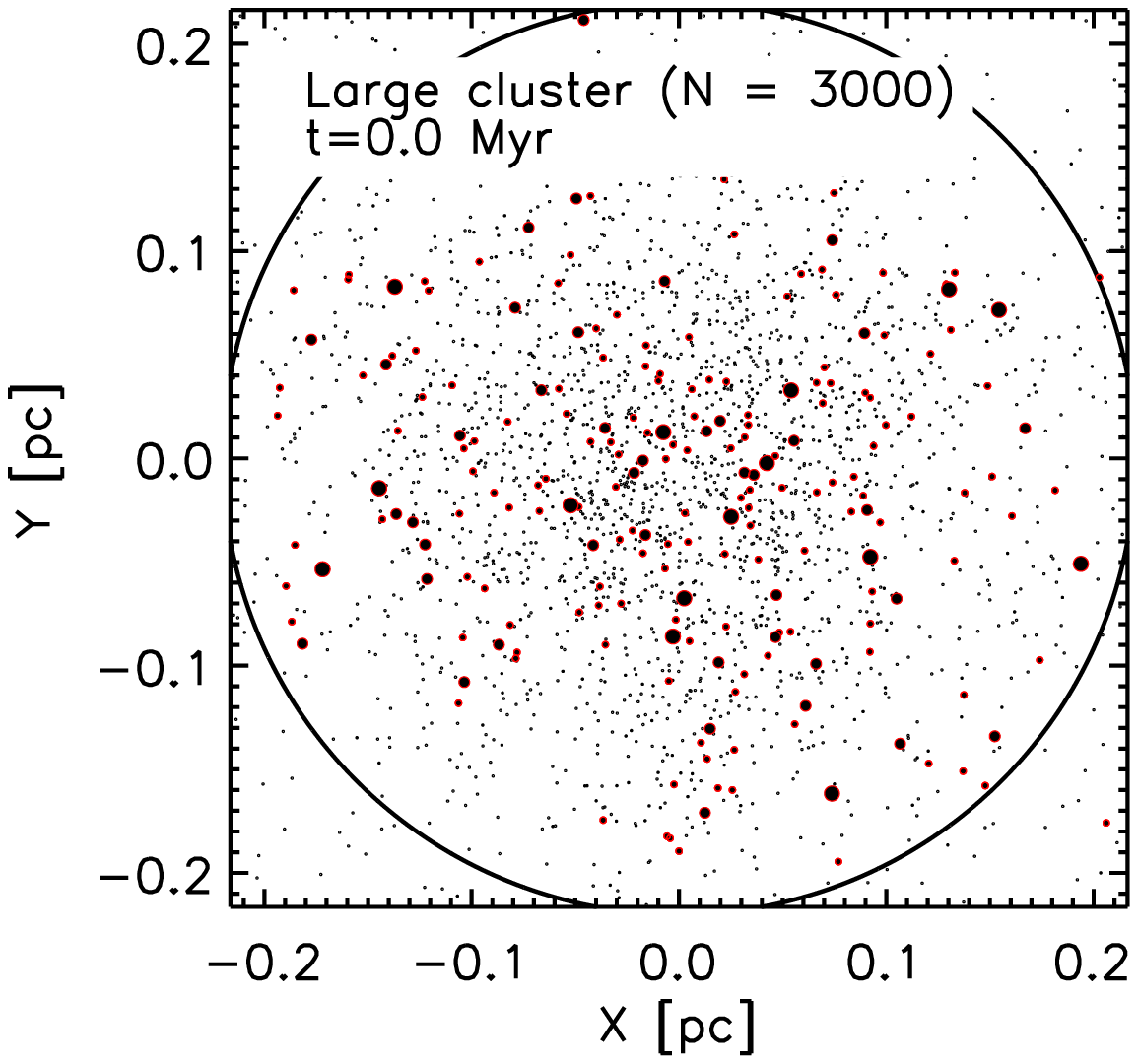}}}
\vskip -0.5 in
\centerline{     \scalebox{0.6}{\includegraphics*[0,  0][450, 400]{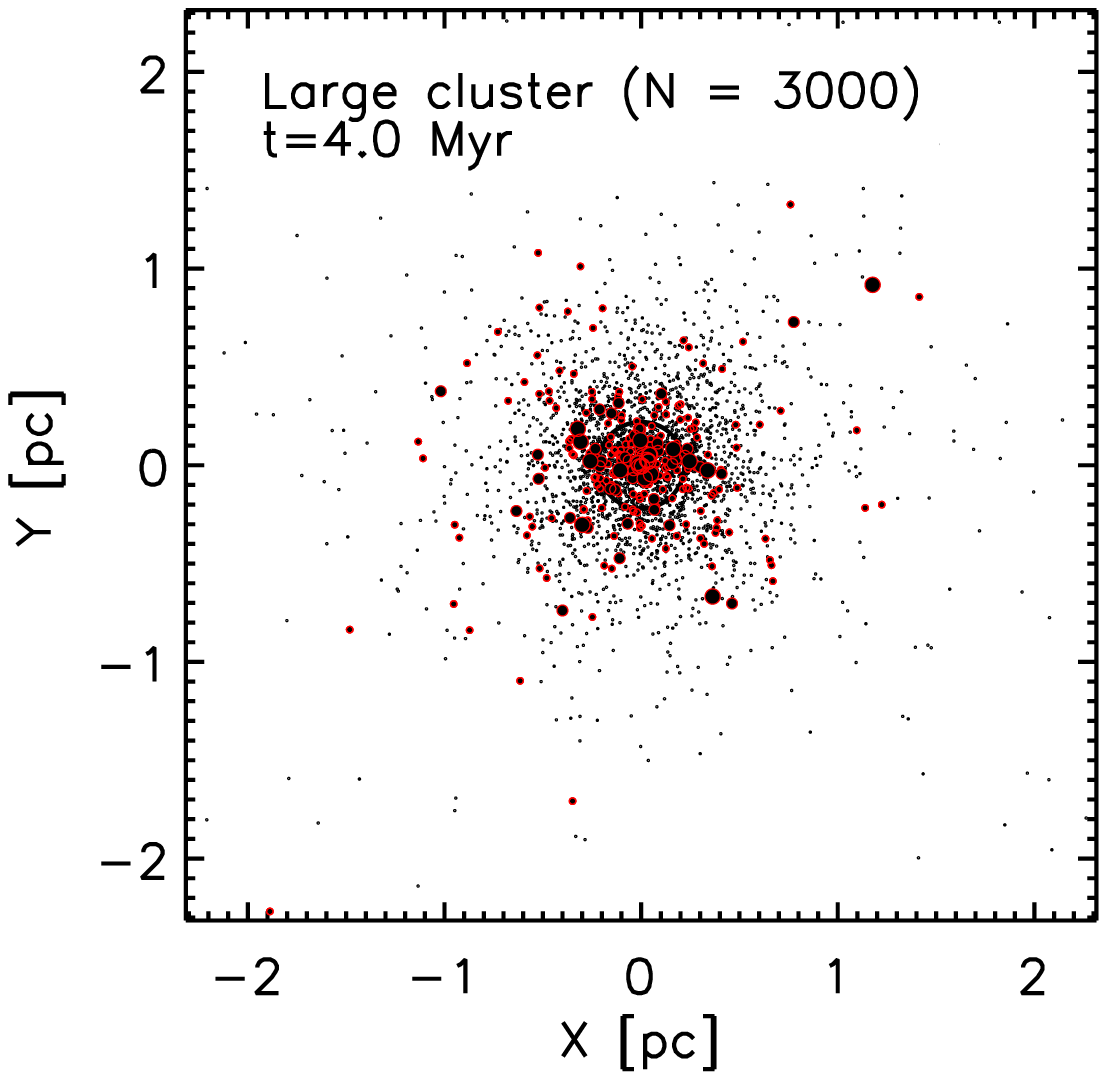}}
\hskip -0.8in    \scalebox{0.6}{\includegraphics*[60, 0][450, 400]{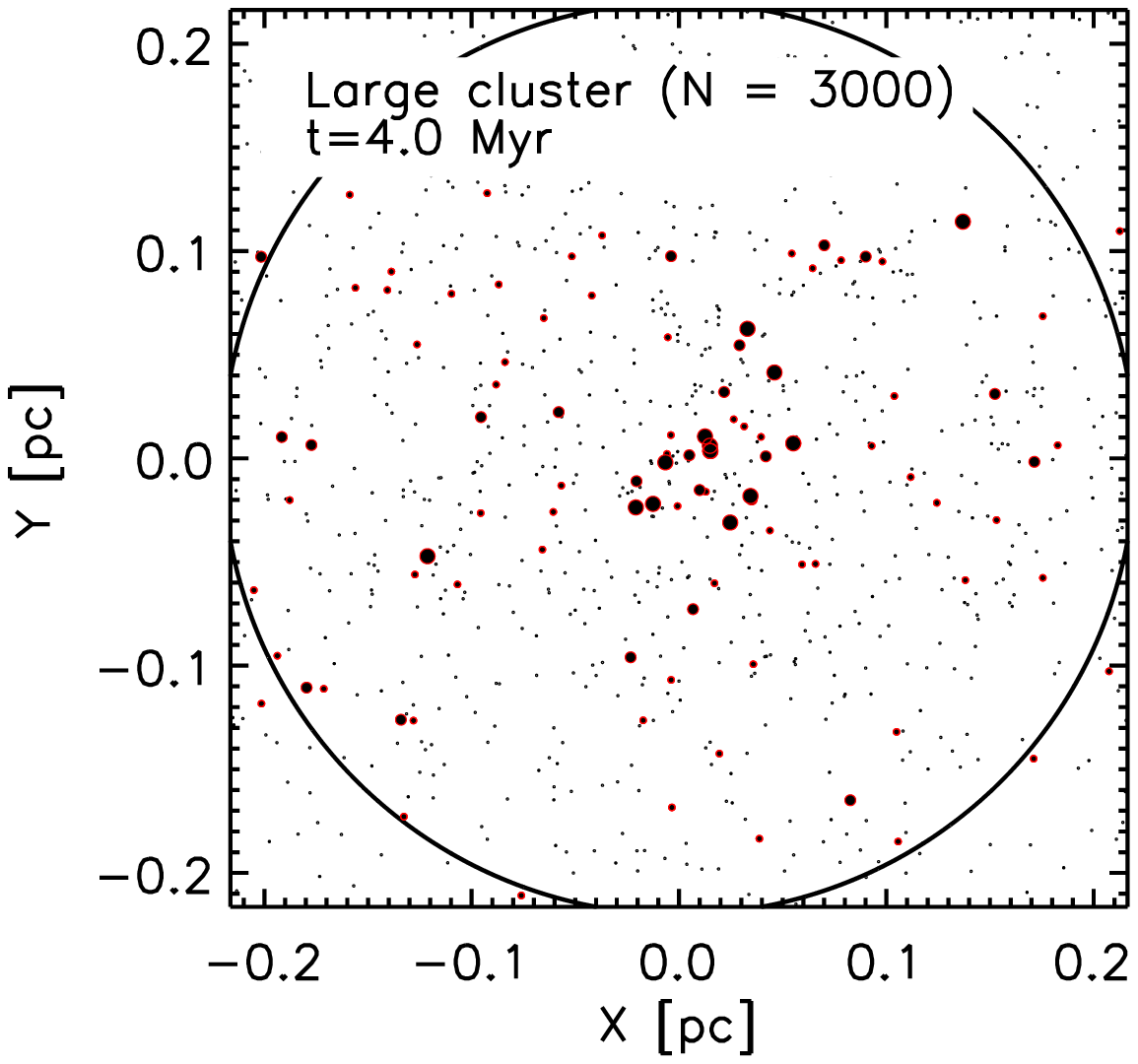}}}
\caption{Overhead views of the {\sc large} cluster simulation, at start (\textbf{top}) and end
(\textbf{bottom}).  The circular aperture indicates the $r_0$ radius of the Plummer sphere; initially, 40\% of the mass
of stars and gas is within this radius.  The two magnified views (\textbf{right}) show just the inner region.  Larger
dots indicate more massive stars.}

\label{fig:cluster_view}
\end{figure}

Figure~\ref{fig:cluster_view} shows images of the {\sc large} cluster at the start and end of the simulation.  During
the first 2~Myr, the cluster expands somewhat, predominantly due to low-mass stars that are given high velocities.  The
expansion speeds up significantly by about 2.5~Myr, when the reduced gas potential begins to be felt.  By the end, the
cluster volume density has decreased by a factor of about 40 inside of 0.1~pc, and increased slightly outward of 1~pc.
About 15\% of the stars are unbound from the cluster; more are lost after the run has stopped.  The {\sc small} and {\sc
medium} clusters evolve generally in a similar way to the {\sc large} cluster.

The simulations have been validated for numerical consistency by checking for conservation of angular momentum and
energy.  Energy is conserved up until the point at which gas loss begins.  After this time, the total energy smoothly
decreases to reflect the central potential's reduced mass.  Angular momentum is similarly conserved up until gas
dispersal, at which point it increases to reflect stars' expanding orbital radii.


\subsection{Stellar motions and Bondi-Hoyle accretion}

\begin{figure}
\centerline{\scalebox{0.55}{\includegraphics*[  0,60][354,335]{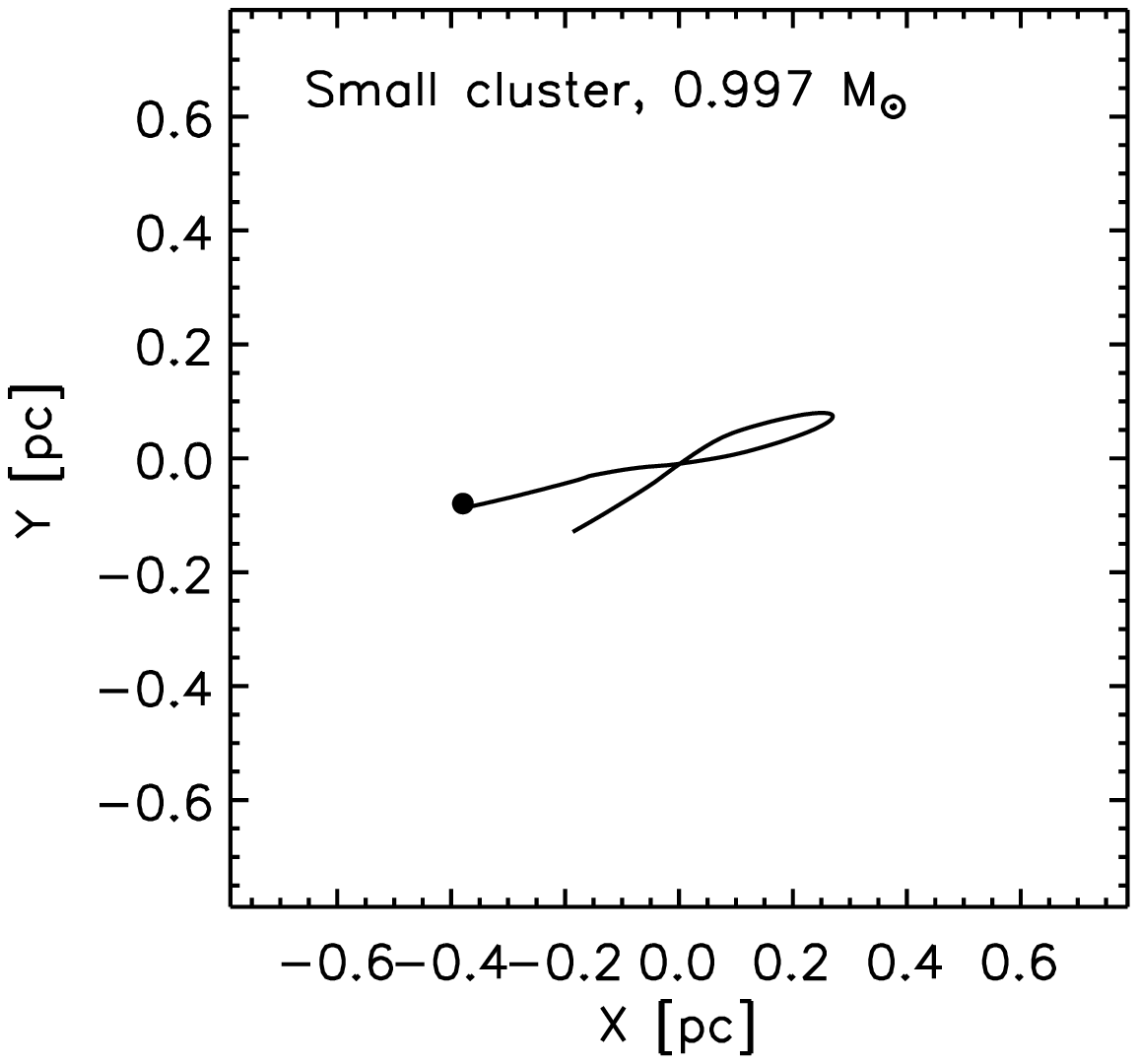}
                           {\includegraphics*[ 85,60][354,335]{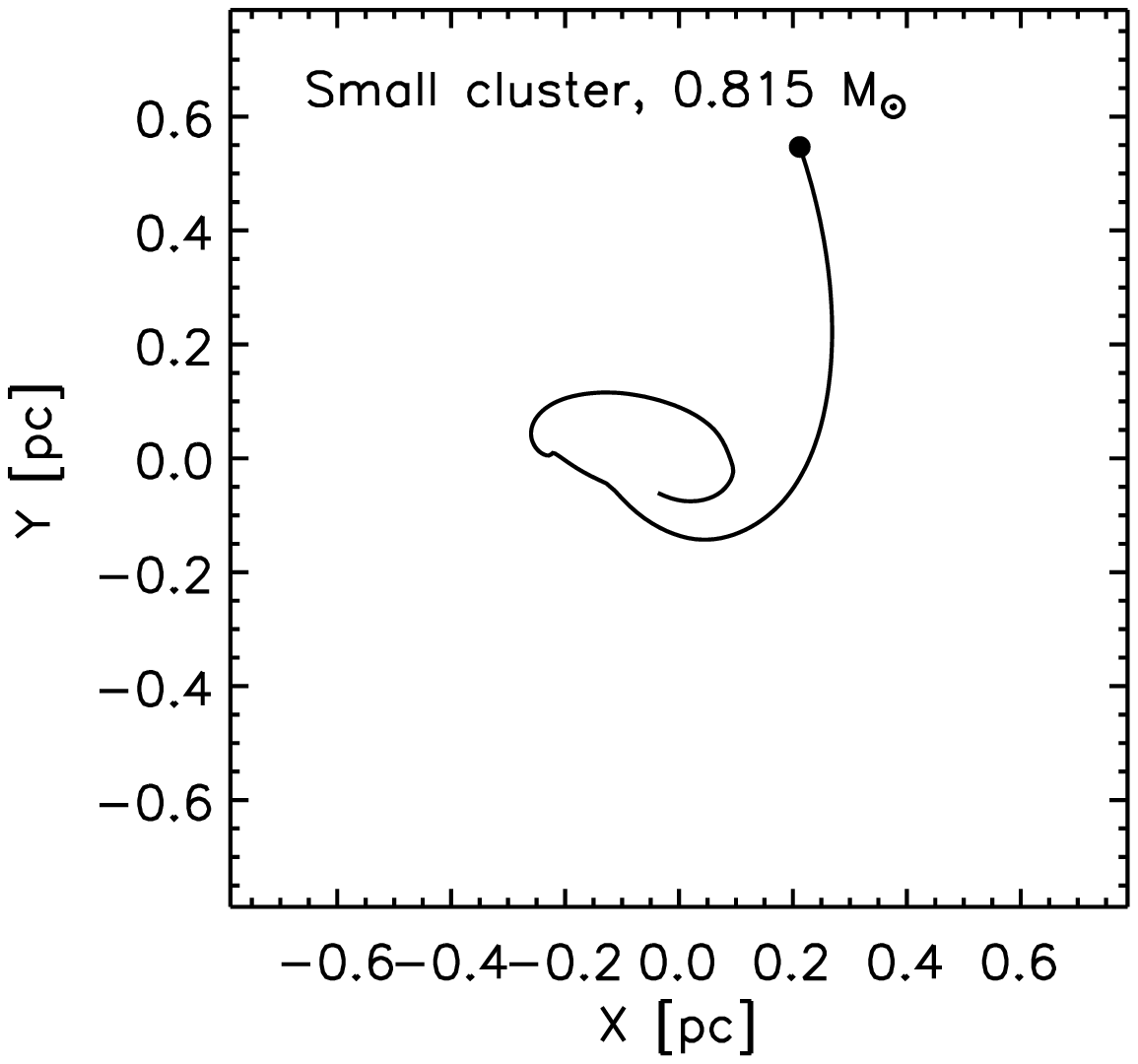}}
                           {\includegraphics*[ 85,60][354,335]{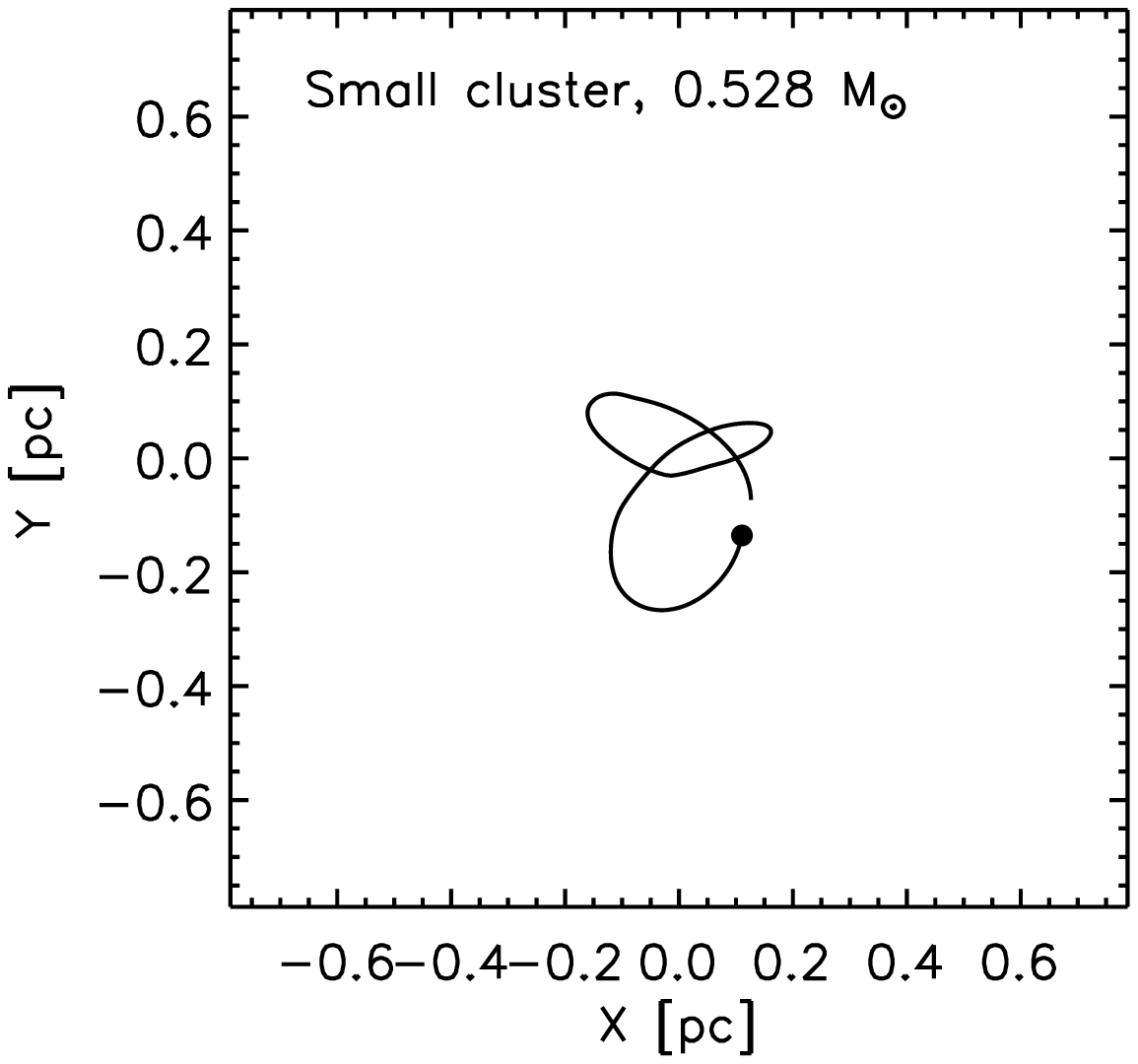}}}}
\centerline{\scalebox{0.55}{\includegraphics*[  0,60][354,335]{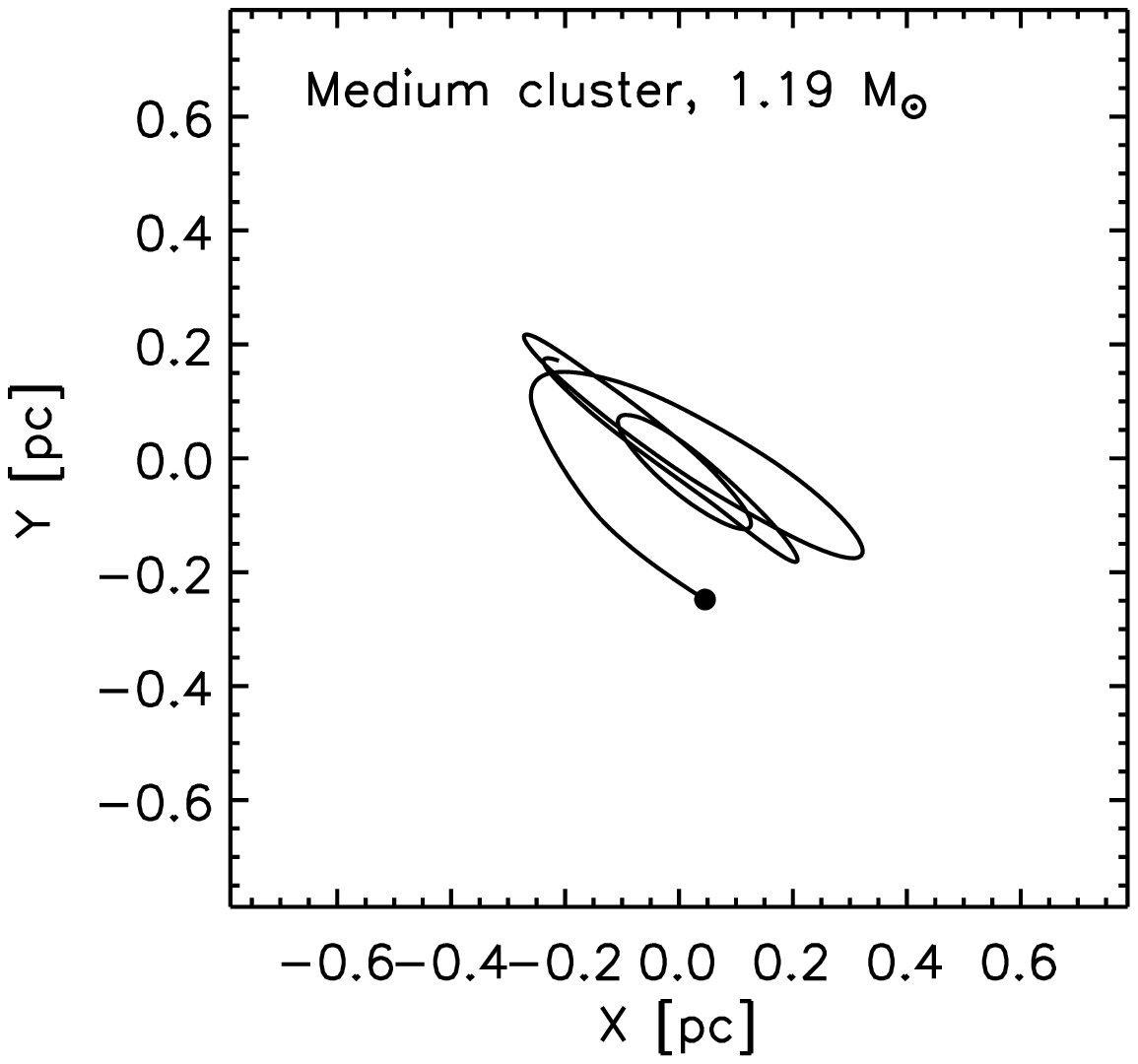}
                           {\includegraphics*[ 85,60][354,335]{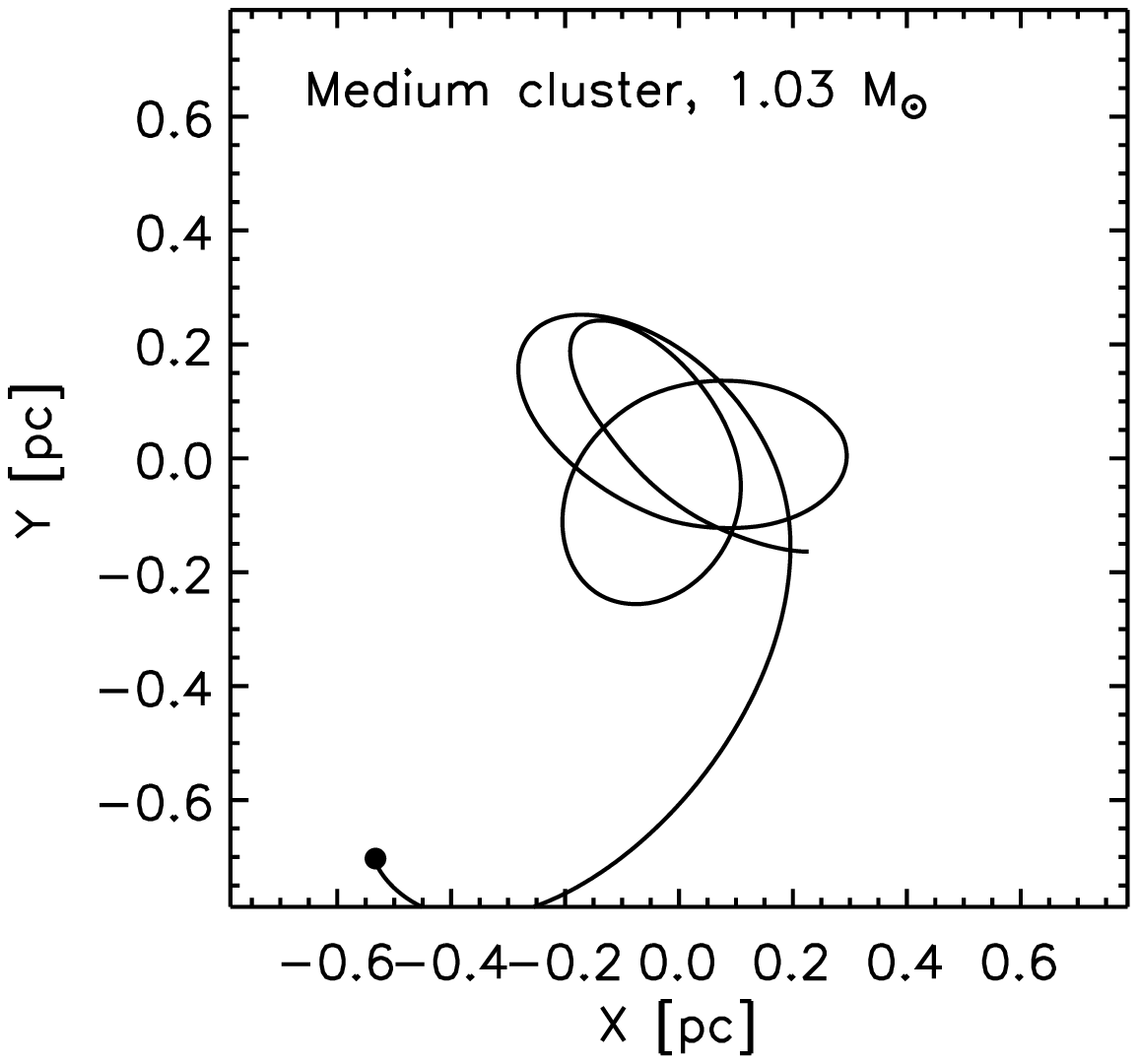}}
                           {\includegraphics*[ 85,60][354,335]{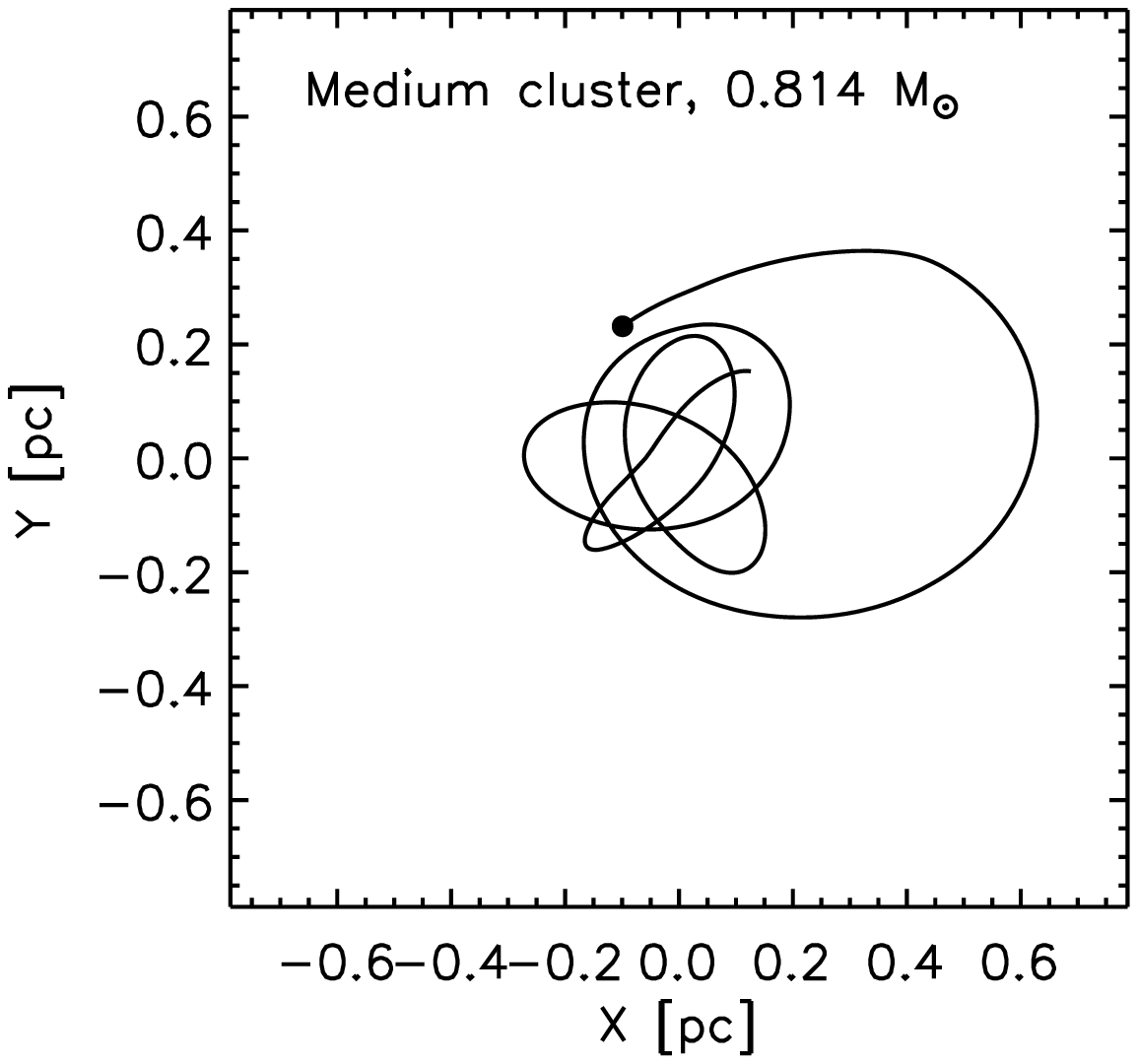}}}}
\centerline{\scalebox{0.55}{\includegraphics*[  0,20][354,335]{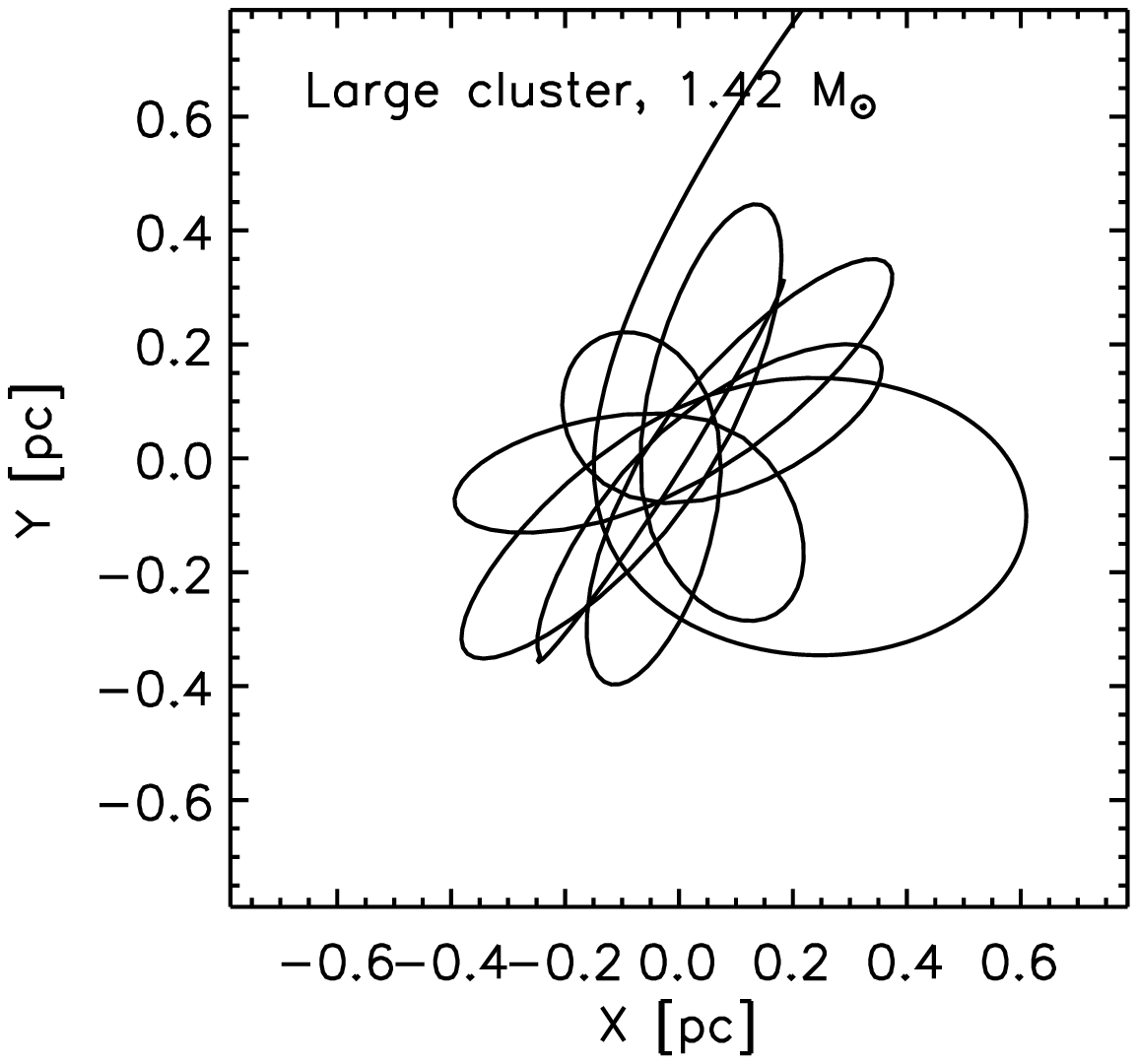}
                           {\includegraphics*[ 85,20][354,335]{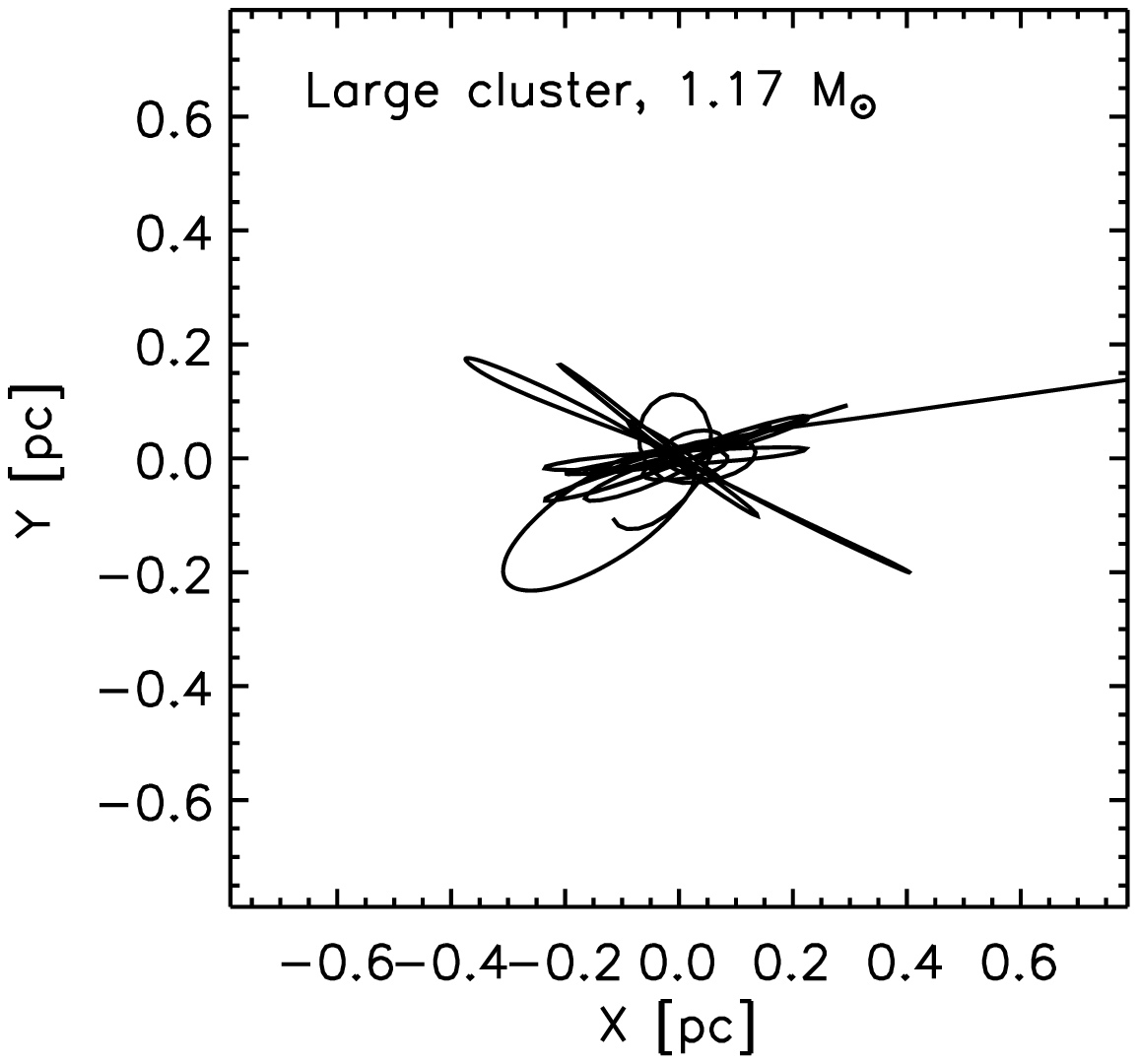}}
                           {\includegraphics*[ 85,20][354,335]{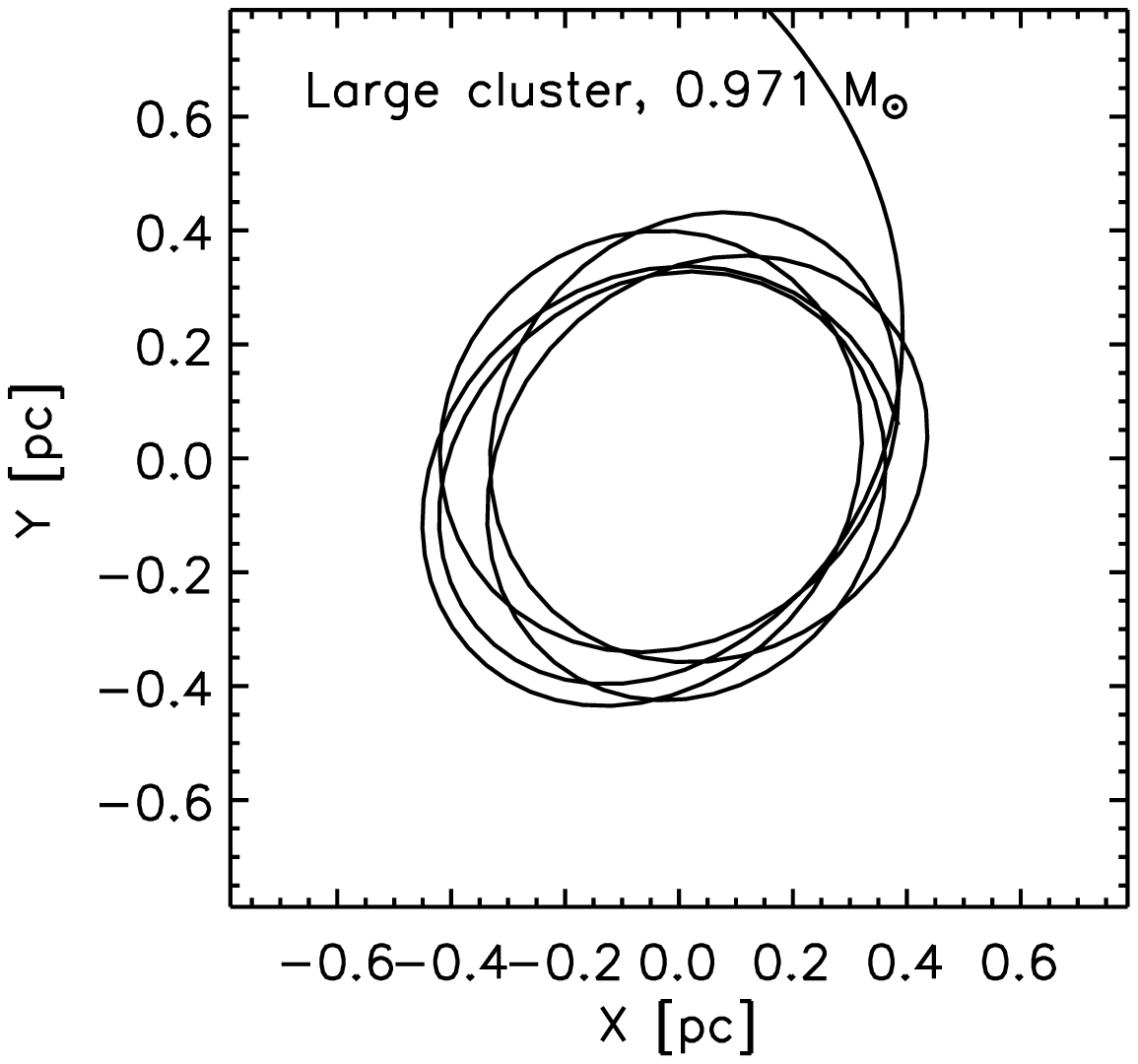}}}}
\caption{Stellar position for nine sample stars, taken from the three simulations.  Top row: three stars from {\sc
small} cluster.  Center: {\sc medium} cluster.  Bottom: {\sc large} cluster.  Stellar masses are indicated on
the plots, and range from $0.5-1.4~\msol$.  Each star's final position is shown with a filled circle.  Because the
orbits are projected into two dimensions, eccentricities appear exaggerated.}

\label{fig:example_position}

\end{figure}

\begin{figure}
\centerline{\scalebox{0.4}{\includegraphics*[ 0,50][478,340]{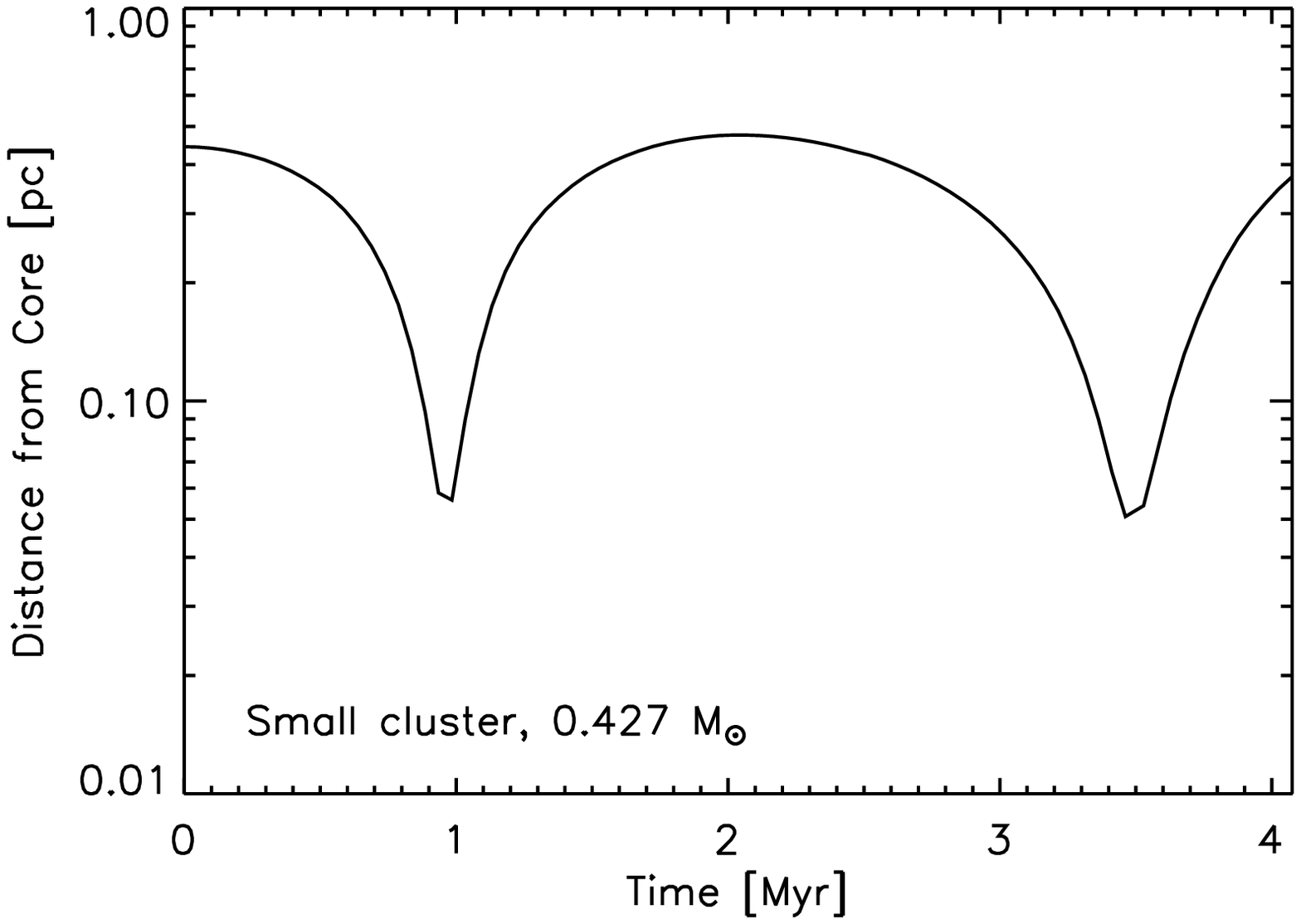}
                          {\includegraphics*[85,50][478,340]{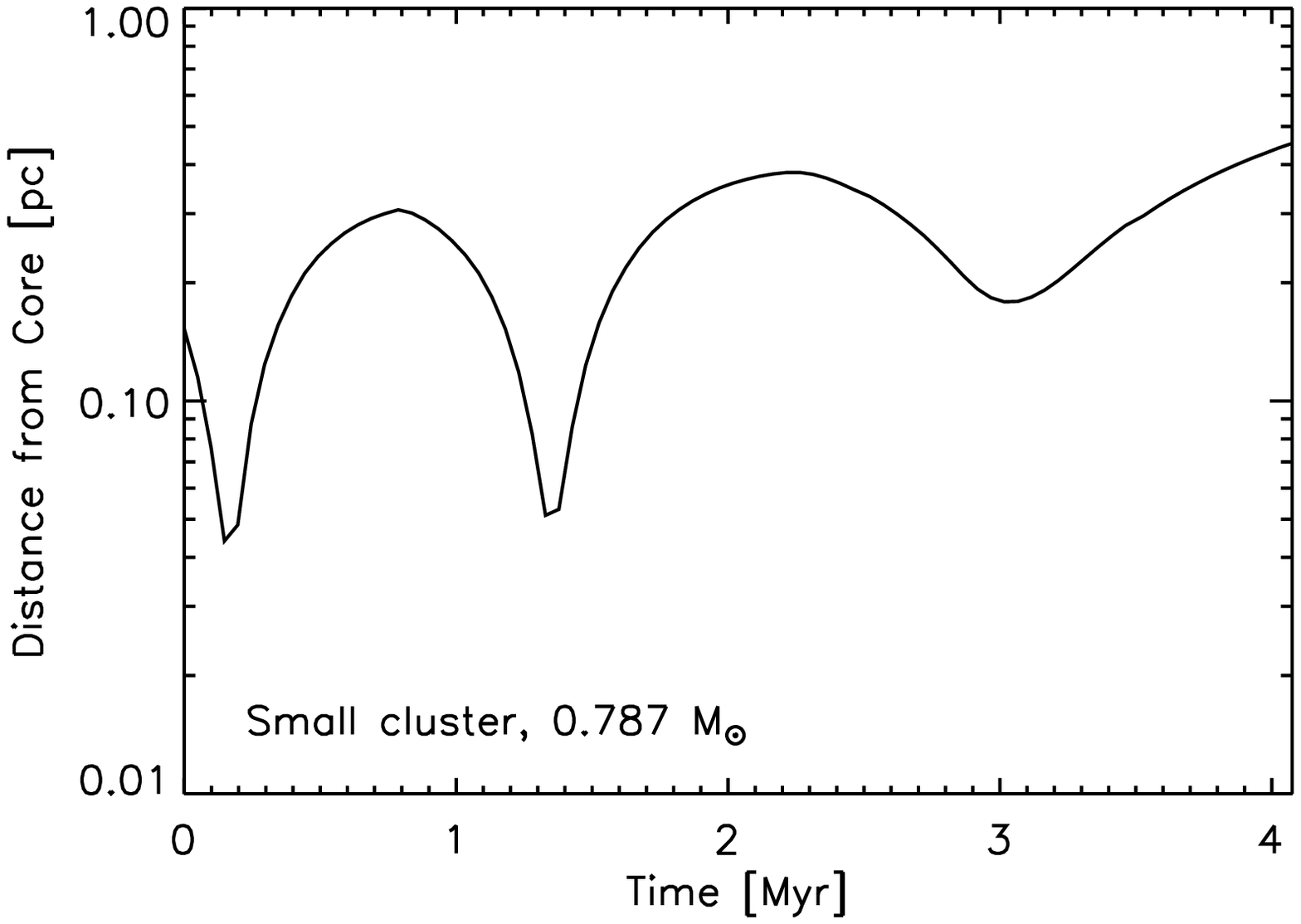}}
                          {\includegraphics*[85,50][478,340]{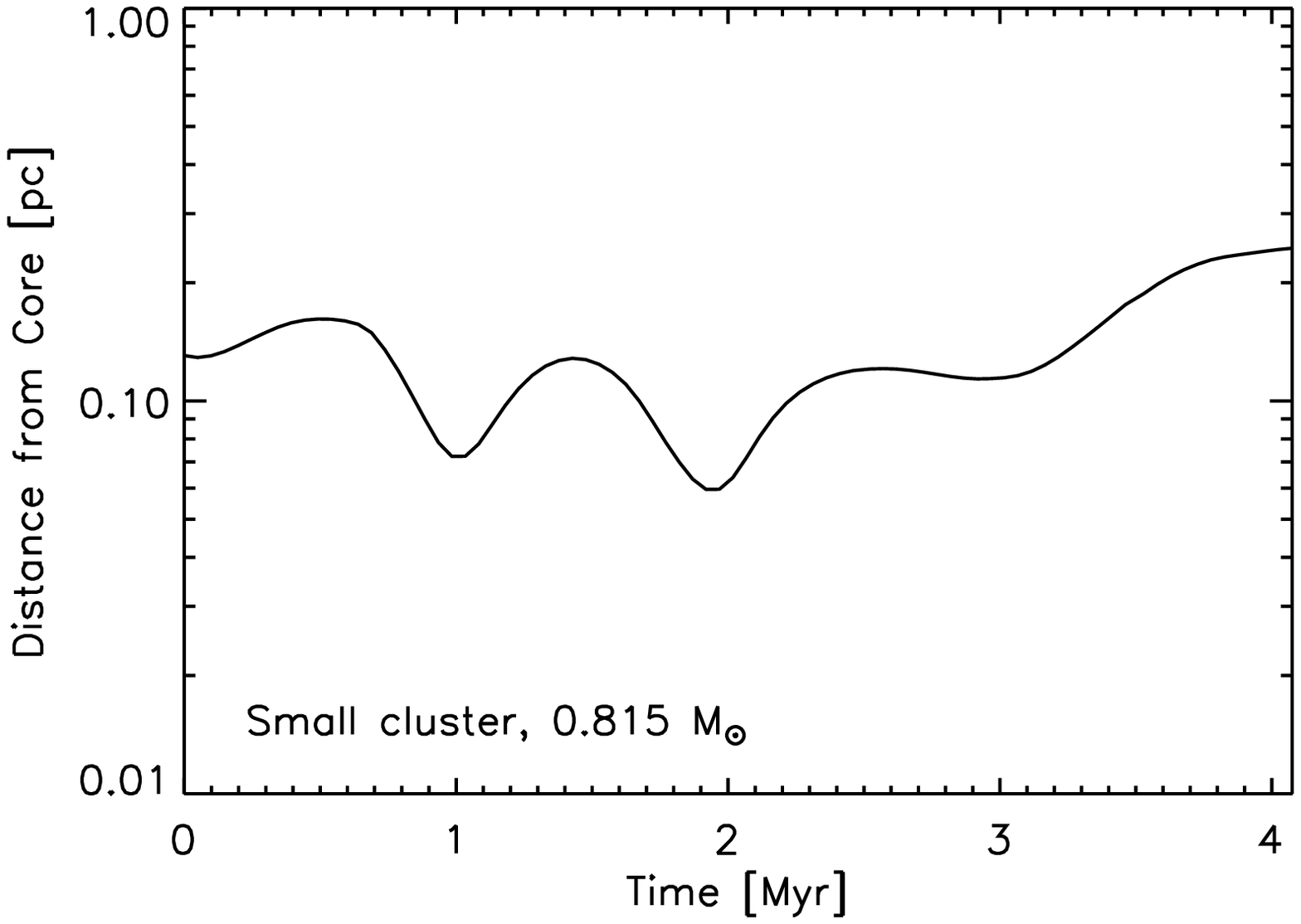}}}\hskip 0.6in}
\centerline{\scalebox{0.4}{\includegraphics*[ 0,50][478,340]{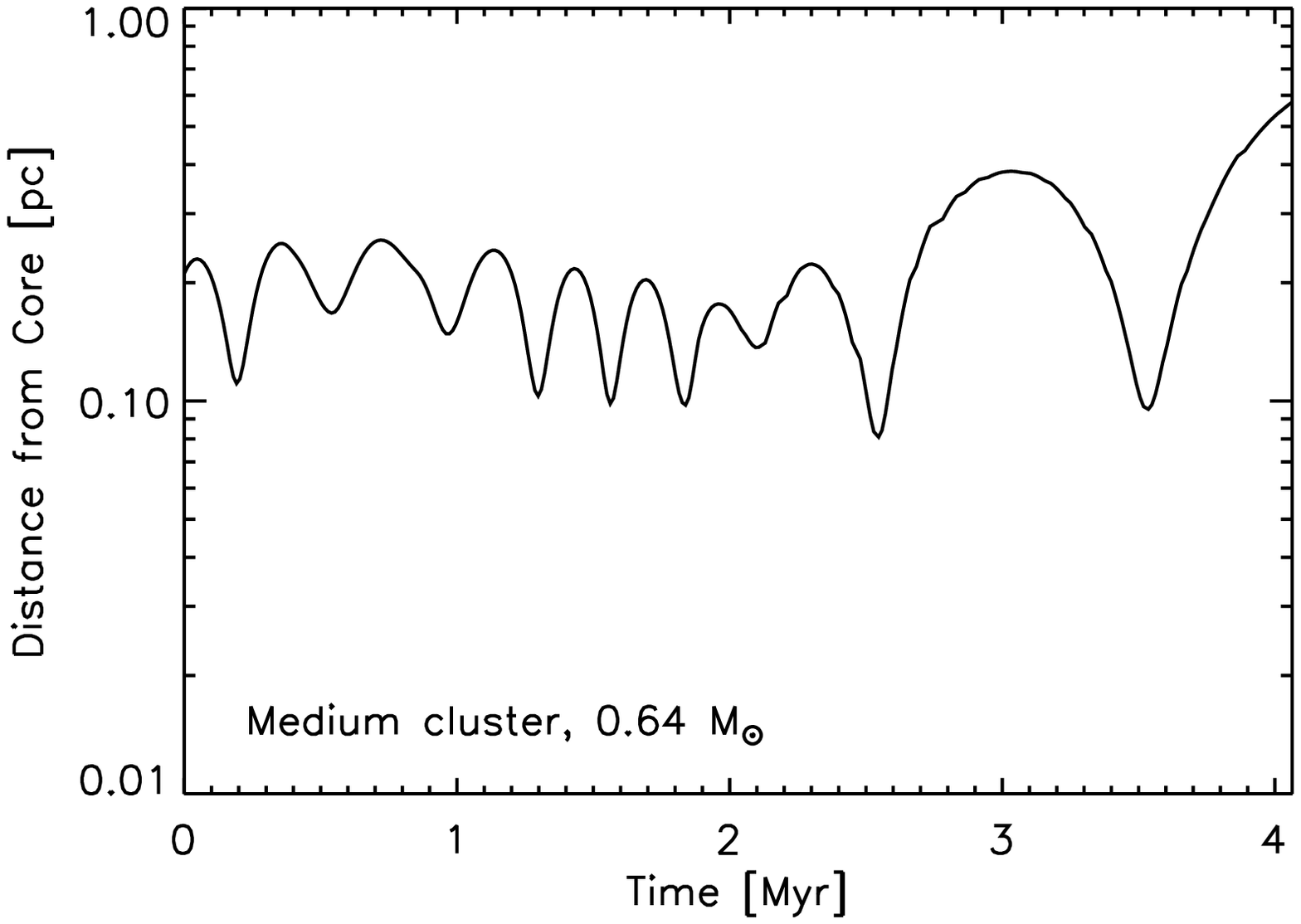}
                          {\includegraphics*[85,50][478,340]{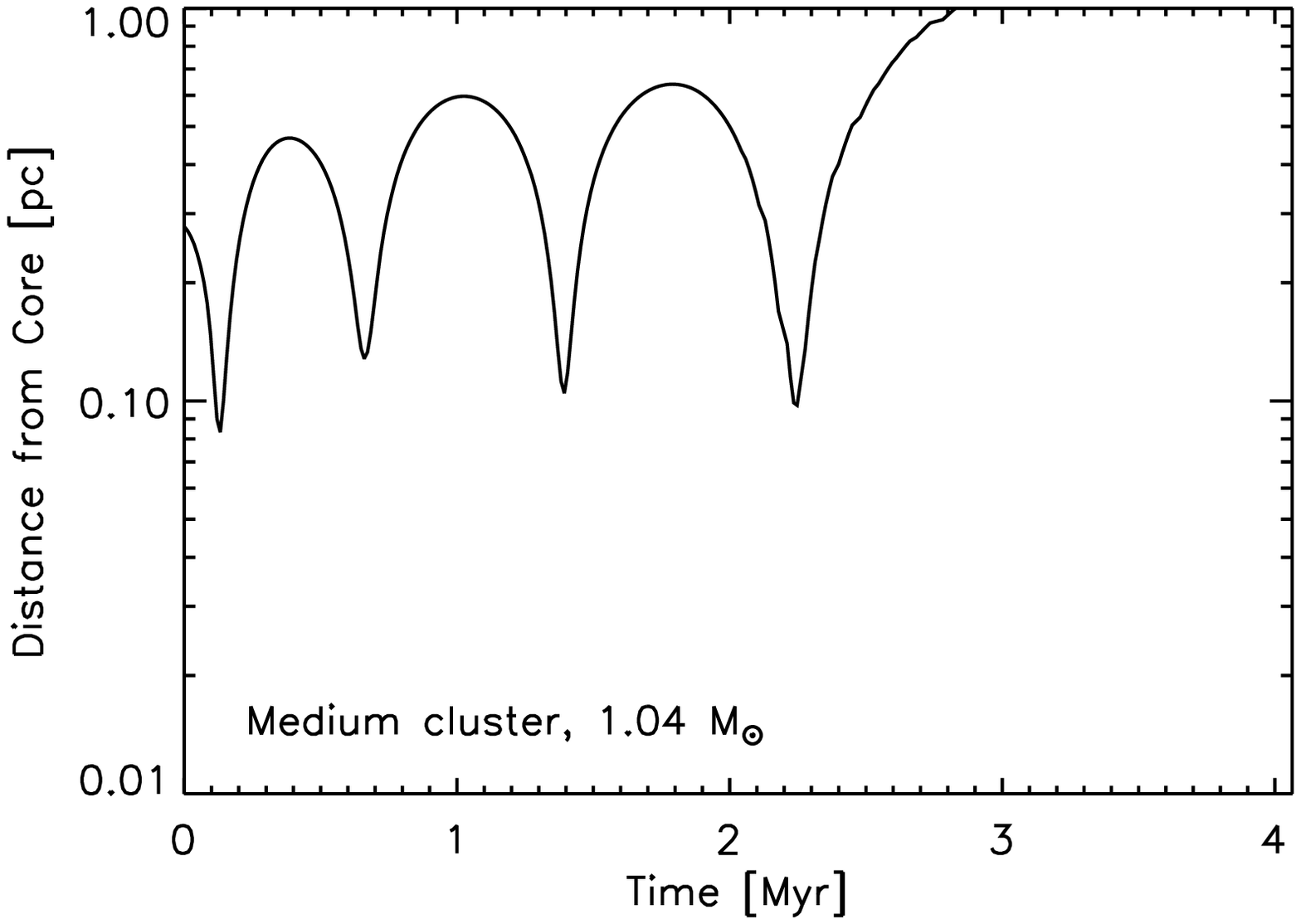}}
                          {\includegraphics*[85,50][478,340]{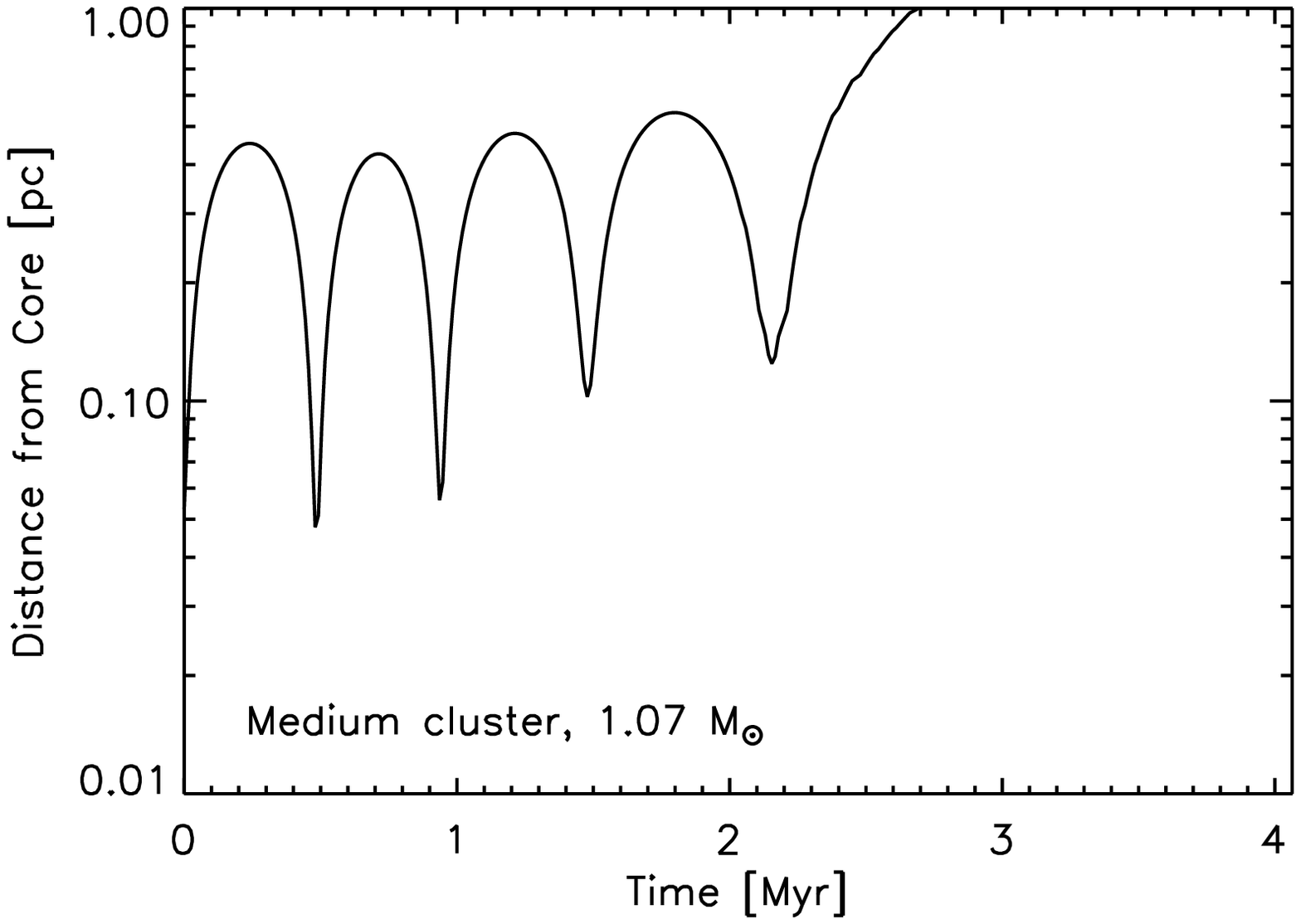}}}\hskip 0.6in}
\centerline{\scalebox{0.4}{\includegraphics*[ 0,00][478,340]{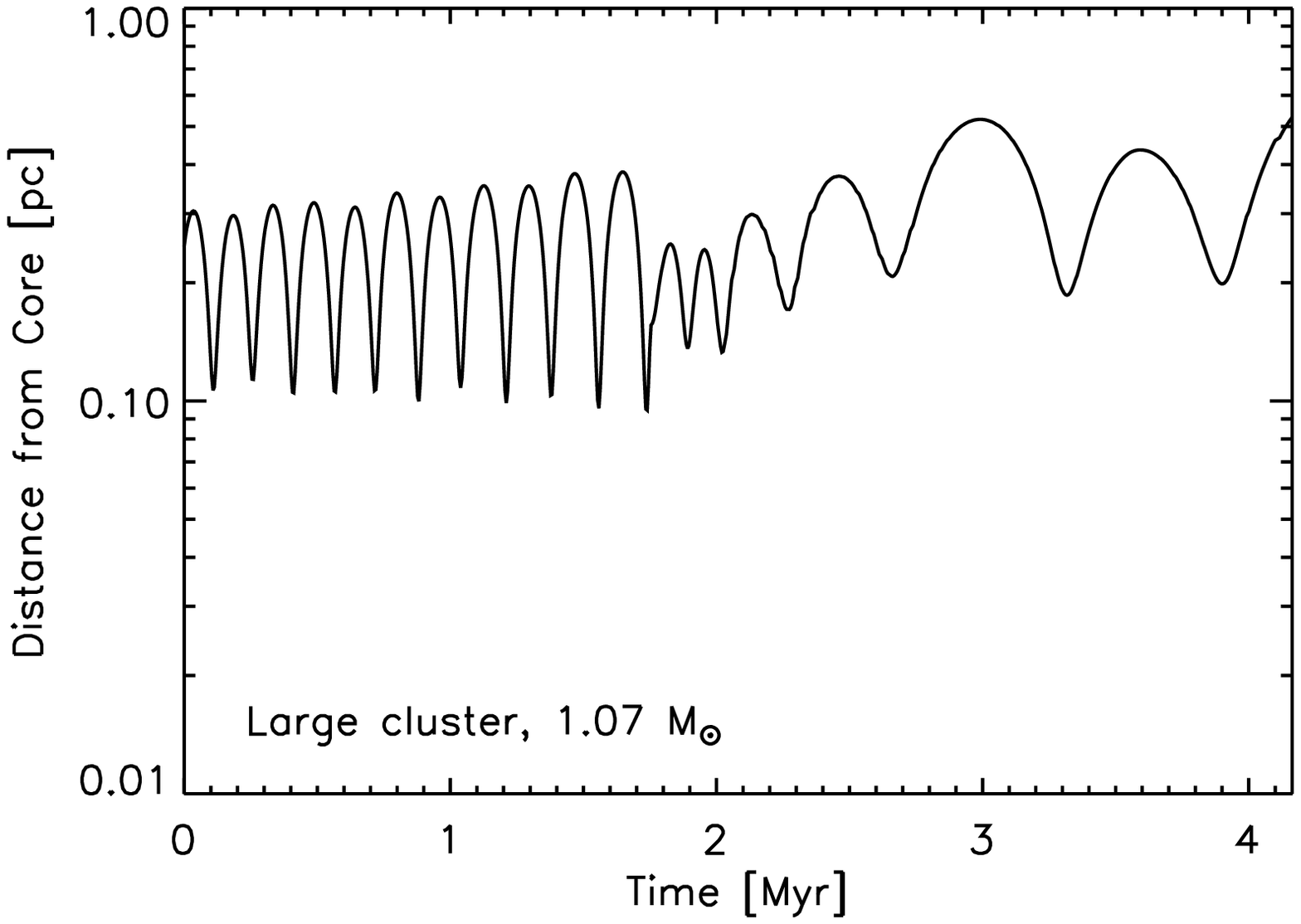}
                          {\includegraphics*[85,00][478,340]{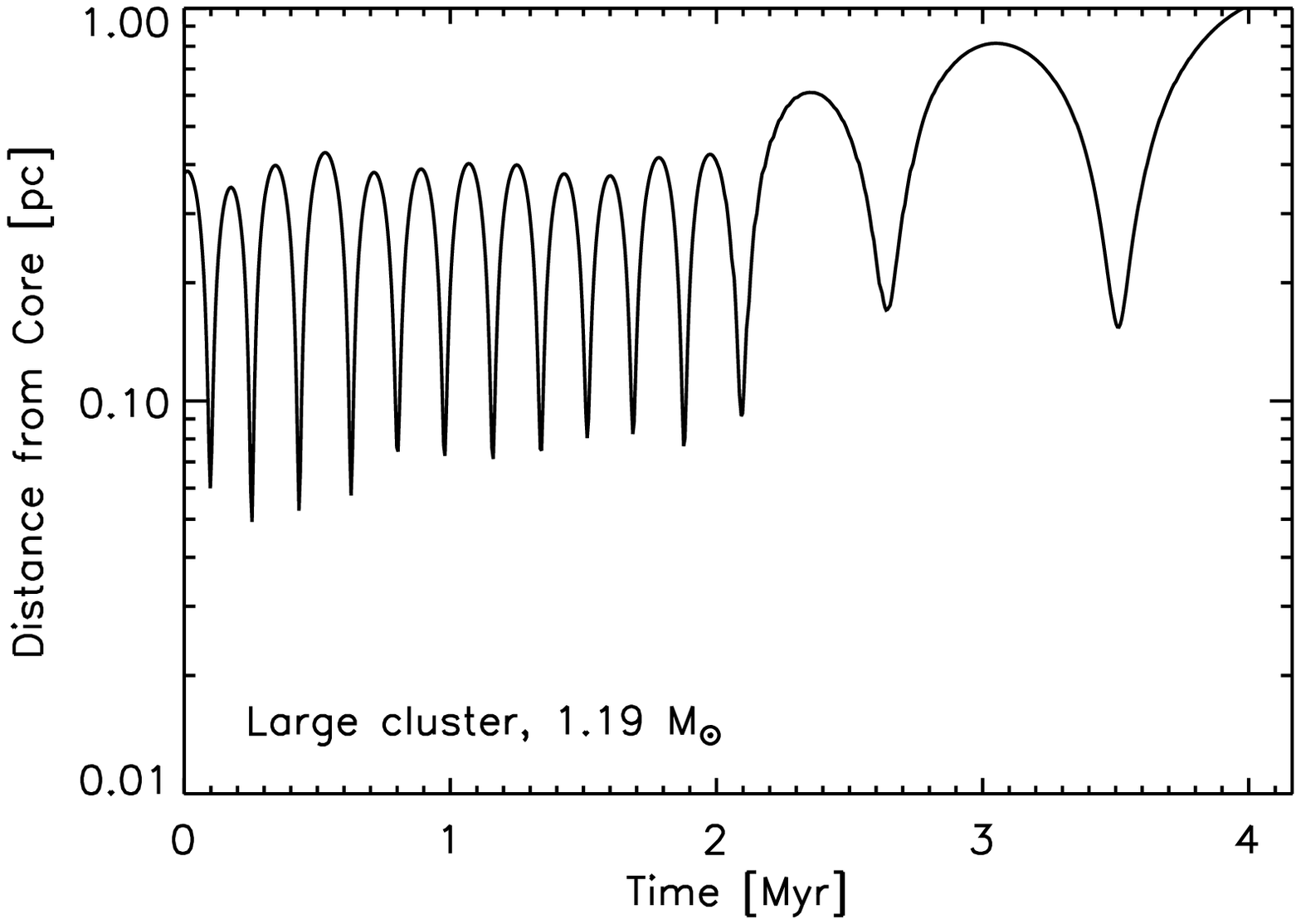}}
                          {\includegraphics*[85,00][478,340]{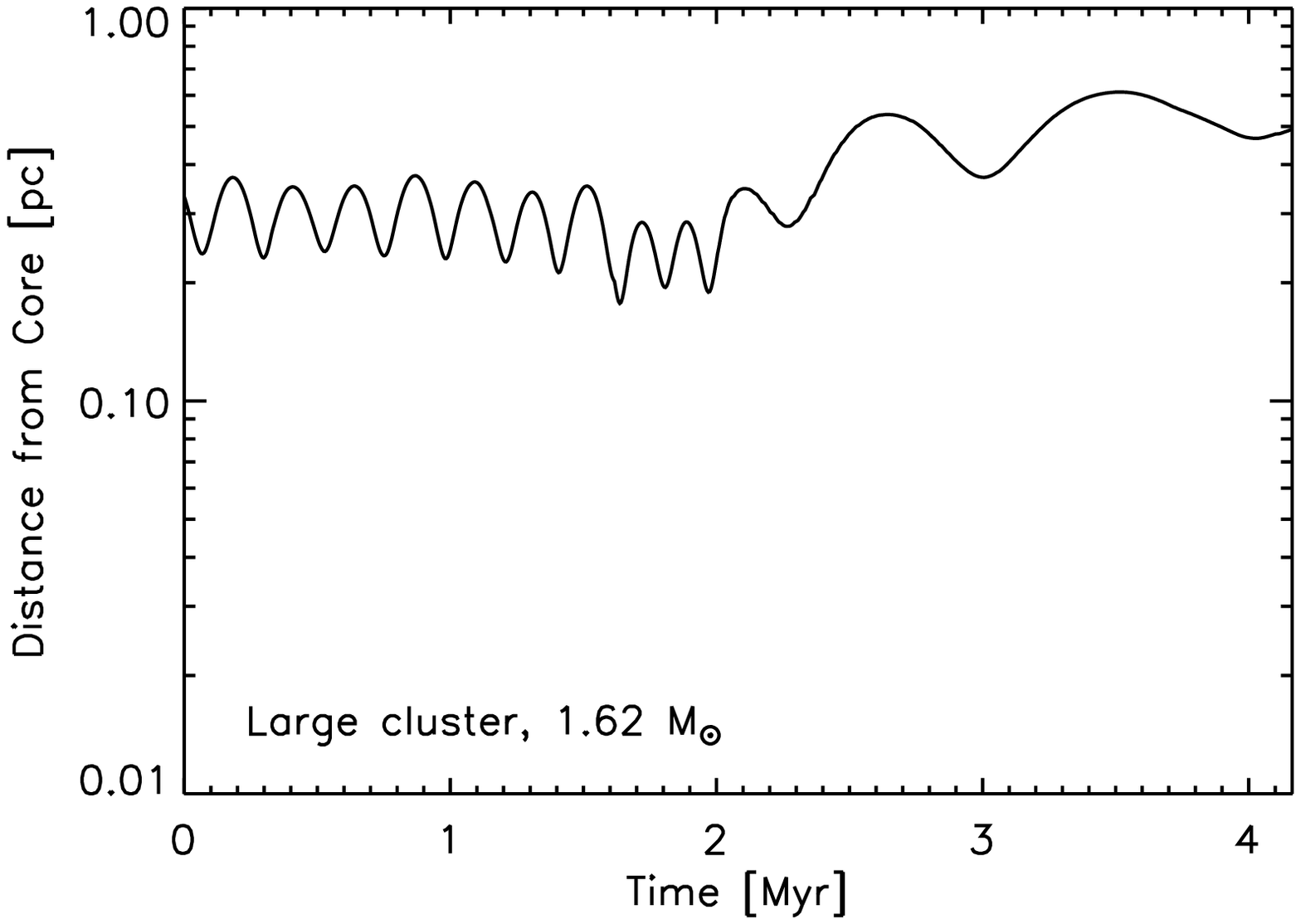}}}\hskip 0.6in}

\caption{Stellar distance from core.  Stars are the same as identified in Figure~\ref{fig:example_position}.  Stars
travel on highly eccentric orbits.  Orbits in the {\sc small} cluster are the slowest, because of the cluster's low
potential.  As the gas disperses, orbital periods and distances increase, and some stars become unbound.}

\label{fig:example_dist}
\end{figure}

Figures~\ref{fig:example_position} and \ref{fig:example_dist} show the positions and distances for nine stars: three
typical examples from each of the three clusters.  Each of the stars takes brief, rapid passages through the cluster
core at $< 0.1\ \rm{pc}$, but spends most of the time at apoapse near $0.5\ \rm{pc}$.  For the first 2~Myr, most stars
in the clusters move on eccentric orbits with $e>0.7$, calculated by measuring periapse and apoapse on a typical orbit.
Stars in the {\sc small} cluster orbit slowly due to the low central potential, and stars have only $\sim1$ crossing
period before the gas is lost.  These stars' orbits are most easily perturbed by other stars, because the central gas
potential is low relative to the stars' individual mass.  In comparison, the {\sc large} cluster has a higher mass and
thus shorter orbital periods, so stars have $\sim10$ crossings during the simulations.  These stars follow more regular
orbits and are less frequently perturbed than stars in the {\sc Small} cluster.  The {\sc medium} cluster is bounded by
the other two.  Expansion of the clusters is clearly evident in all three runs: when gas begins to be lost at 2~Myr,
stellar orbits expand, and their periods increase.  Because the gravitational potential is not a point mass, stellar
orbits are not keplerian and tend to precess, most visibly in the {\sc large} cluster.  As the gas is lost, 
orbits become less periodic, and some stars are lost from the cluster.

\begin{figure}
\centerline{\scalebox{0.4}{\includegraphics*[ 0,50][418,340]{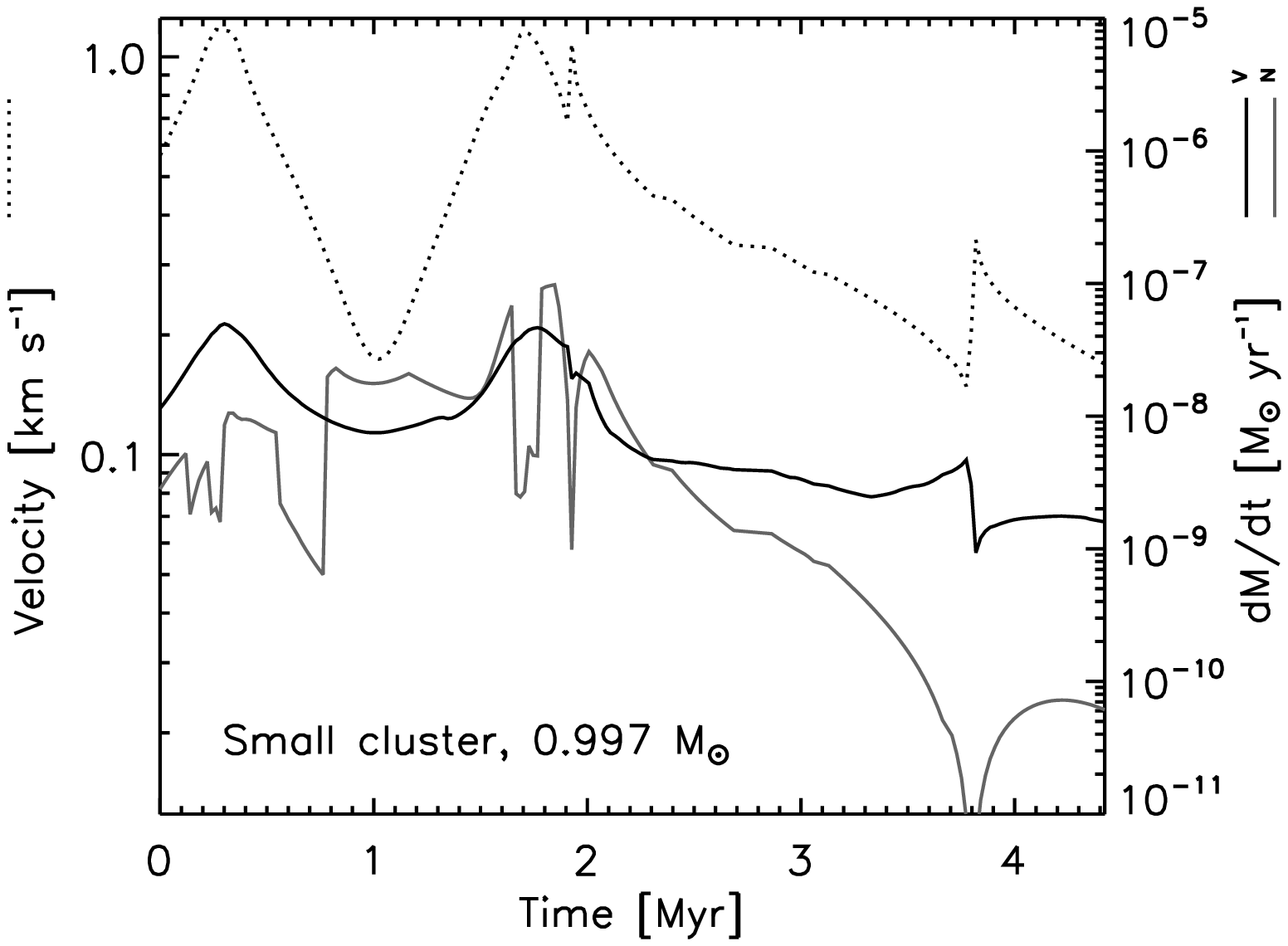}
                          {\includegraphics*[85,50][418,340]{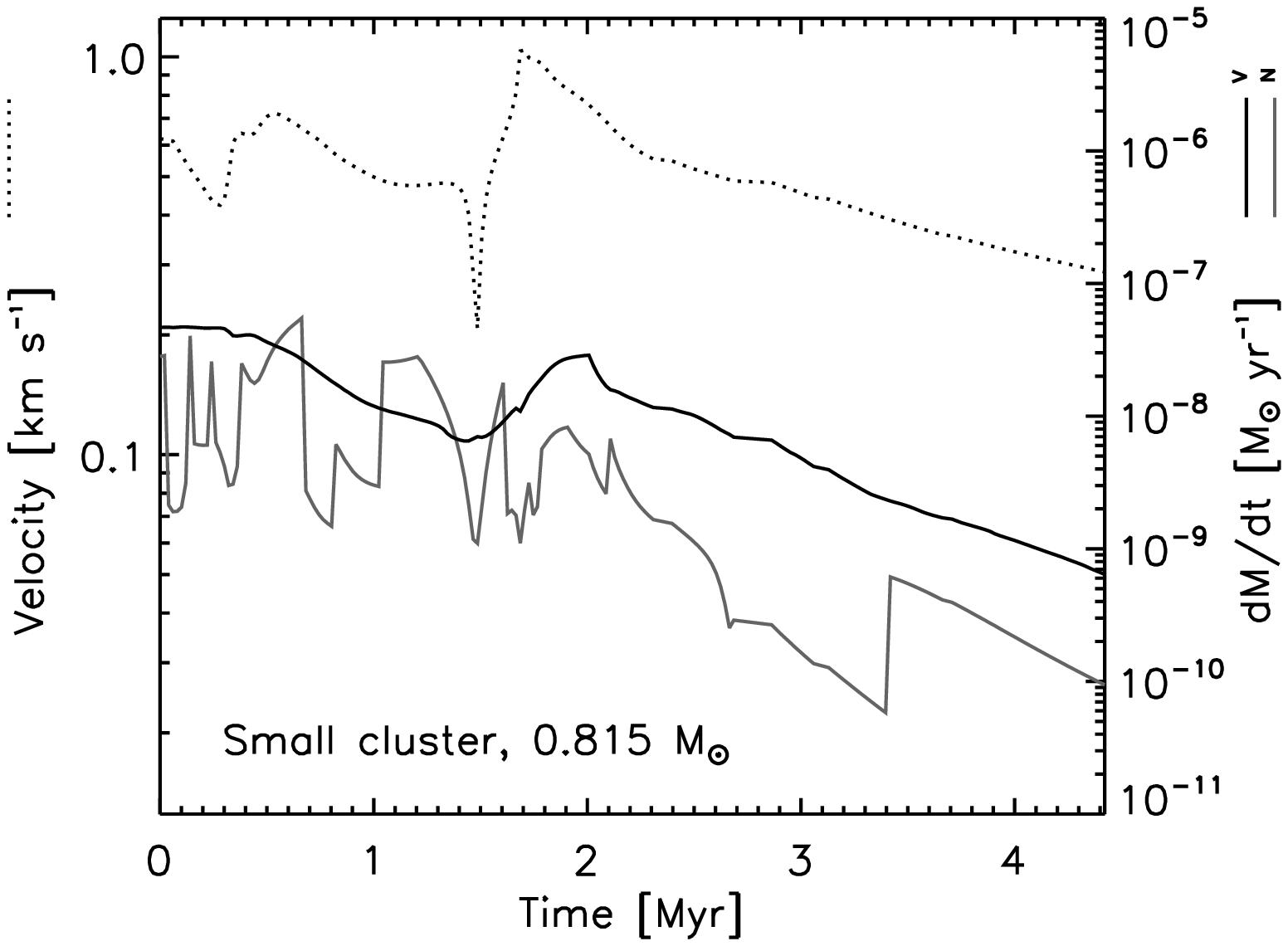}}
                          {\includegraphics*[85,50][478,340]{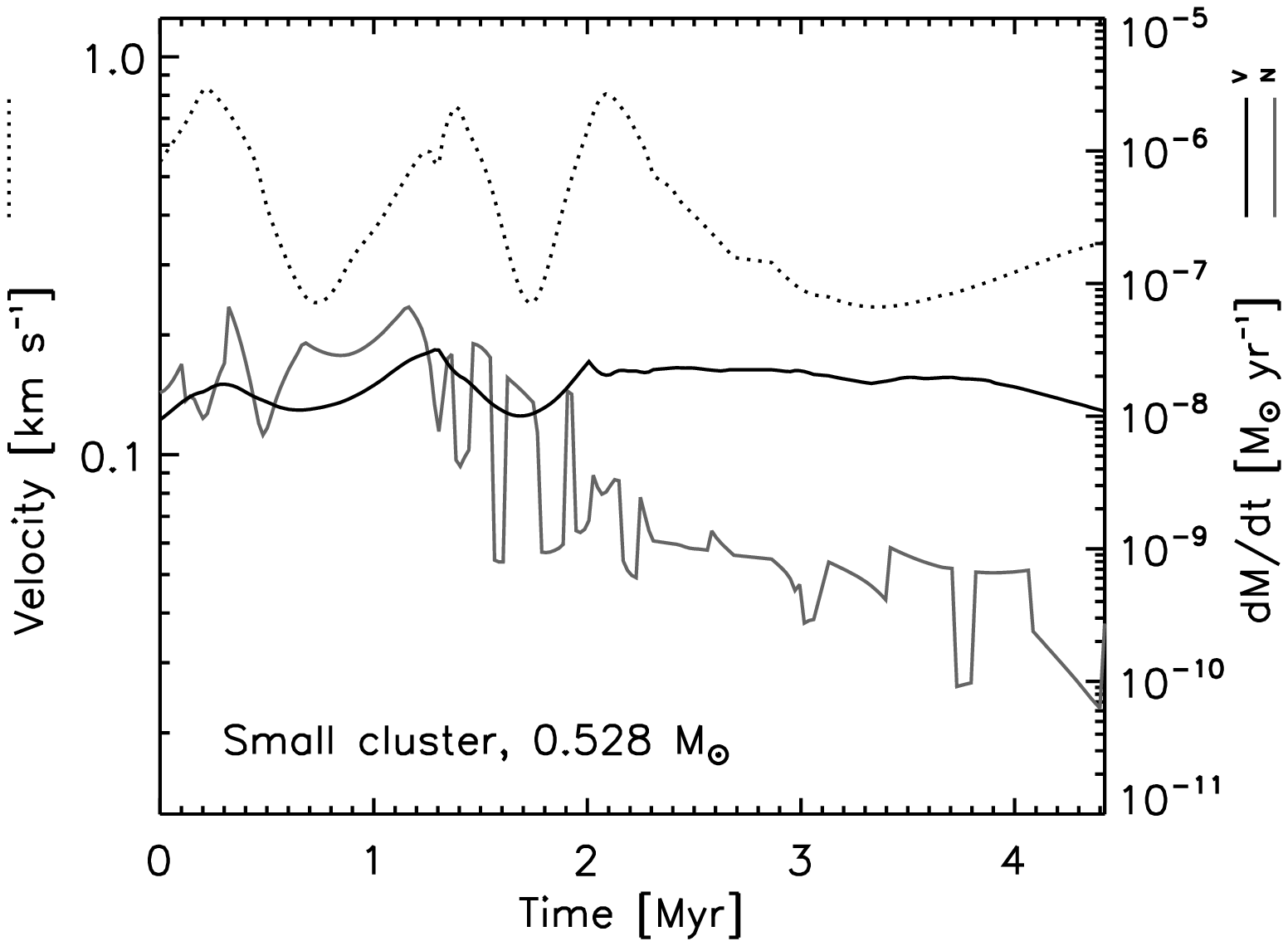}}}\hskip 0.6in}
\centerline{\scalebox{0.4}{\includegraphics*[ 0,50][418,340]{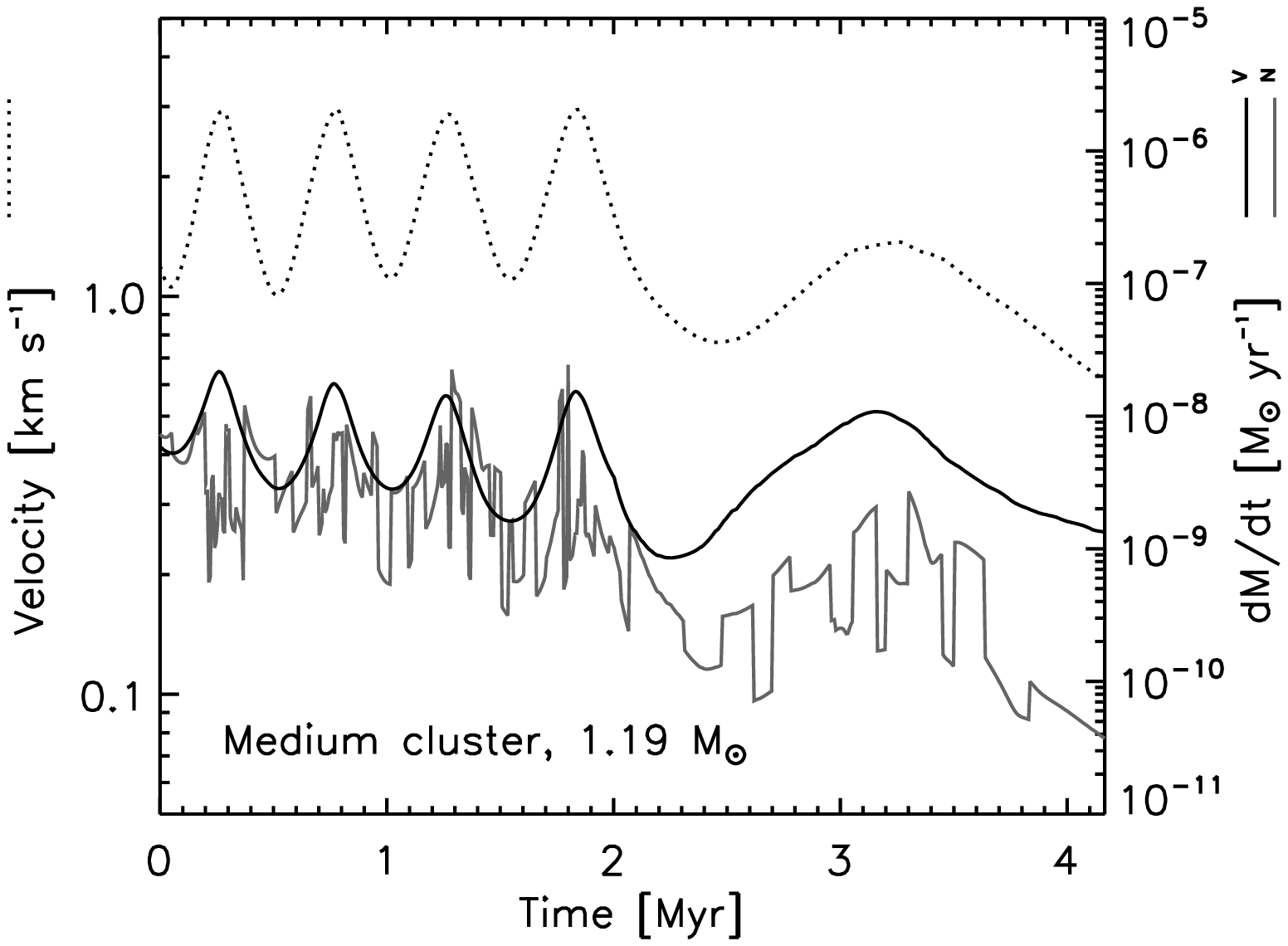}
                          {\includegraphics*[85,50][418,340]{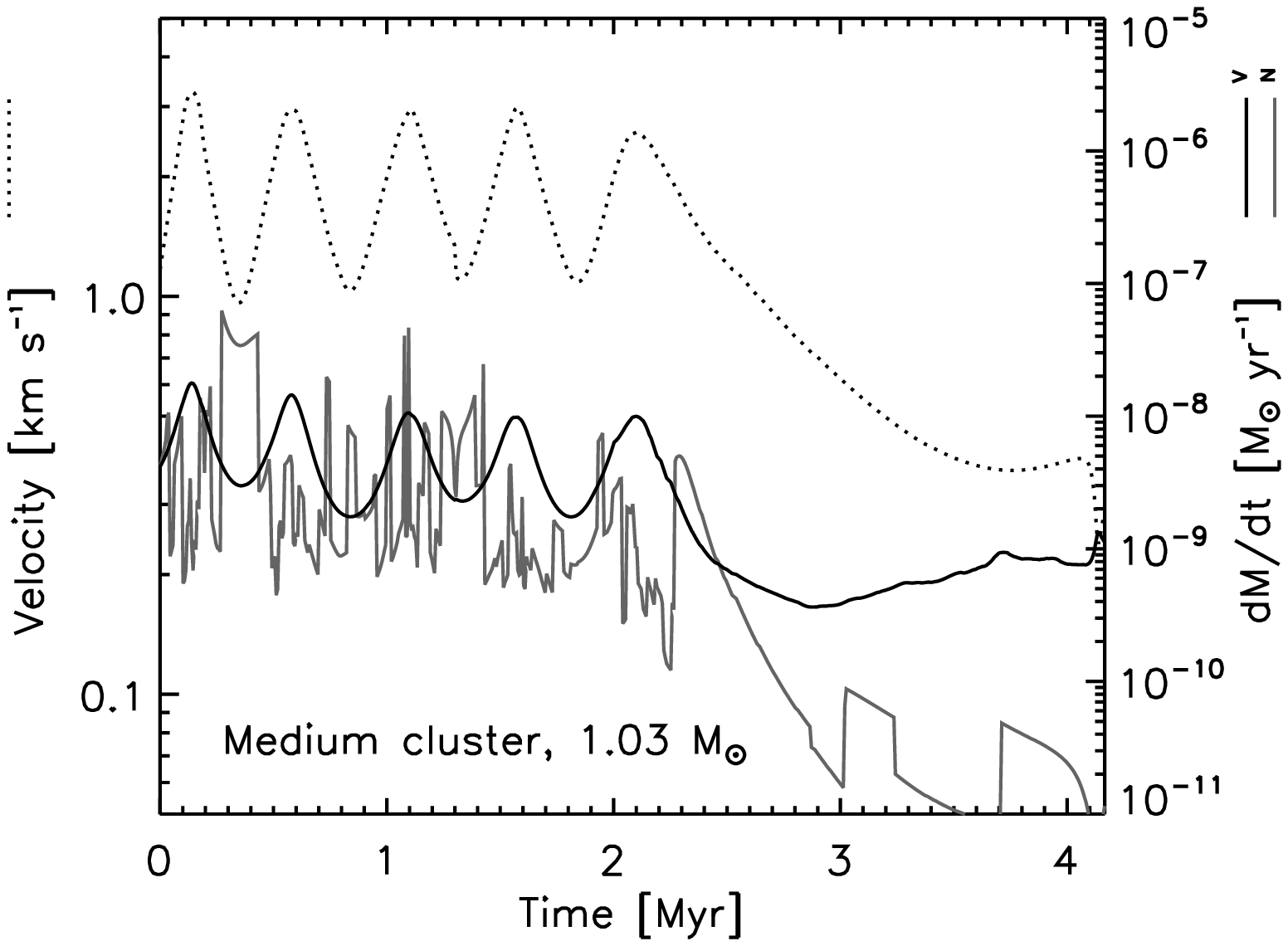}}
                          {\includegraphics*[85,50][478,340]{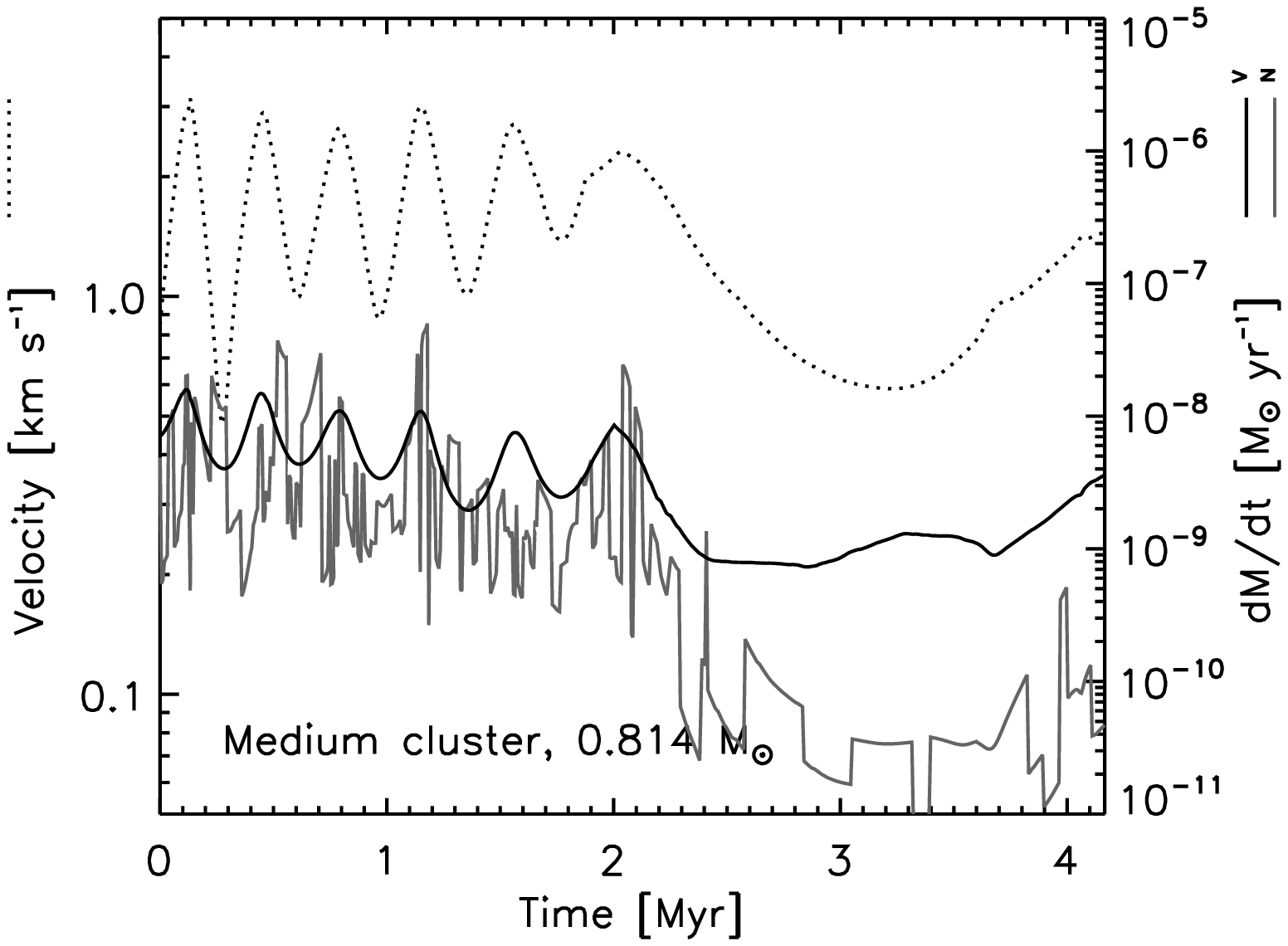}}}\hskip 0.6in}
\centerline{\scalebox{0.4}{\includegraphics*[ 0,00][418,340]{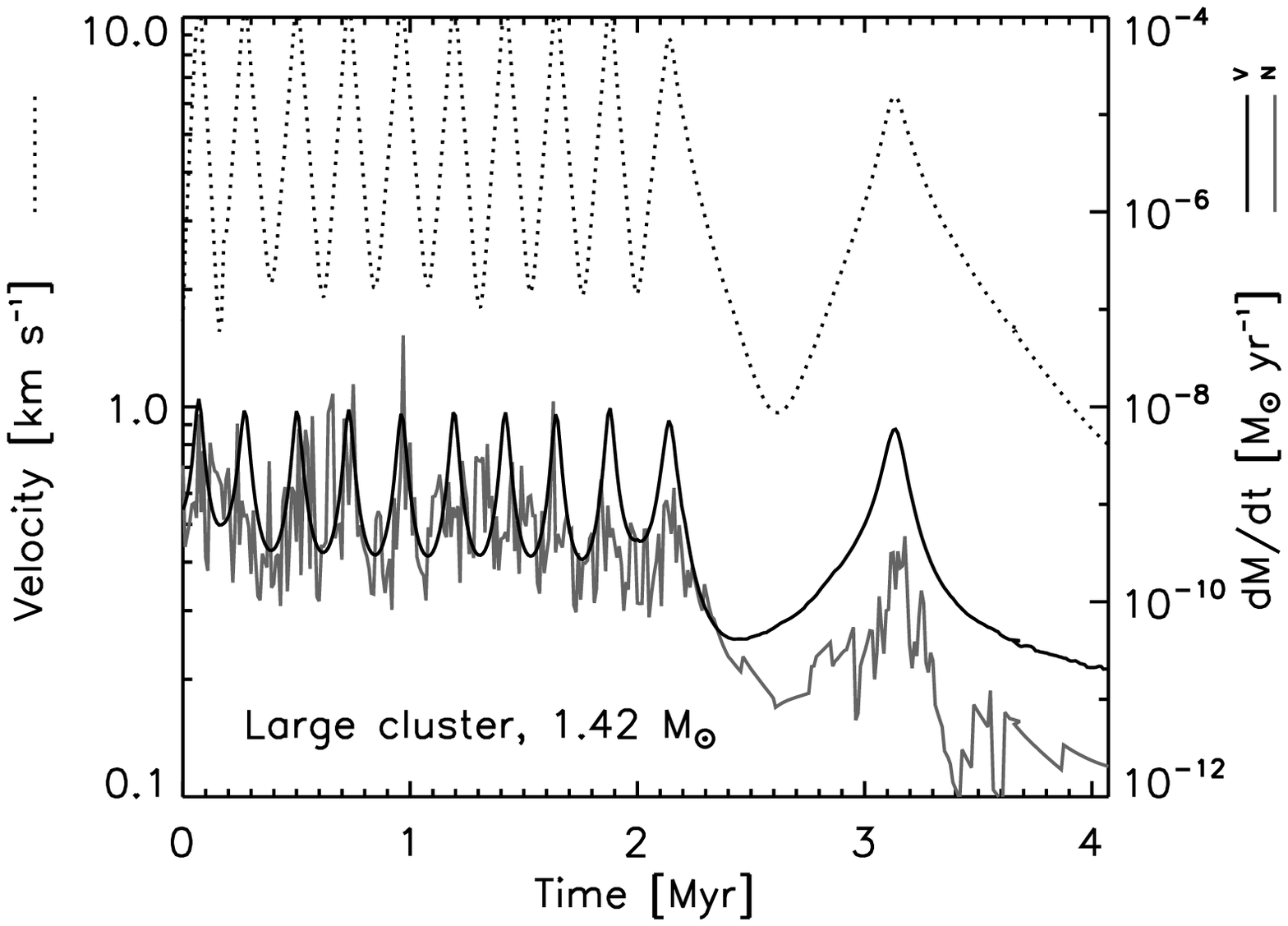}
                          {\includegraphics*[85,00][418,340]{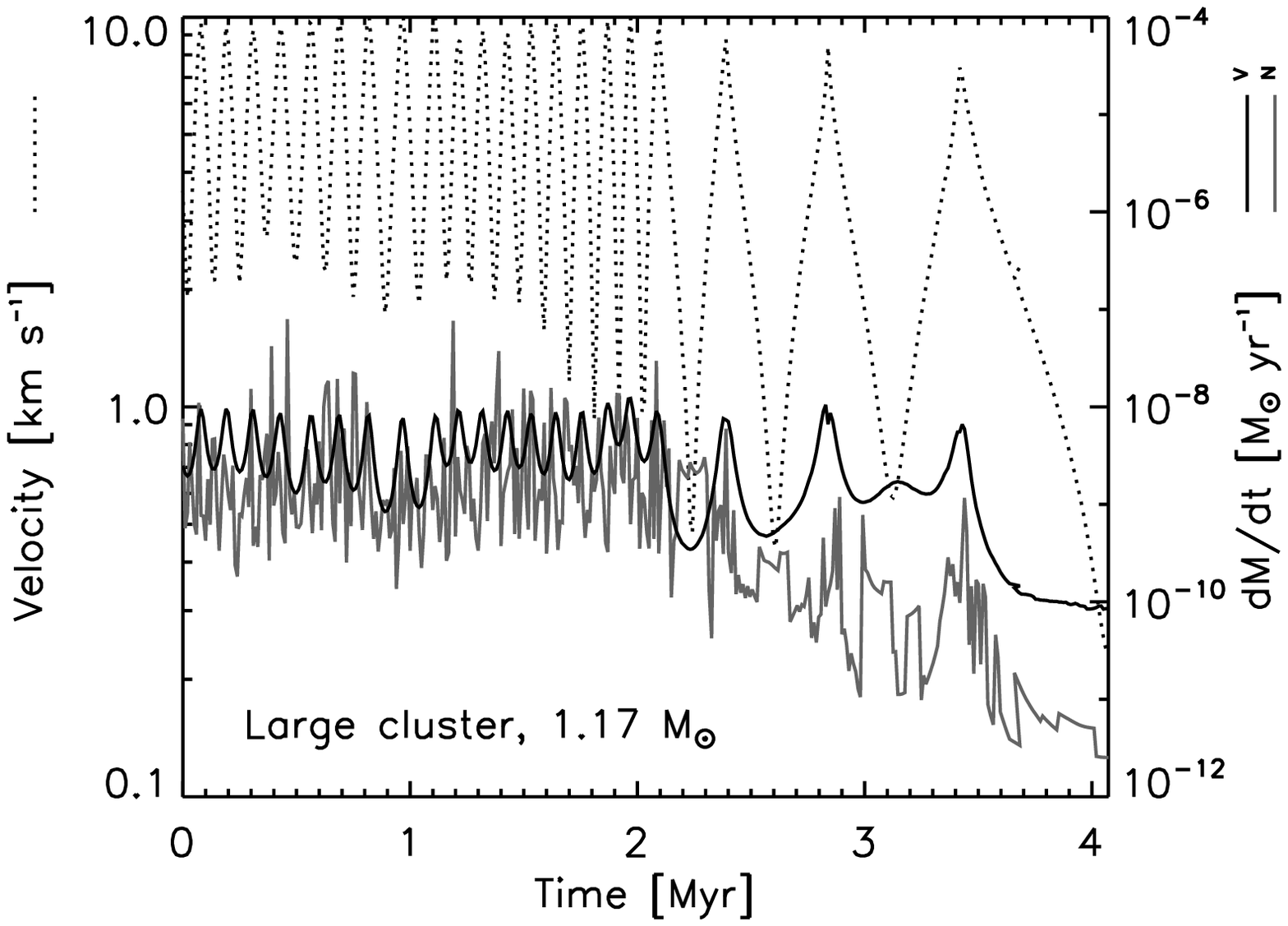}}
                          {\includegraphics*[85,00][478,340]{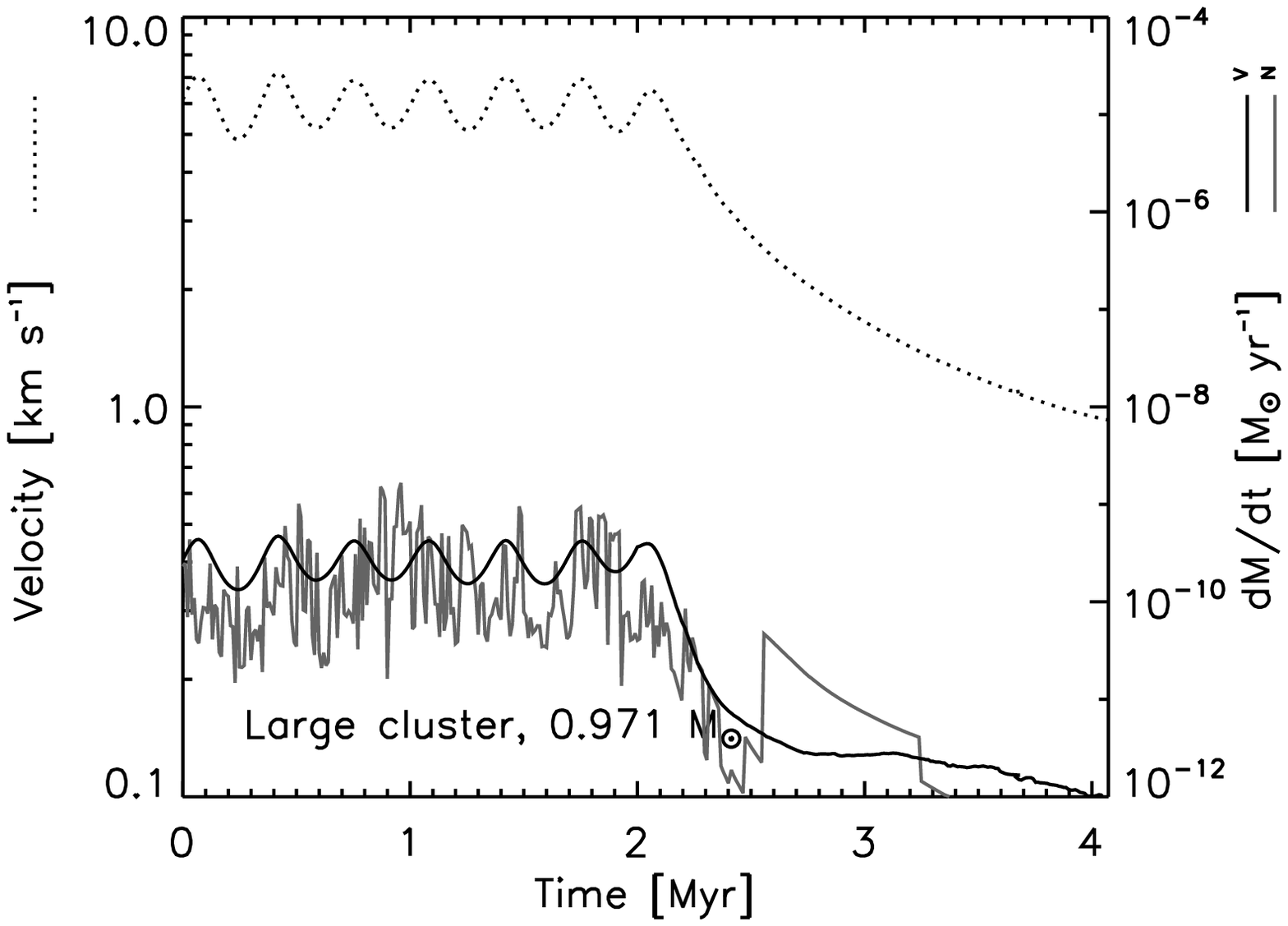}}}\hskip 0.6in}
\caption{Stellar velocity and BH accretion rate \dmdtbh.  \textbf{Dotted lines:} stellar velocity.  Velocity increases 
as stars approach the cluster core, and is fastest for the {\sc large} cluster.  \textbf{Black lines:} BH accretion
rate, based on the `virial' velocity method (see text).  Accretion rate usually correlates with velocity,
because regions near the core have a higher gas density.  Accretion slows after gas loss begins at 2~Myr, due to reduced
gas density and larger stellar orbits.  Small high-frequency perturbations are due to close encounters with other stars.
Stars are the same as identified in Figure~\ref{fig:example_position}. \textbf{Grey lines:} BH accretion rate, based on
the `nearest neighbor' velocity method.  This is more accurate, but the correlation with stellar velocity is not as
obvious.}
\label{fig:example_velocity_dmdt}
\end{figure}

Stellar velocities $v_*$ for the same nine stars are plotted in Figure~\ref{fig:example_velocity_dmdt}.  Velocities are
lowest in the {\sc small} cluster ($0.1 - 1\ \kms$) and highest in the {\sc large} ($1-10\ \kms$), because of the higher
gravitational potential.  Velocities increase at periapse and are slowest at apoapse.  Because of the dispersed central
potential, the velocity change through a star's orbit is not as strong as for regular keplerian motion.  As the gas
disperses, the velocities decrease.


The accretion rate \dmdtbh\ is also plotted in Figure~\ref{fig:example_velocity_dmdt}.  The two curves of \dmdtbh\
assume two different methods (eqs.~\ref{eq:v_v}--\ref{eq:v_n}) of transforming the stellar velocity $v_*$ into a
stellar-gas velocity $v$ to be used to calculate \rbh.  The curve for $v_v$ (virial method) is smooth because the virial
velocity changes only slowly.  On the other hand, $v_n$ (nearest-neighbor method) changes rapidly.  The values for $v_n$
are similar to $v_v$, but with a high-frequency noise added due to the rapidly changing velocity of a star's actual
nearest neighbor.

Typical accretion rates are $10^{-9}$ to $10^{-8}~\msolyr$ in the clusters.  The {\sc small} cluster has the lowest gas
densities (which decreases $\mdotbh$), but also has the lowest stellar velocities (which increases $\mdotbh$).  These
two terms nearly cancel out: accretion in the {\sc large} cluster is only a factor of 3 lower than the {\sc small} case,
even though the gas densities differ by a factor of 100.

The change in accretion rate through an orbit can be most easily seen in the `virial' \dmdtbh\ plots.  The accretion
rate for an individual star changes by up to a factor of 10 through its orbit.  The highest accretion rates usually are
near the cluster core, where the density is highest.  Velocity is also highest at the core, but the effect of the higher
density dominates.  The accretion rate for all stars drops as gas is lost from the cluster, although the drop in gas
density is (once again) partially offset by the lower velocity.

The `neighbor' \dmdtbh\ curves show the same effect, but the accretion in effect has a noise added to it due to the
velocities of its rapidly changing nearest neighbors.  The accretion peak often occurs near periapse, but can occur
elsewhere if a star has a fortuitous encounter with a slow-moving neighbor star.  



\begin{figure}
\centerline{\scalebox{0.37}{\includegraphics*[ 0,50][488,340]{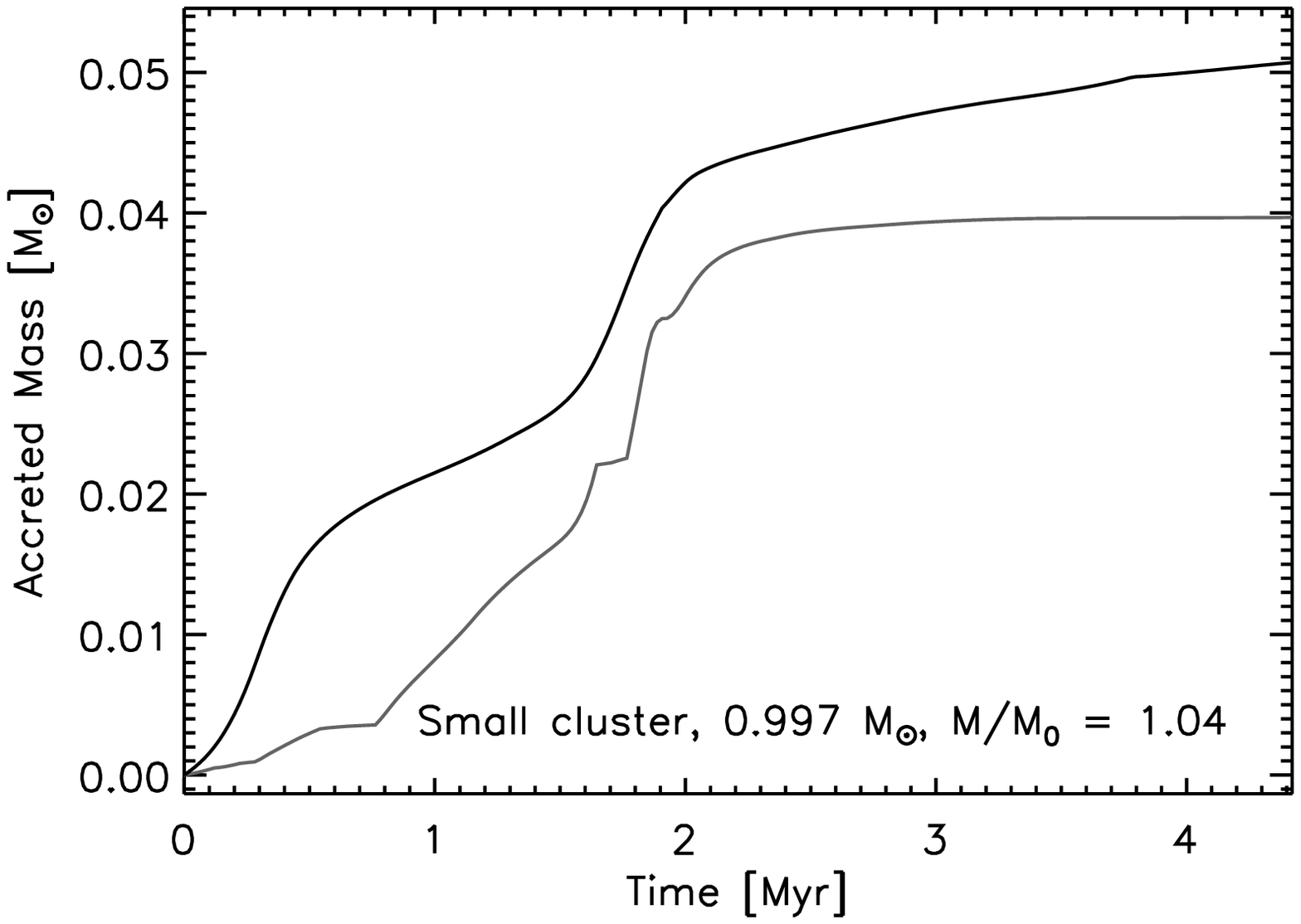}
                           {\includegraphics*[43,50][488,340]{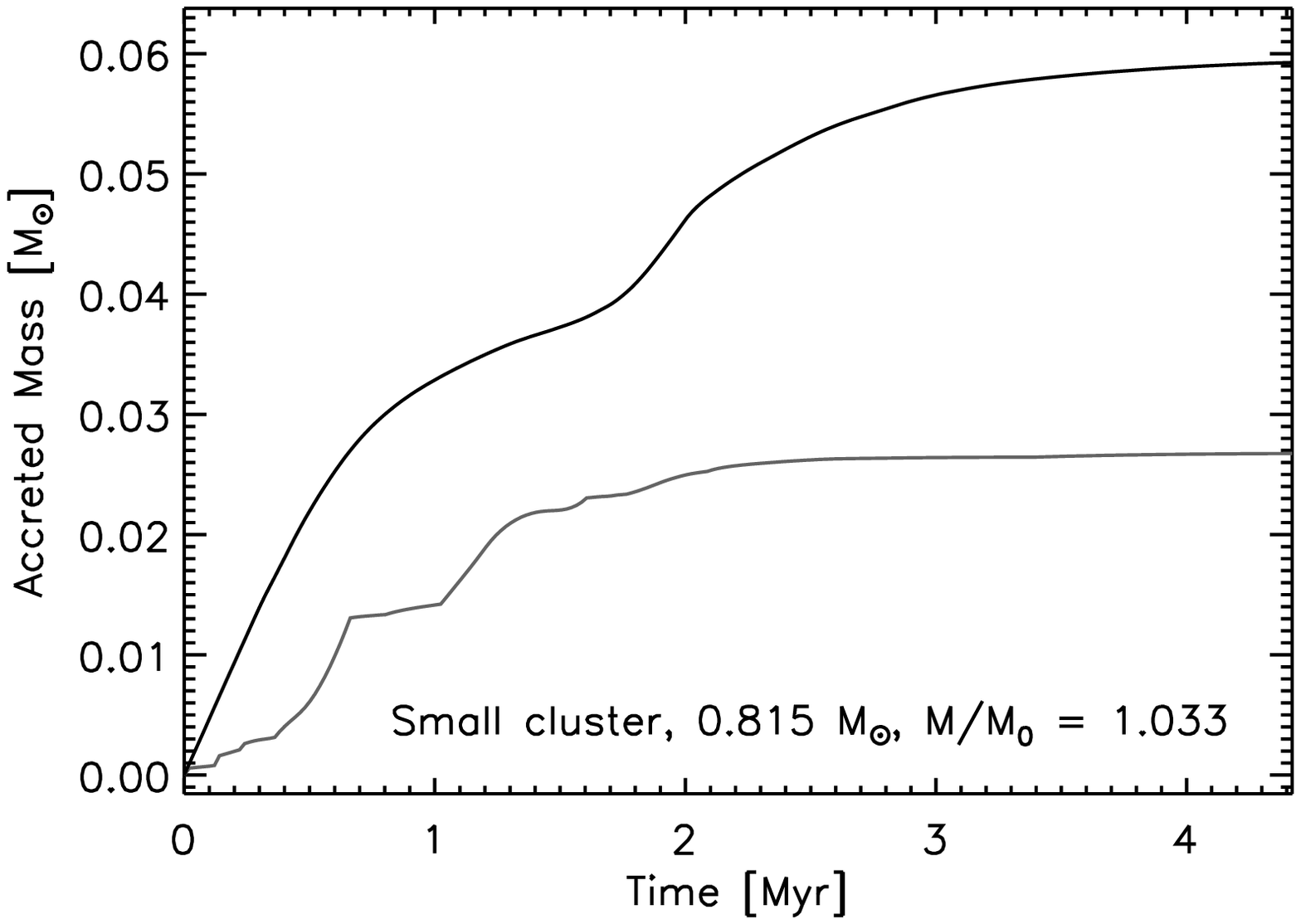}}
                           {\includegraphics*[43,50][488,340]{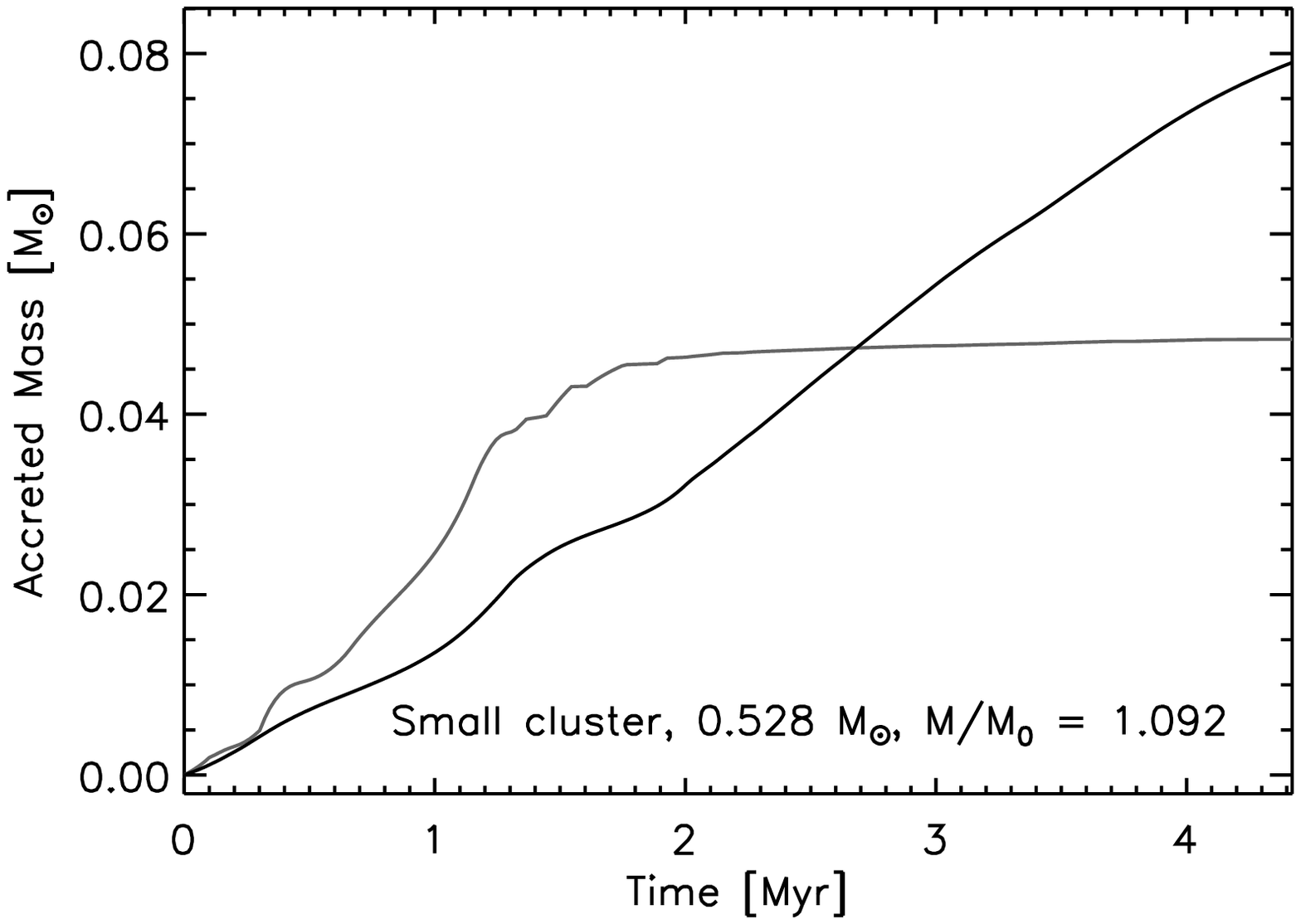}}}\hskip 0.6in}
\centerline{\scalebox{0.37}{\includegraphics*[ 0,50][488,340]{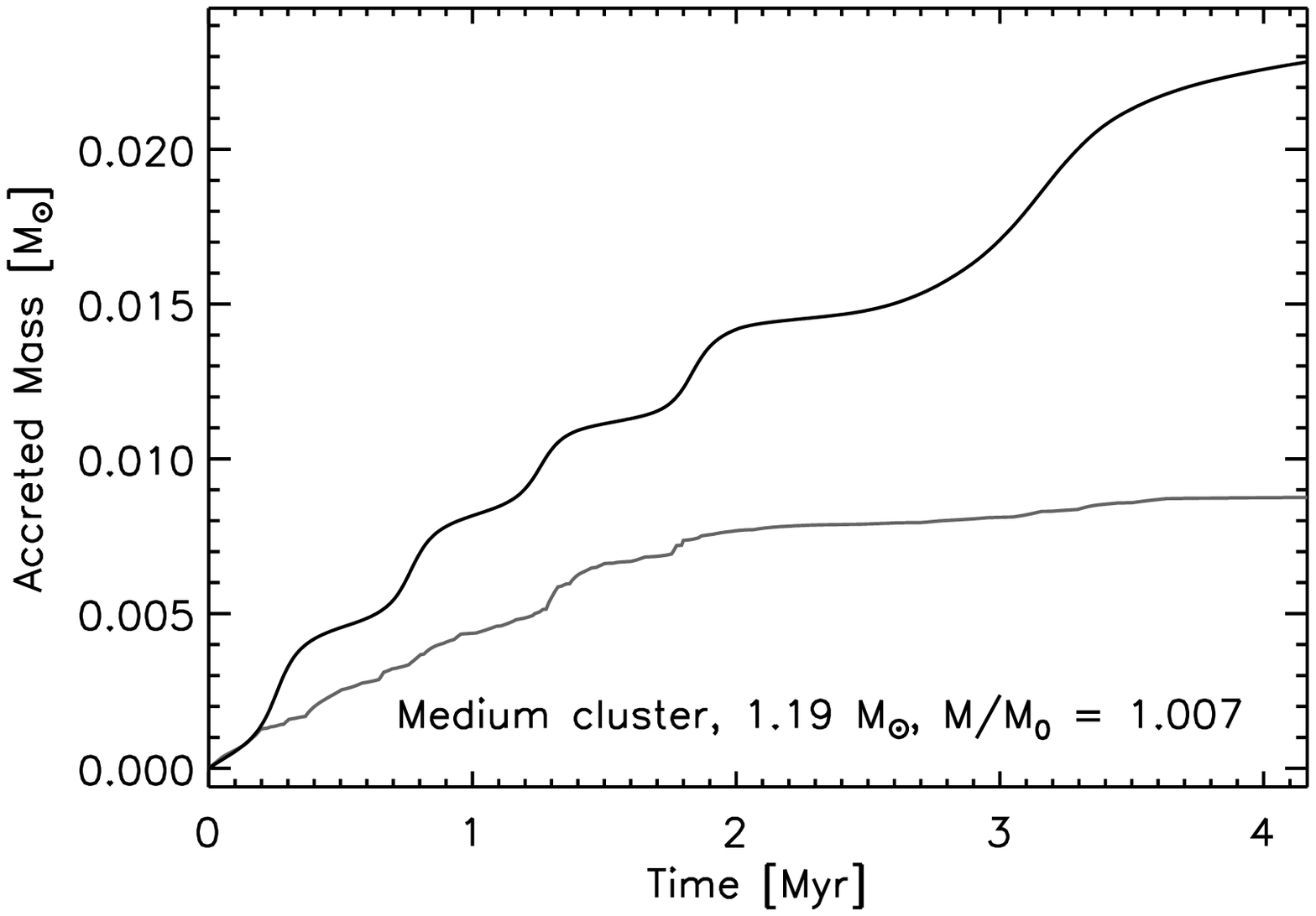}
                           {\includegraphics*[43,50][488,340]{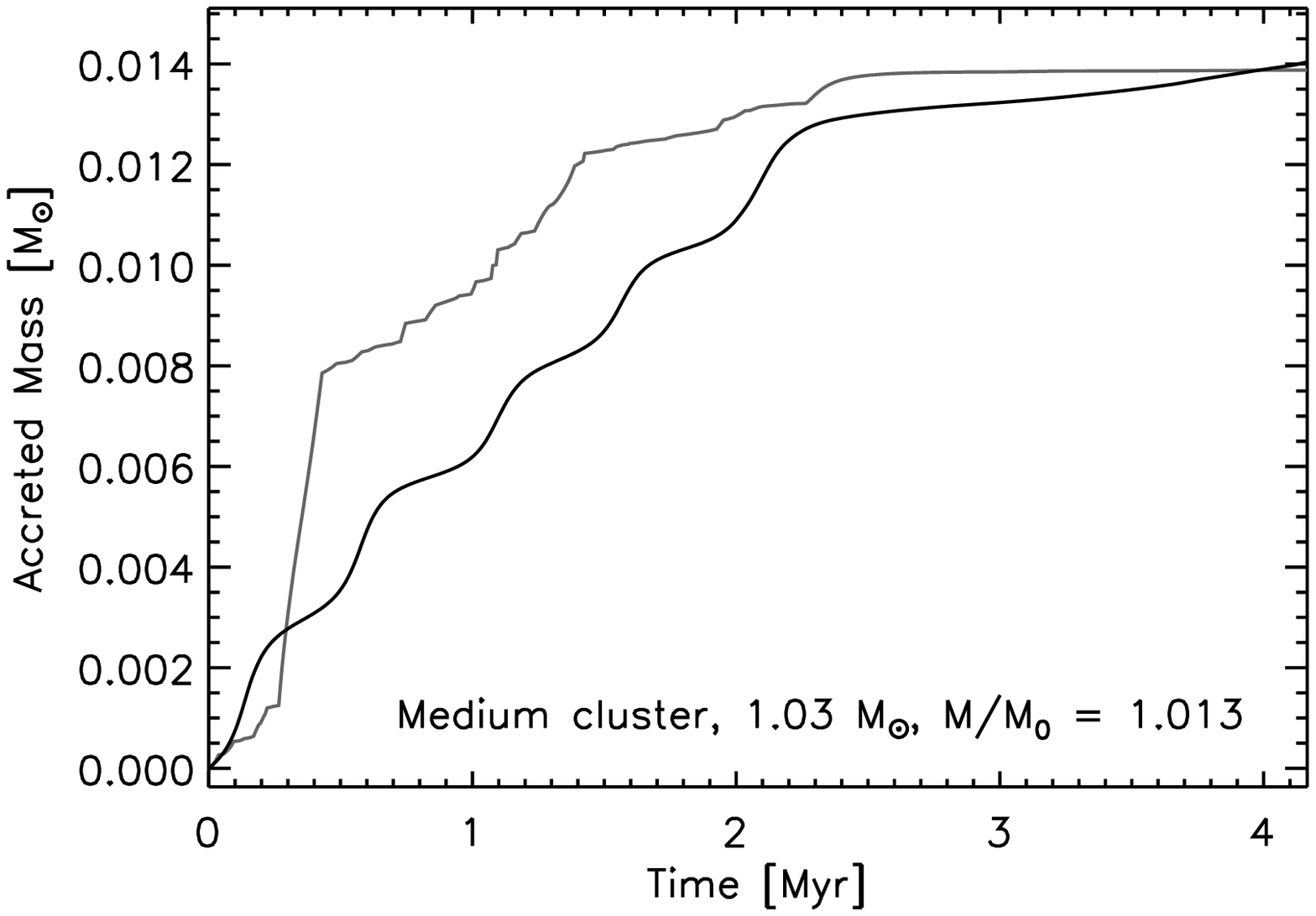}}
                           {\includegraphics*[43,50][488,340]{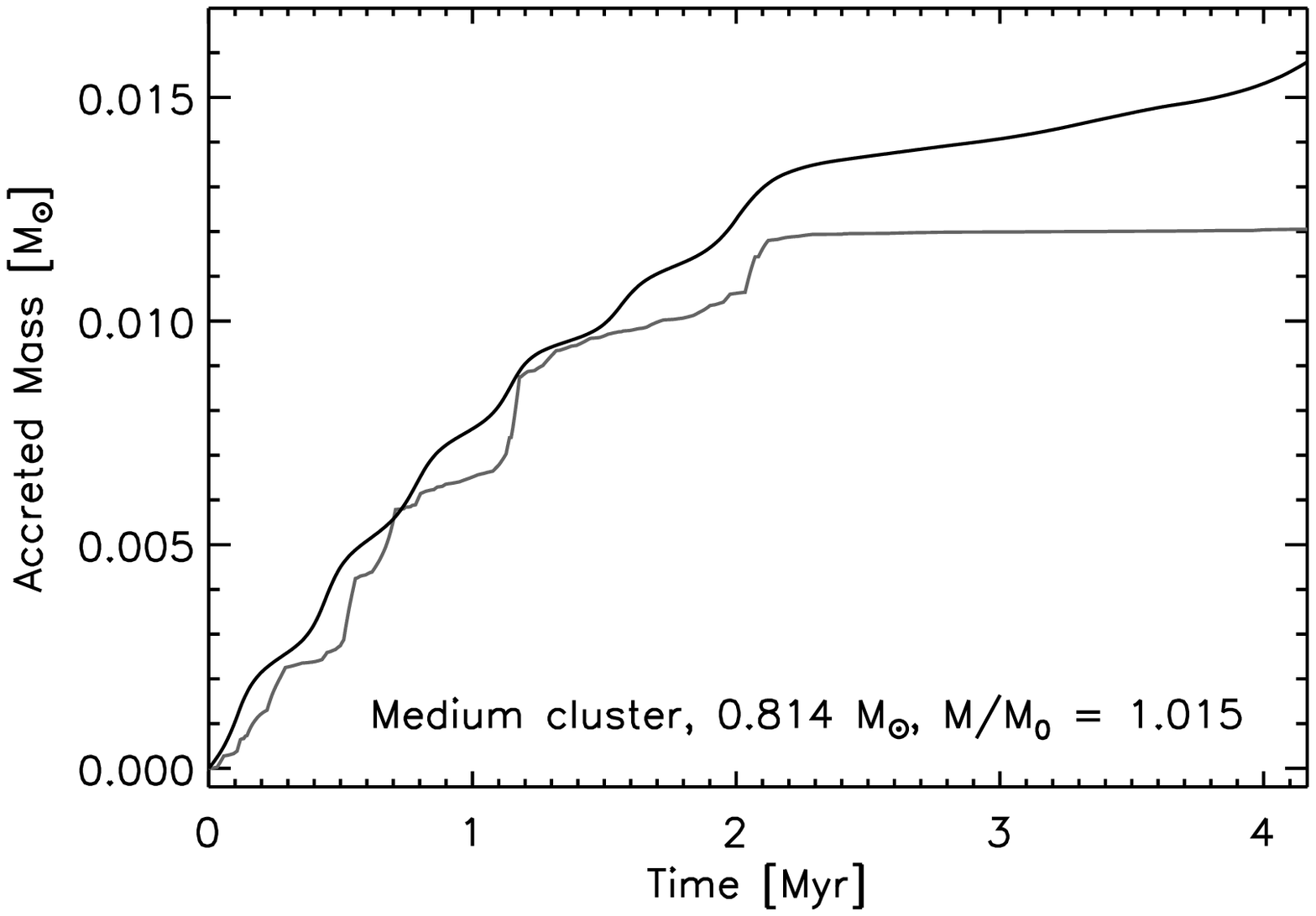}}}\hskip 0.6in}
\centerline{\scalebox{0.37}{\includegraphics*[ 0,00][488,340]{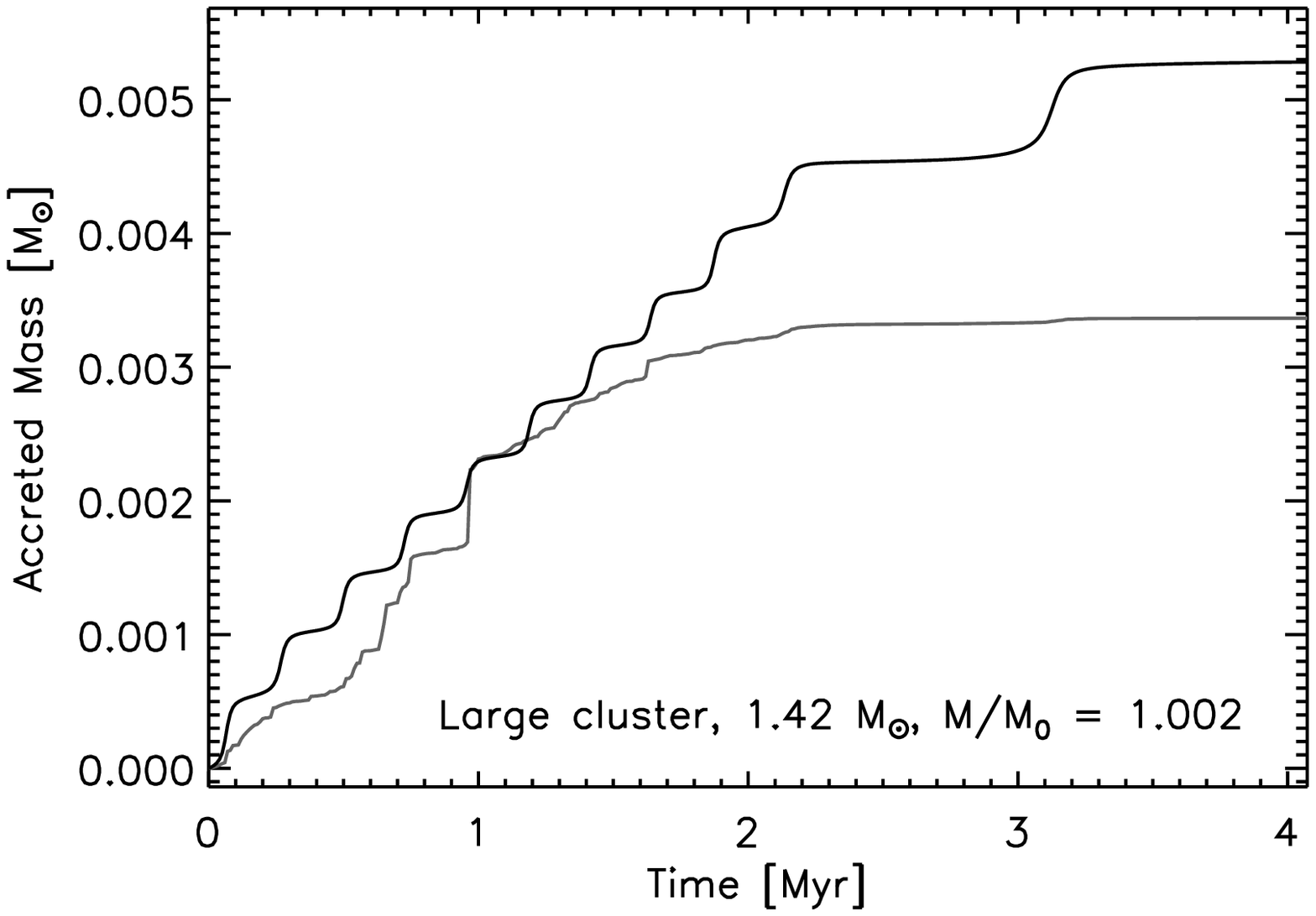}
                           {\includegraphics*[43,00][488,340]{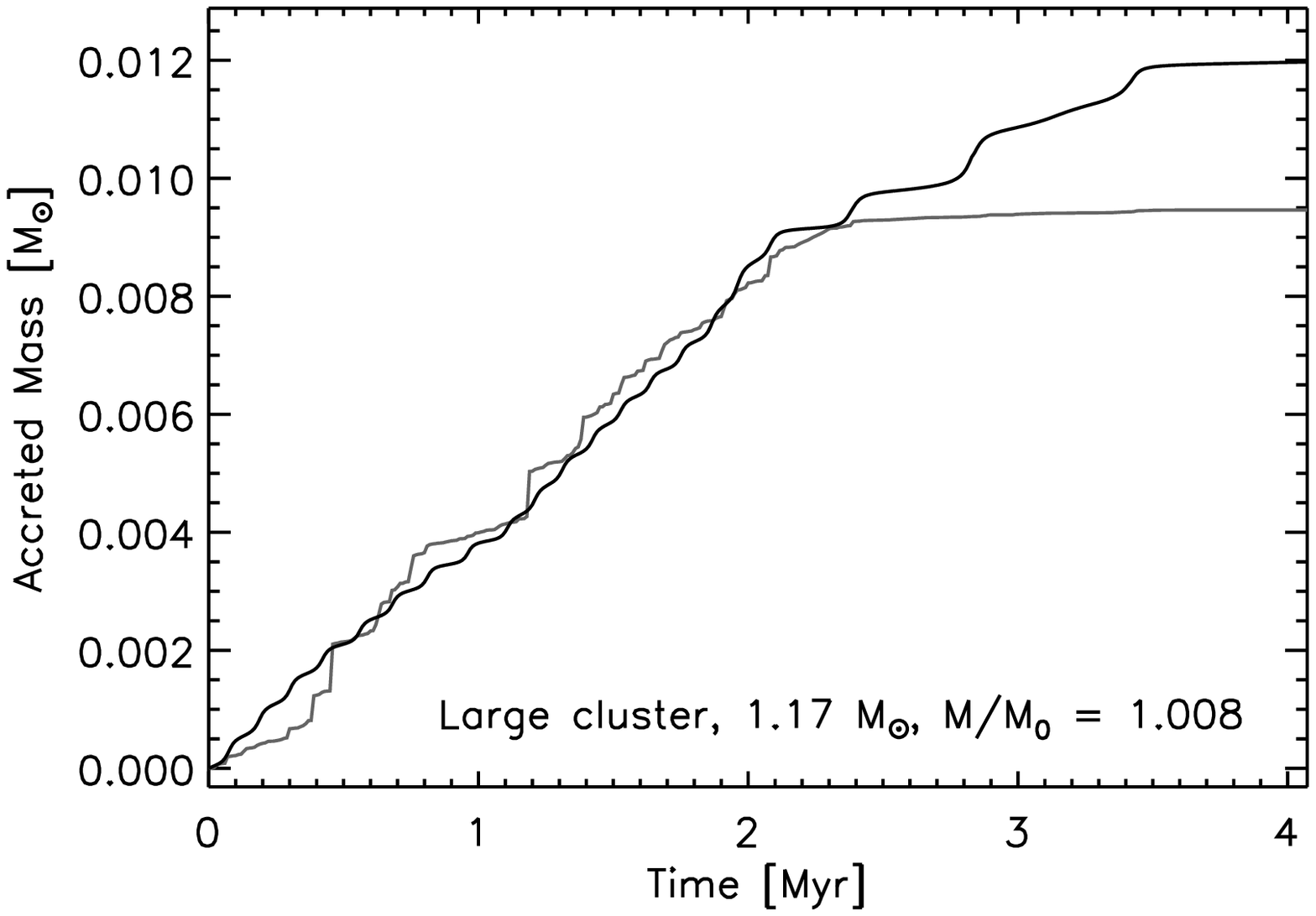}}
                           {\includegraphics*[43,00][488,340]{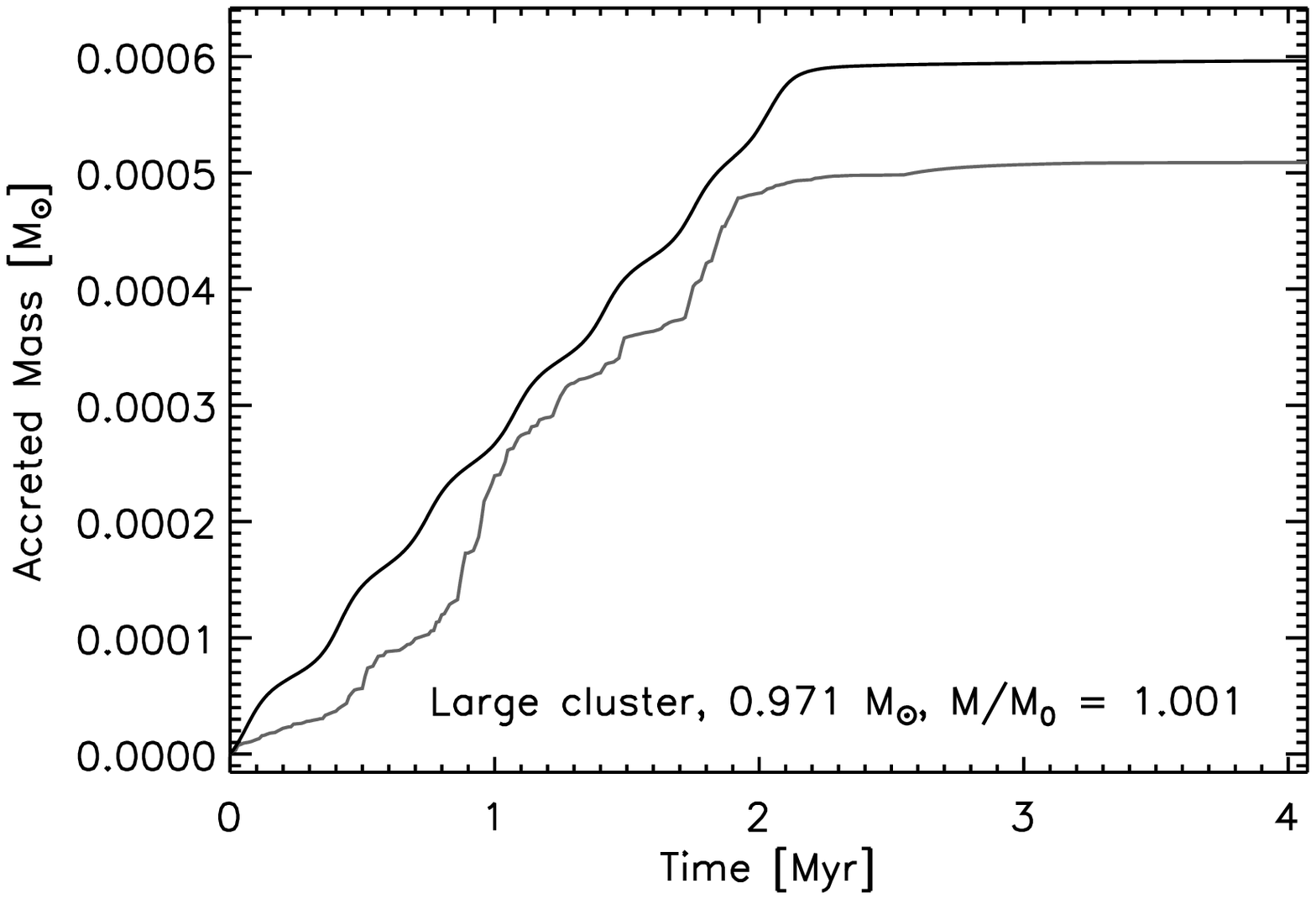}}}\hskip 0.6in}
\caption{Total mass accreted per star (\ie, integral of the solid line in Figure~\ref{fig:example_velocity_dmdt}.)
Accretion is often episodic, with most of the mass being deposited during short periods.  Accretion in each `stairstep'
can be many Jupiter masses ($1 M_J \simeq 0.001 \msol$). Stars are the same as identified in
Figure~\ref{fig:example_position}. \textbf{Black lines:} Virial method.  \textbf{Black lines:} Nearest-neighbor method.}

\label{fig:example_dm}
\end{figure}

The instantaneous accretion rate \dmdtbh\ can be summed at each timestep to compute the total mass accreted for each
star.  This quantity \dmbh\ is plotted in Figure~\ref{fig:example_dm} for each of the nine stars.  By the end of the
simulation, the stars have accreted between 0.1\% and 9\% of a stellar mass.  The accretion is generally episodic, and is 
most easily visible in the `virial' case (grey lines), where the velocity is not influenced by the particular nearest
neighbor.  The episodic accretion is easily visible as stairsteps in the accretion.  These stairsteps each deliver
$\sim$~0.5--10~Jupiter masses, in $\sim 100\ \rm{kyr}$--$300\ \rm{kyr}$ steps ($1 M_J = 0.001 \msol$).  As cluster size
increases, the episodes of accretion are shorter, but more frequent.  There is a diversity in how episodic the accretion
is: while the stairsteps are visible for some stars, several on more circular orbits show an accretion rate that is
nearly constant, until it drops at the end.

For these nine stars, the largest individual mass gains (both in absolute and relative mass) are in the {\sc small}
cluster, where one YSO has increased its mass by nearly 9\%.  Examination of this star's particular orbit
(Figures~\ref{fig:example_position}, \ref{fig:example_dist}, upper right) indicates that it was formed near the cluster
core, and stayed inward of 0.2~pc where it was immersed in high-density gas for nearly the entire 4~Myr simulation.  On
the other extreme, one star shown in the {\sc large} cluster accretes less than 0.1\% of its mass.  This star's initial
position is far from the cluster core at 0.4~pc, and it never once passes through the center before the gas starts
to disperse.  

\begin{figure}
\vskip -0.4 in
\centerline{ \scalebox{0.5}{\includegraphics{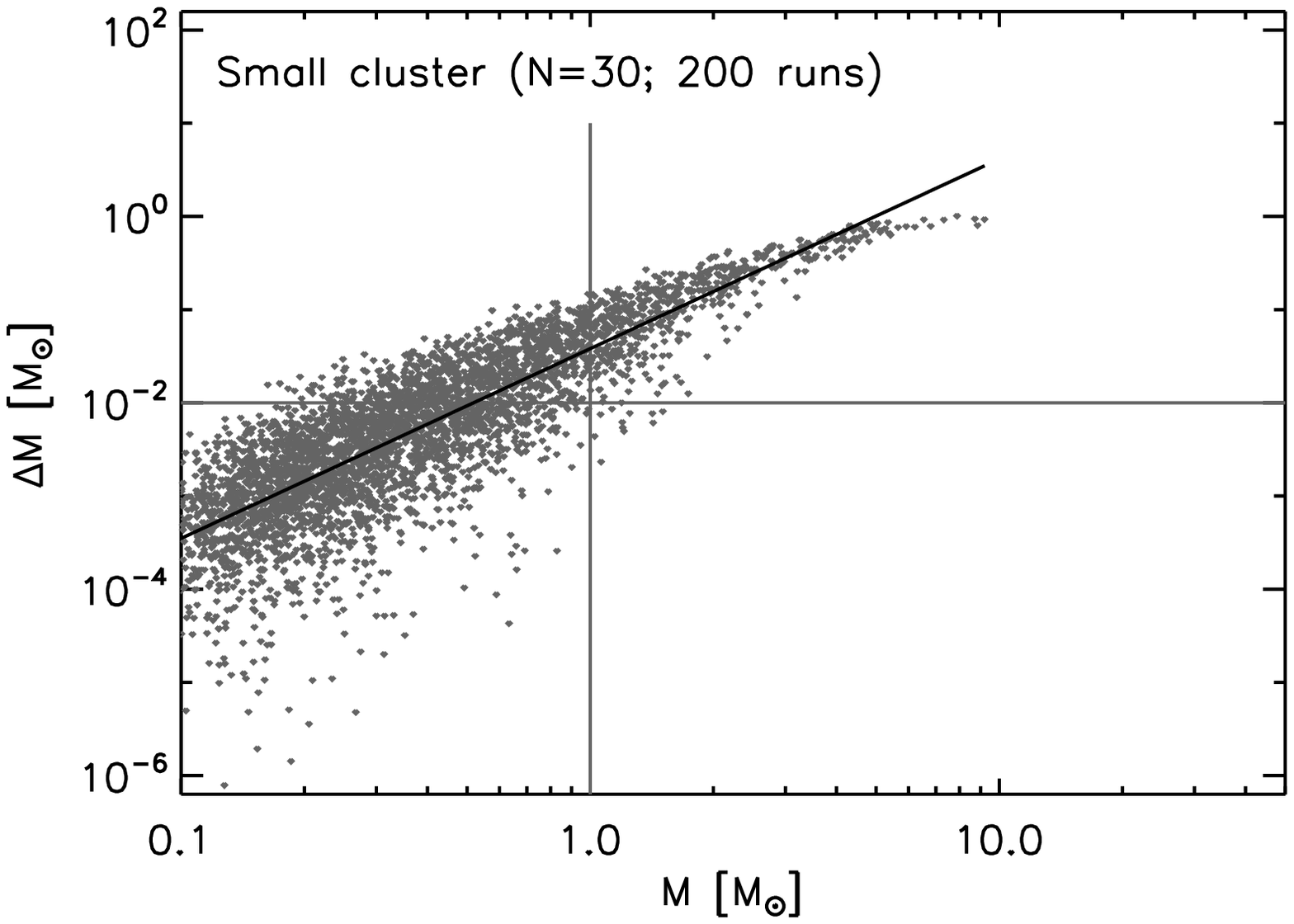}}}
\centerline{ \scalebox{0.5}{\includegraphics{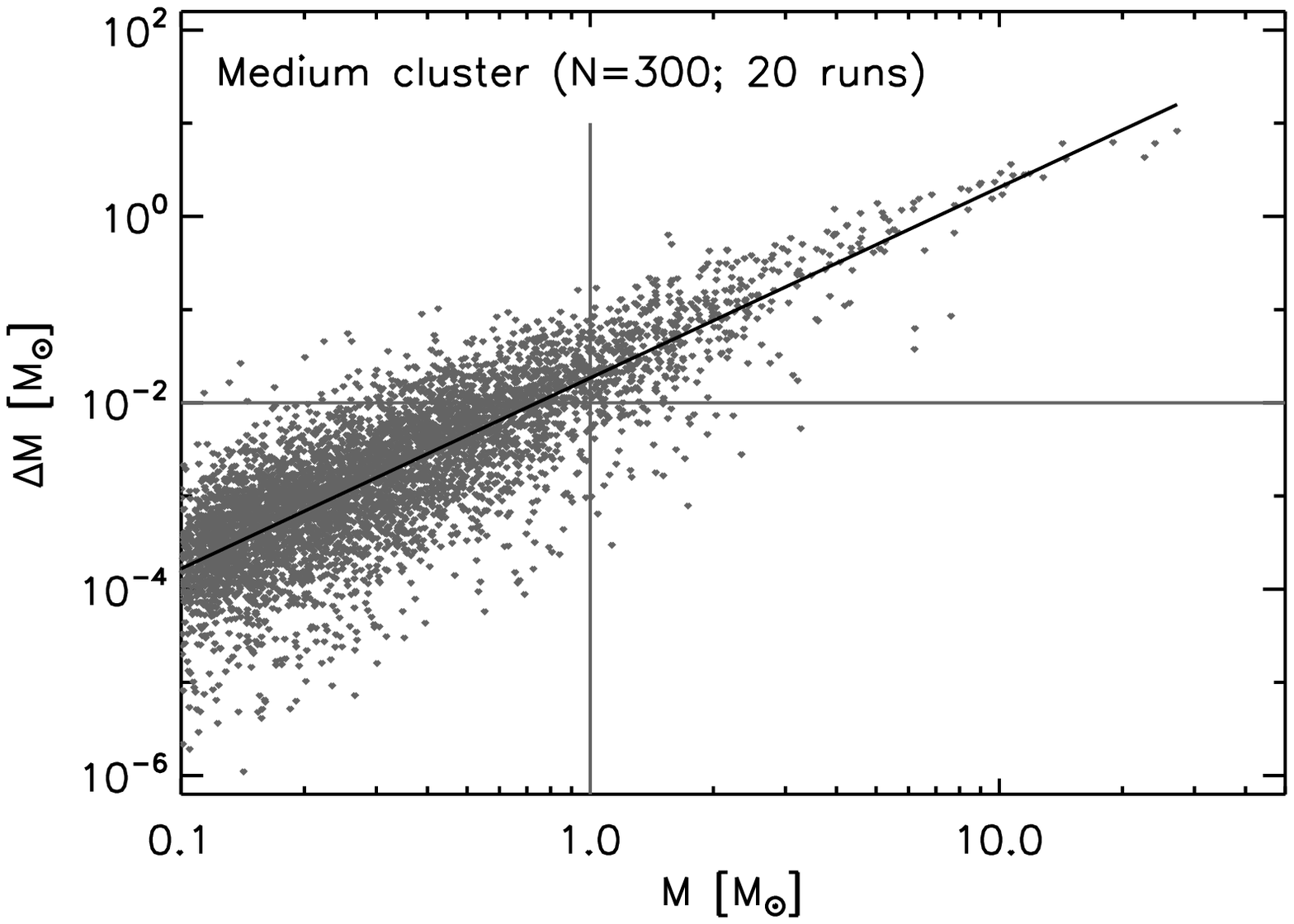} }}
\centerline{ \scalebox{0.5}{\includegraphics{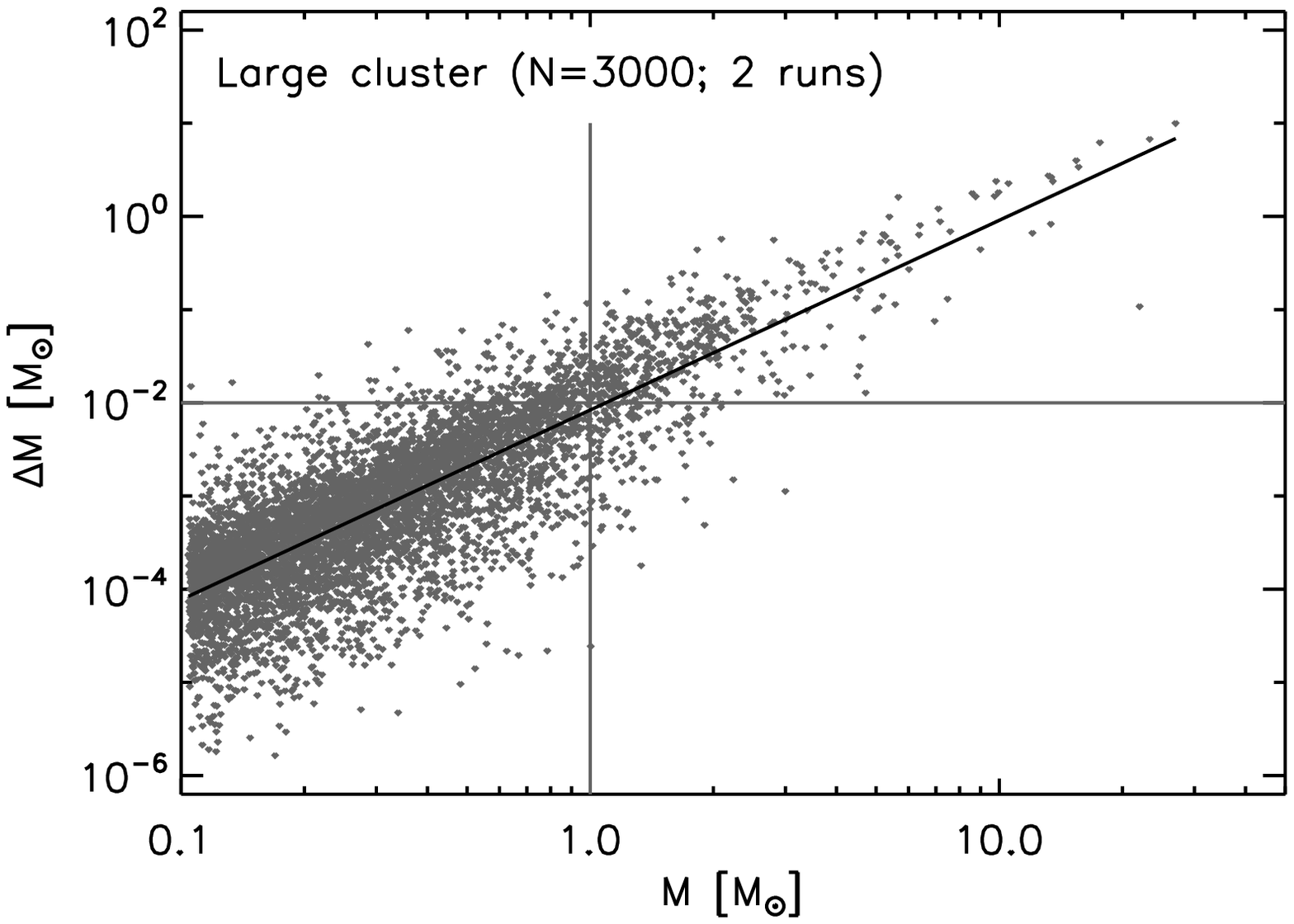}}   }
\caption{Total mass growth for stars in the three model clusters.   Each point corresponds to one star in the
simulation, and measures its total mass growth by $t=4~\rm{Myr}$.  The solid lines are fits to the model results, which
find $\dmbh \propto M^{2.1\pm0.1}$ across a range of 100 or more in mass.  The grey grid-lines indicate growth of one
MMSN (0.01~\msol) for a solar-mass star.  Mass growth is highest for the {\sc small} cluster, although the difference
between the three clusters is small compared to the scatter within the cluster.  Radiation pressure (ignored here) cuts
off accretion for the few stars above 10~\msol.}  

\label{fig:histogram_dm}
\end{figure}

Figure~\ref{fig:histogram_dm} plots the total \dmbh\ as a function of $M$ for every star in the three simulations.
Typical solar-mass systems accrete $\sim1$ MMSN during the runs, with stars in the {\sc small} cluster accreting
slightly more than those in the {\sc large}.  A small fraction accrete $10^{-3}\ \msol$ or less; these are stars that
are ejected before even one passage through the cluster core.  On the other extreme, about 1\% of the solar-mass systems
accrete as much as 0.1~\msol.  The full range of accretion for solar-mass stars spans four orders of magnitude!  This
diversity in accretion amounts may contribute to the observed diversity of disk sizes; for instance, in Orion many stars
have disks that are not resolvable at the resolution limit of $\sim20\ \rm{AU}$, while other stars likely of similar
mass have disks hundreds of AU across. 



A clear trend between \dmbh\ and $M$ is visible across a factor of $\gtrsim 100$ in stellar mass.  Fits to the results
in Figure~\ref{fig:histogram_dm} are computed using a least-squares method in logarithmic space (\ie, a linear fit to
the log of the mass and accretion rate).  The mass accreted is listed in Table~\ref{table:results} and is

\begin{equation}
\dmbh\ = \left\{ \begin{array}{r@{\quad:\quad}l}
0.038\, {\left({M \over \msol}\right)}^{2.0}\  \msol\ & \hbox{\sc small} \\
0.018\, {\left({M \over \msol}\right)}^{2.0}\  \msol\ & \hbox{\sc medium} \\
0.008\, {\left({M \over \msol}\right)}^{2.0}\  \msol\ & \hbox{\sc large} \ ,
\end{array}\right.
\label{eq:fit_dm}
\end{equation}
or dividing by 4~Myr, a mean rate of

\begin{equation}
\dmdtbh\textrm{(mean)} = \left\{ \begin{array}{r@{\quad:\quad}l}
9.3 \times 10^{-9}\, {\left({M \over \msol}\right)}^{2.0}\  \msolyr\ & \hbox{\sc small} \\
4.3 \times 10^{-9}\, {\left({M \over \msol}\right)}^{2.0}\  \msolyr\ & \hbox{\sc medium} \\
2.1 \times 10^{-9}\, {\left({M \over \msol}\right)}^{2.0}\  \msolyr\ & \hbox{\sc large} \ .
\end{array}\right.
\label{eq:fit_dm_msol}
\end{equation}

We find that solar mass stars on average accrete slightly less than $\sim$1--4 MMSN worth of material, depending on the
cluster size.  Stars in the {\sc small} cluster accrete more material because of the lower stellar velocities there,
which more than cancels out the effect of that cluster's lower gas density.  Likewise, gas densities in the {\sc large}
cluster are high, but this is more than canceled out by the higher encounter speeds.  The difference between the {\sc
large} and {\sc small} clusters is a factor of $\sim5$, relatively small considering the total range in accretion rate of
$10^5\rm$ times during the simulation.  We assume here that stars accrete for 4~Myr; if a cluster is longer- or
shorter-lived than this, or its accretion is inhibited by the birth of massive stars and an \hii\ region, then the total
should be adjusted.

The effect of the different temperature assumptions makes minimal effect on the results (Table~\ref{table:results},
bottom).  In the {\sc small} cluster, increasing the temperature from 5~K to 25~K decreases \dmbh\ by about 30\%.  In
the {\sc medium} and {\sc large} models, the difference is $<$5\%; these clusters have higher stellar velocities in the
first place, so the effect of $c_s$ is expected to be much weaker.  Therefore, most of the difference between the three
clusters is a result of the change in density and velocity, not temperature.



Because of the large time-variability of \dmdtbh, the \textit{mean} rate calculated above is not representative of the
\textit{median} instantaneous rate that an observer would be likely to measure in a cluster at a given instant.  Because
most of the mass is deposited during brief passages through the dense core, the instantaneous \dmdtbh\ should be lower
than the mean \mdotbh.  We therefore performed a similar fit as above, but using the entire ensemble of data points
(\eg, 3000 stars at 200 timesteps for 2 runs for the {\sc large} cluster).  Best fits to these rates yield


\begin{equation}
\dmdtbh\textrm{(median)} = \left\{ \begin{array}{r@{\quad:\quad}l}
4.7\times10^{-9}\, {\left({M \over \msol}\right)}^{2.1}\ \msol\ \rm{yr^{-1}} & \hbox{\sc small}\\
1.6\times10^{-9}\, {\left({M \over \msol}\right)}^{2.2}\ \msol\ \rm{yr^{-1}} & \hbox{\sc medium}\\
0.53\times10^{-9}\, {\left({M \over \msol}\right)}^{2.1}\ \msol\ \rm{yr^{-1}} & \hbox{\sc large}.
\end{array}\right.
\label{eq:fit_dmdt}
\end{equation}

These median rates are 2--4 times lower than the mean rates; \ie, an observer would be biased to underestimate the net
accretion rate, because most of the accretion happens during brief events that the observer is less likely to see!



\begin{figure}
\centerline{ \scalebox{0.7}{\includegraphics{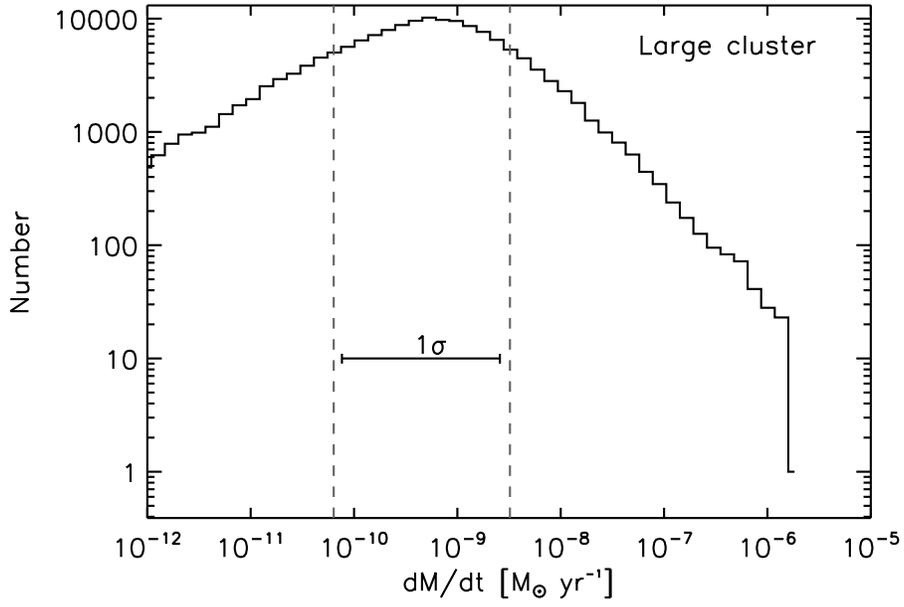}}}
\caption{Spread in \dmdtbh, for solar-mass stars in the {\sc large} cluster.  Histogram shows the range of values of
\dmdtbh\ for the entire simulation, with one data point per star, per timestep.  The $1\sigma$ spread in accretion rates
is a factor of 50.} 

\label{fig:spread_dmdt}
\end{figure}

The plots appear to show that the range in \dmdtbh\ between individual \textit{stars} at the same mass is greater than
the systematic variation between the individual \textit{clusters}.  To quantify this range, Figure~\ref{fig:spread_dmdt}
shows the spread of \dmdtbh\ for just the solar-mass stars in the {\sc large} cluster.  We calculate the $1\sigma$ width
in median \dmdtbh\ to be a factor of $50$; that is, 67\% of the instantaneous \dmdtbh\ values for solar-mass stars lie
within the range $5\times10^{-10}$ to $2.5\times10^{-8}\ \msolyr$.  (Or, at any given time, 1/6th of the stars in
the cluster are accreting more than 2.5~MMSN per Myr!)  This range reflects changes in accretion rate along a star's
orbit, in addition to each star's varying initial conditions.  The spread is comparable in all three clusters.  Thus,
wide variations in accretion rate are expected to be observed in clusters, even for stars of the same mass, observed at
the same time, and even in our highly idealized case of a smooth gas distribution.  Local variations in gas density
(which we do not model) would add an additional spread on top of this. 

\begin{figure}
\centerline{ \scalebox{0.7}{\includegraphics{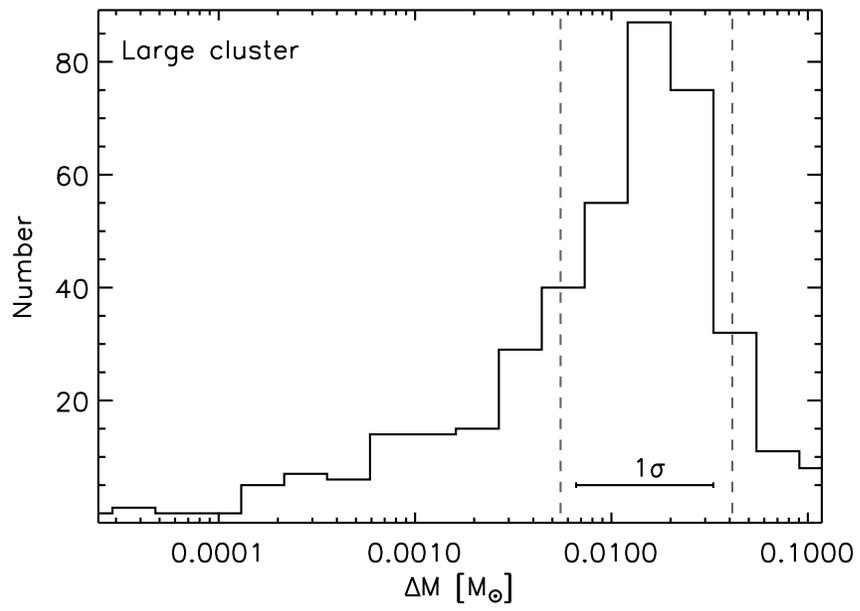}}}
\caption{Same as Figure~\ref{fig:spread_dmdt}, but plot shows total mass accreted by solar-mass stars.
$1\sigma$ spread is a factor of 7.}

\label{fig:spread_dm}
\end{figure}

Figure \ref{fig:spread_dm} shows the range in total accretion amount \dmbh\ for solar-mass stars, where we calculate the
$1\sigma$ width to be a factor of $7$ (\ie, 1/6th of the disks accrete more than 4~MMSN during the simulation).  This
range is less than the $1\sigma$ width of 50 in \dmdtbh\ because it is time-averaged, but still exceeds the difference
between the three clusters.

In all cases, the exponents are very close to the value of 2 for BHA through a uniform medium at a fixed speed.  This is
not entirely surprising, because the orbit of an individual star is virtually independent of its mass.  That the value
is slightly above 2 may be a result of dynamical friction, which causes the most massive stars to slow and begin
settling toward the cluster center.  This mass segregation is a well-studied effect on long timescales \citep{bt08}.  A
more detailed model which included the dynamical effect of increasing stellar mass on the N-body simulations (which we
ignore) would probably increase this exponent.




\section{Observational effects}

\label{sect:observations}

\subsection{Comparison with observed accretion rates}
\label{sect:observed_rates}

%
%


Accretion directly onto young stars can be detected through signatures in their optical continuum veiling or H$\alpha$
emission line profiles.  Recent observations have found the unexpected result that stellar accretion scales with stellar
mass as $\mdot \propto M^{2\pm0.2}$, with $\mdot \approx 10^{-8}\ \msol\ \rm{yr^{-1}}$ for solar-mass stars.  The
relationship has been observed for over 200~stars spanning the mass range $0.01\ \msol - 3\ \msol$
\citep[\eg,][]{ntr06,mlb05,shh05,mjb05,mbj05}.  There is a large scatter, with values of \mdot\ for the same
$M$ varying by a factor of $10^2$.  This scatter is intrinsic and not an observational uncertainty.  The samples are
pre-main sequence stars in young clusters.  The cluster environment appears to have little effect on accretion: stars in
the young, compact $\rho\ \rm{Oph}$ cluster show very similar accretion rates to stars in older, smaller Taurus and
Cha~I, although none of the regions surveyed are as large and dense as is Orion today.

\begin{figure}
\centerline{ \scalebox{0.7}{\includegraphics{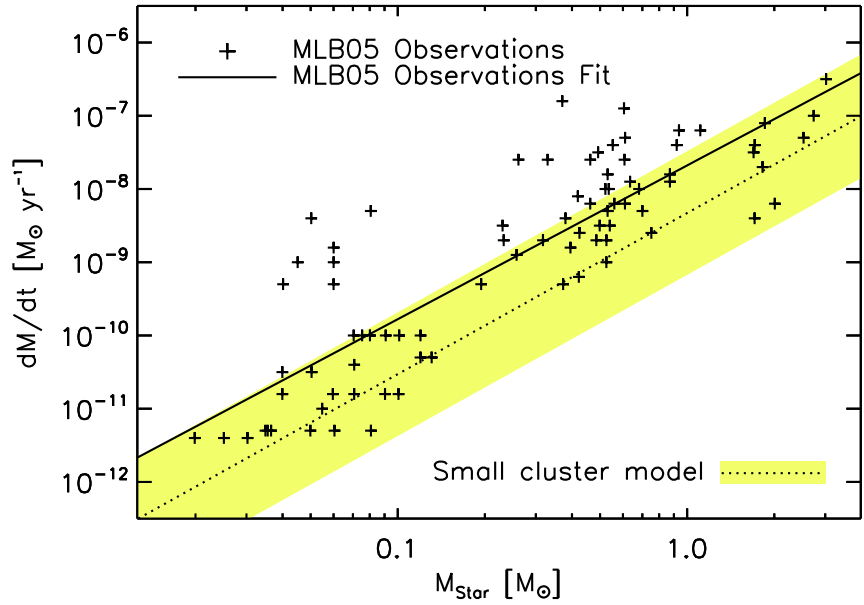}}}
\caption{Comparison between observation of accretion onto young stars, and calculated BH accretion rates.
Data shown is from \citet{mlb05} (MLB05); solid line is our fit to their data.  The dotted line is the result from BH
accretion in the {\sc small} model.  The shaded area shows the $1\sigma$ scatter of \dmdtbh\ in the model.  The scatter
from the two distributions clearly overlap, although the center of the model is several times lower than the data.}
\label{fig:obs} \end{figure}

Figure \ref{fig:obs} compares the observations of \citet{mlb05} in Taurus and Cha~I (both consisting of small cores
analogous to our {\sc small} cluster) to our calculated \dmdtbh.  We have fit their observational data in the same way
as eq.~[\ref{eq:fit_dmdt}] and find

\begin{equation}
\dmdtbh(\rm{observed})  \approx 21\times10^{-9}\, {\left({M \over \msol}\right)}^{2.1}\ \msol\ \rm{yr^{-1}},
\end{equation}
or about 5 times higher than our computed rate in the {\sc small} cluster.  The slope and scatter clearly match well.


Standard models of disk accretion do not explain the $\dmdt \propto M^2$ relationship, but many new models have been
proposed to address it.  \citet{gjs06} improves the standard layered disk model \citep{gam96} by assuming a temperature
structure that depends on the central star's luminosity.  \citet{hdc06} assumes a complex magnetic field geometry that
introduces a mass dependence in the accretion rate.  Although these models both move in the correct direction, neither
provides a sufficiently steep mass dependence in the exponent.  \citet{aa06} develop a model which reproduces the $M^2$
dependence, but requires certain disk initial conditions: namely, that the disk mass scale as the stellar mass (or
faster), and that the disk radius be highest for low-mass stars.  Alternatively, \citet{cp06} argue that the $\mdot
\propto M^2$ dependence may be an observational bias caused by the masking of low accretion rates by the bright continua
of high-mass stars because the detectability of accretion signatures scales as $M^2$.  

\citet{pkn05} proposed that the observed luminosity could be due to BH accretion from the molecular cloud environment.
Indeed, they noted that the functional form of BH accretion reproduces the $\dmdt \propto M^2$ relationship, and the
scatter in the observations can be explained by variations in stellar speed and gas density.  They performed 3D adaptive
mesh refinement (AMR) calculations of a single star traveling through gas filaments in a Taurus-like region, and found
that the BHA rate could roughly match the observed accretion rates.

Our calculated rates are $\sim$5 times lower than those observed, although within the errorbars of both, many points
overlap.  The rough similarity we find between the data and the BHA rates is interesting, but it must be interpreted
with caution because the observations detect \textit{disk-to-star} accretion, while the BH rates we calculate are for
\textit{cloud-to-disk} accretion.  These are two distinct physical processes.  The two accretions may be ultimately
related, buffered through transport within the disk.  The disk viscous timescale can be estimated by $t_v \approx
r^2/\nu$ for disk radius $r$ and viscosity $\nu$.  Typical viscous times range from $10^5\ \rm{yr}$ ($r=10\ \rm{AU},
\alpha=10^{-2}$) or less, up to $10^7\ \rm{yr}$ ($r=100\ \rm{AU}, \alpha=10^{-3}$; \citealt{hyj00}).  If BHA deposits
mass at distances of 10--50~AU, then this material would be accreted onto the star and cause an observable signature
during the timescale of star formation.  If mass is deposited much further out, or onto disks of low $\alpha$, then BHA'
effects would not be observed while the star was still in the cluster.  (Or alternatively, the effects of BHA may be
manifested as accretion onto the star long after a star has left its molecular cloud environment.)  

We raise the possibility that accretion onto the disk may
trigger a larger accretion onto the star, thus linking these two rates.  In particular, because the accreted material
can have essentially arbitrary angular momentum and can be highly episodic, it may have a destabilizing effect on local
regions of the disk, causing rapid inward migration and accretion.  Almost certainly, multiple process contribute to the
observed accretion rate and disk dispersal (such as those mentioned above), and BHA is only one of these.  More detailed
modeling of the accretion flow must be performed to constrain the dynamics and timescales, and determine to what extent
the observed and BH accretion rates may be related.




\subsection{Observational signatures of BH accretion}


Bondi-Hoyle accretion will impact the disk surface with speeds of order the higher of the stellar velocity or
keplerian velocity, typically several \kms.  Over most of the surface area of a disk, this impact velocity will result
in shock speeds that are small and post-shock layers that are cool.  Ongoing accretion from the inner disk onto the
central star manifests itself as UV excess, veiling at visual wavelengths, and H$\alpha$ emission; on the other hand,
BH accretion onto the outer disk is likely to produce a signature visible only at far-infrared and sub-millimeter
wavelengths due to the accretion-shock velocity.  Shock speeds of a few \kms\ will result in post-shock temperatures of
only a few times $10^2$ to $10^3$ K.  This post-shock layer may be detectable in the pure rotational transitions of
H$_2$ (such as the 28, 17, and 12 $\mu$m lines), in vibrationally excited transitions of species such as CO and CS, or
high-lying pure rotational states of common molecules such as CO and OH.  The fine-structure lines of common ions and
atoms such as \ion{O}{1}, \ion{O}{3}, \ion{N}{2}, and \ion{C}{2} may also be excited.  High angular resolution
observations should reveal that these tracers originate from an extended region of the disk comparable to the
gravitational accretion radius.  Thus, the emission produced by BHA should be distinguishable from processes related to
the accretion of matter from the inner disk onto the star, or from shocks produced by outflows.

\section{Discussion}
\label{sect:discussion}

BH accretion can provide a substantial amount of material to young disks at a critical time.  The implications of this
process have not been considered by standard models for the formation and evolution of disks, planets, or stars.  In
this section, we discuss several possible implications of BH accretion.

\subsection{Bondi-Hoyle accretion vs. other environmental effects}

Historically, models for disk evolution (including the proto-Solar nebula) have not considered the influence of the
environment.  In the last decade, models have begun to study photo-evaporation and close stellar encounters as two
potentially significant effects of the environment.  First, photo-evaporation is a disk dispersal mechanism that removes
gas and small dust grains by a UV-heating mechanism at rates of $10^{-7}$--$10^{-9}\ \msolyr$
\citep[\eg,][]{tb05,mjh03,tbe01,jhb98}.  In large clusters such as Orion the UV source can be external O/B stars (which
remove the disk outer edges), while in small clusters a star's own UV luminosity is the dominant removal source (from
the inner edge outward).  Second, close approaches between stars can strip protoplanetary disks to roughly 1/3 of their
miss-distance.  These encounters rarely affect the inner disk, but can be important for the largest disks in the densest
clusters: typical close-approach distances are $\sim 500\ \rm{AU}$ after several Myr in an Orion-like cluster
\citep{apf06}.

Unlike these two processes which remove disks, BHA causes an initial mass \textit{increase} to the disk.  BHA may have a
wider diversity of manifestations than the other environmental effects.  For instance, the photo-evaporation rate for
far-ultraviolet (FUV) flux depends on disk size, but is essentially independent of stellar distance.  Thus, even though
the intercepted flux changes quickly for stars, they experience minimal change in photo-evaporation through their orbits
\citep{sc01}.  The BHA rate, however, varies by 10 times or more through a star's orbit, by a few times more based on
the star's initial position and velocity, and by another factor of $(M/\msol)^2$ for different stars.  BHA also occurs
in clusters of all sizes, unlike photo-evaporation and close approaches, which both occur primarily in the largest and
densest of clusters.  Initially BHA causes a mass gain, but depending on the flow dynamics, the mismatch in angular
momentum between the disk and ISM may trigger a disk instability, resulting in net mass \textit{loss} from the disk onto
the star.  The total mass accreted by BHA is comparable to (or slightly less than) the mass loss due to
photo-evaporation; both of these are in general much larger than mass loss from close encounters.  

Depending on the environment, photo-evaporation and BHA may both occur together, but at alternate times.  For instance,
young stars near an OB association may routinely pass between an \hii\ region (where photo-evaporation removes the
disk), and a cool molecular cloud region outside the Str\"omgren sphere (where BHA can restore the disk).  This
`batter-fry-batter' scenario would be possible today in a region like Orion.  The Orion Molecular Cloud (OMC) is a
region of dense non-ionized gas ($n = 10^5\ \rm{cm}^{-3}$) lying just 0.1--0.2~pc behind the Trapezium cluster and the
Orion nebula.  Stars orbiting through the Trapezium region may regularly pass through the OMC, causing alternate
episodes of photo-evaporation and disk accretion.


Because of the diversity of effects of tail-end BHA, the full consequences on the formation of stars, planets, and disks
cannot yet be judged.  Most important is an understanding of how the inflow interacts with the existing disk, including
where and when mass and angular momentum are deposited onto the disk.  This is likely to be a complex function of (at
least) the disk size and orientation, and the speed, density, and angular momentum of the local medium.  For instance,
accretion in a filamentary molecular cloud may have a different episodic nature than a much smoother cloud or one with
lower turbulence.  The rotation of a cloud may make a difference, as could its particular shape and temperature
structure.






\subsection{Stellar mass growth} 

As mentioned in \S~\ref{sect:bh_imf}, the role of BHA in the origin of the IMF is the subject of some debate.  While we
have argued that the accretion is predominantly onto the disk and not the star, much of the accreted mass will
eventually accrete onto the star, because in the absence of photo-evaporation, inward accretion is the dominant clearing
mechanism for material inward of 100~AU on $10^7\ \rm{yr}$ timescales \citep{hyj00}.  An upper limit to the stellar mass
increase is simply the mass \dmbh\ accreted onto the disk.

Our {\sc small} model is of a similar mass and diameter cluster as that in the `competitive accretion' model studied by
\citet{bb05}.  They found that stars grew via BHA from small initial condensations to normal IMF in a few $10^5\
\rm{yr}$.  In contrast, our results find that BHA is generally insufficient to change the mass of most stars: only a
handful of solar-mass stars increased their mass by as much as 10\% during 4~Myr.  However, the two simulations are
really not comparable, because we have simplified dramatically the interaction between gas and the youngest stars.  The
SPH simulations of \citet{bb05} model the gas dynamics and filamentary structure to a high resolution.  Most of the
accretion occurs in these filaments, where fragmentation begins and the relative velocity between stars and gas is very
low.  The star stays in these dense filaments during most of its formation.  Once it is formed, it is often ejected by
interaction with another young star, or the filament itself disperses.  We model a later phase, where stars have
dynamically decoupled from structure in the local gas.  We have taken more conservative estimates for the gas
temperature and the gas-stellar relative velocity, and we allow for gravitational expansion as the gas is removed.  We
believe that our values are very safe lower limits for accretion rate, but we find no reason to believe that the stellar
mass function changes substantially during the `tail-end' accretion phase.  

The competitive accretion model predicts very rapid star formation -- at most a few $10^5\ \rm{yr}$ -- which must be
followed soon by cluster dispersal in order to stop accretion from continuing.  If stars are shown to form on such short
timescales, then the tail-end accretion amounts computed here would necessarily be affected; however, measurements for
cluster lifetimes do currently favor long timescales \citep{ll03}.


\subsection{Stellar accretion and metallicity enhancement}

\label{sect:stellar_metallicity}

For solar-mass stars and smaller, BH accretion is not enough to substantially increase the stellar mass.  The accreted
mass can affect the stellar metallicity, however, if the composition of the accreted material differs from that of the
star.  This effect can be particularly important for stars which are not fully convective.  While the lowest mass stars
remain fully convective for tens of Myr, stars above a few solar masses will develop radiative cores surrounded by a
thin convective shell within a few Myr of birth.  For example, at an age of about 2 Myr, a 2.5~\msol\ star's convective
envelope contains only the outer 10\% of its mass \citep{la97}.  Accretion of high- (or low-) metallicity `polluted'
material can lead to enhancement (or depletion) of the star's metallicity.  

Metallicity variations between a star and the accreting material could occur for several reasons.  First, the cloud itself
can vary in metallicity both spatially and temporally.  A single supernova can significantly alter the metallicity of
the surrounding ISM impacted by its ejecta.  For example, depending on the explosion energy, the demise of a star with
an initial mass of 23 \msol\ can produce 3 to 4 \msol\ of oxygen, 0.5 \msol\ of carbon, 0.5 \msol\ of silicon, 0.001
\msol\ of $^{26}$Al, and 10$^{-4}$ \msol\ of $^{60}$Fe \citep{yf07}.  If this material were to be mixed with $10^4
\msol$ or less of ISM, the abundances of the most common elements can be more than doubled and the abundances of
short-lived radioactive species can be enhanced by orders of magnitude.  However, X-ray studies of supernova remnants
such as Cass-A show that the supernova ejecta is highly inhomogeneous and contains clumps with highly non-uniform
composition.  When such clumps impact the surrounding molecular cloud, they can induce even larger local variations in
composition.  The passage of a young, low-mass star surrounded by a disk through such a chemically enhanced region of a
cloud can result in the accumulation of a veneer of freshly synthesized products of supernova nucleosynthesis.  This
accretion can introduce abnormally high concentrations of common elements such as O and C, and short-lived radioactive
species such as $^{26}$Al and $^{60}$Fe.  Such elements can then contaminate the disk, the stellar photosphere, or both.

Contamination by supernovae ejecta has been suggested to explain metallicity variations of stars in Orion.  Observations
by \citet{csp00,csl98} and \citet{cl94} of 29 B, F, and G stars found abundance variations up to 4 times between stars
within the same Orion subgroup.  These metallicity variations were spatially correlated, lending credence to the idea
that recent supernovae have contaminated distinct regions of the cluster.  The abundance variations were seen in O and
Si (which are produced by massive stars and type II supernovae), but not in Fe, C, and N (which are produced to a much
lesser degree, and thus would not be expected to contaminate the stellar atmospheres).  The observations alone cannot
say whether the metallicity variation was incorporated into the stars after their formation, or whether it is indicative
of heterogeneous mixing in the parent cloud, but both scenarios argue for recent injection of large amounts of metals
into the cloud \citep{har96}.

Spatial variation of a molecular cloud's gas:dust ratio can also lead to metallicity enhancements in accreted material.
\citet{pcj06} used measurements of visual extinction and $^{13}$CO abundance in Taurus to infer variations in the
gas:dust ratio of up to 2--5 times within individual clouds, on size scales $\sim\rbh$.  They also showed that turbulent
concentration processes within the ISM could lead to such clustering, and that the expected enhancement factors rose
quickly across smaller distances, to $\sim$~100 times on AU scales.  Accretion over long scales could reduce this
concentration, but spatial heterogeneities may persist in the disk for long timescales if radial mixing of small grains
within the disk can be inhibited \citep[\eg,][]{jom07}.

Supernova contamination requires consideration of the evolution of OB associations.  The most massive stars can explode
within 3 Myr of their birth; the more common massive stars have main-sequence lifetimes of 10--30 Myr.  A variety of
observations have shown that OB associations are assembled on a timescale of about 10--20 Myr.  Orion presents a nearby
example \citep{bal01}.  The oldest subgroup, Ori-OB1a, has an age of 12 to 15 Myr, while the 1b and 1c subgroups have
ages of order 3 to 8  Myr.  The OB1d subgroup is actively forming today and includes the Orion Nebula, NGC 2024, and a
number of other smaller clusters.  These subgroups form a well-ordered sequence in both space and time, indicating
sequential star formation.

An important feature of most OB associations, including Orion, is that all of the subgroups, and the surviving portions
of associated molecular clouds, are located in the interiors of their respective supershells and bubbles.   The
superbubbles are created by the combined effects of ionization, stellar winds, and supernova explosions, and are
therefore heavily contaminated with supernova ejecta.  Gamma-ray observations have provided direct evidence for the
presence of large amounts of live $^{26}$Al and $^{60}$Fe in the interior of the Orion / Eridanus feature
\citep{dhk07,die02}.  The Orion / Eridanus Loops form a 100 by 300 pc H$\alpha$ emission-line feature that marks the
current ionization front of the Orion OB association.  The energetics of the HI shell that lies outside the
H$\alpha$-emitting rim indicates that at least a dozen supernovae have exploded in Orion over the last 5 to 10 Myr.  All
of the surviving molecular clouds and active sites of star formation are embedded within the interior of the supershell,
and are therefore exposed to the contents of the superbubble.    

Two sites in Orion provide tangible examples of regions that could become SN-contaminated.  First, the Orion Nebula
itself is located directly behind the Ori-OB1c sub-group that contains the 3--5~Mya clusters NGC1980 and NGC1981.
Although there is no evidence for a recent supernova in Ori-OB1c, there are massive stars in this complex.   When they
do explode, they will be in excellent position to contaminate the Orion Nebula region and the proximal portions of the
L1641 cloud.  Second, the $\sigma$-Ori sub-group of Ori-OB1c is located about 30\arcmin\ or 5 pc west of the southern
part of the Orion B molecular cloud.  Although there is an HI and H$\alpha$ bubble surrounding this region, there is no
concrete evidence for a recent SN.  However, as seen from the center of the $\sigma$-Ori cluster, the cloud subtends at
least 1 steradian, and when an explosion happens, it will pollute the western side of the Orion A cloud.  Young stars
forming from the polluted material, or older stars with disks that undergo BHA, will become contaminated with supernova
debris.

Tail-end accretion has a testable consequence for large clusters where metallicity enhancement from supernovae is
expected to occur after several Myr.  Veneers of metal-rich ejecta accreted onto stars in the cluster will have the
largest effect on the oldest moderate mass stars which have the smallest convective zones.  In this scenario, we expect
a correlation between increasing metallicity, increasing mass, and decreasing stellar age.

\subsection{General effects on Solar System formation}

The planets in the Solar System are believed to have begun formation soon after formation of the disk.  Models build the
gas giants within 5--10~Myr \citep{ls07}, consistent with the strong observational requirement that by 10~Myr, virtually
all stars have lost their observable disks \citep[]{chs00}.  For the terrestrial planets, observations indicate that
grains begin to grow rapidly in even the youngest disks \citep{dbc07,tbe01}, and models consistently predict runaway
growth to planetesimals of size $\sim0.1\, M_{\rm{Earth}}$ within 1--10~Myr \citep[][and references therein]{ntk07}.
Because this is in the same time that tail-end BHA is occurring, accretion can affect not only the initial conditions of
planet formation, but also the physics \textit{during} the process.  

Tail-end accretion provides a fundamentally new input to the process because it delivers material to the disk well after
the disk has formed.  The amount of material accreted onto and processed by the disk might exceed (by a few times) the
original disk mass, or its instantaneous mass.  (Model disk masses for our Solar System are usually greater than 1~MMSN,
but rarely exceed 10~MMSN because such large disks become rapidly unstable.)

This new material may replenish -- either locally or globally -- gas and dust in the disk that have already been used.
It may have dynamical effects on the existing disk, because of the mismatch in angular momentum.  And, because accreted
material comes from elsewhere in the ISM, any composition heterogeneities in the nebula may be inherited by the disk.

One `cartoon' scenario provides an interesting taste of the possibilities.  Consider the case of a pre-planetary system
that begins to form in the Trapezium region, but before O/B stars turn on.  Its initial disk is several MMSN, and in its
first 1--2 Myr it forms a giant planet core.  When the O/B stars are born, photo-evaporation begins and rapidly removes
its remaining gas disk, before the planet can build its gas envelope.  Subsequently, the star's orbit takes it through
the nearby OMC region, where BH accretion rapidly replenishes the disk: the 1~$M_J$ (0.1~MMSN) atmosphere for the planet
can accrete onto the disk in $10^5\ \rm{yr}$.  This gas can then accrete onto the planet, where it is safe against
photo-evaporation even after it re-enters the \hii\ region.  Thus, a birth environment thought to be destructive to giant
planet formation \citep{thr00} may prove instead to have regenerative powers.  The architecture and compositional
history of such planets may lie far outside those considered by standard models.

\subsection{Compositional heterogeneities in the Solar System}

Although it has historically been assumed that the protosolar nebula was well-mixed, recent discoveries have shown that
this is decidedly not the case.  Variations in isotopic and bulk chemical abundances in the Solar System strongly argue
that it formed from distinct reservoirs of material.  

Isotopic measurements of various species provide extremely precise determination of the chemical components that formed
the young solar system.  Isotopic differences between the Earth, Mars, and numerous diverse classes of asteroids has
provided consistent evidence that multiple reservoirs of material were present during formation of the terrestrial
bodies.  To date, isotopic anomalies have been measured for barium, chromium, sulfur, titanium, zirconium, molybdenum,
and oxygen \citep[][and references therein]{tba07,rj06,dmr02}.  The anomalies are small ($\lesssim$3000--10,000 ppm) but
definite.  They are in general not explainable by chemical processes, with the possible exception of oxygen
\citep{yk04}.  Importantly, the variations are not simply a random scatter, but are correlated.  The correlations can be
easily matched with species produced in nucleosynthetic reactions.  Thus, these measurements are usually interpreted as
evidence for heterogeneous mixing between multiple nucleosynthetic-produced reservoirs during the Solar System's
formation (c.f. previous references).  While some of these species are s-process, and thus produced during the lifetime
of massive stars, others are r-process and probably produced only in supernovae.

The observed heterogeneity could be indicative of material accreted onto the disk (\eg, from a site spatially distinct
from the formation location), or it could be present in the original disk.  However, studies of short-lived
radionuclides such as $^{60}$Fe suggest that at least some of the heterogeneity was in fact accreted after the solar
system formed.

\subsection{Short-lived radionuclides in the Solar System}

\citet{but07} compared radioactive decay products present in the Solar System's primordial bodies.  Iron meteorites and
pallasites were formed at $<$ 1~Myr after the start of the solar system (where $t=0$ is the formation time of the
calcium-aluminum-rich inclusions, CAIs).  These bodies show no evidence for short-lived $^{60}$Fe ($t_{1/2} = 1.5\
\rm{Myr}$) incorporated at their formation, based on present-day Ni isotopic abundances.  However, bodies that formed
slightly later at $t = 2-3\ \rm{Myr}$, such as the Earth, Mars, and carbonaceous chondrites, consistently show
signatures implying they \textit{were} formed in the presence of $^{60}$Fe.  Because the only natural sources of
$^{60}$Fe are supernovae, and $^{60}$Fe can be incorporated into an exposed disk \citep{odh07}, \citet{but07}
conclude that this species was not present at initial formation of the protoplanetary disk.  Rather, it was absorbed
later from the environment, presumably a dense cluster with stars of $M > 8\ \msol$ which can produce supernovae.
Further, the authors find that $^{26}$Al, another short-lived species ($t_{1/2} = 0.7\ \rm{Myr}$) primarily produced by
massive stars, was present from the Solar System's origins, and not added later.  This suggests that multiple supernovae
in the Solar System's first several Myr contributed material to the young protoplanetary disk.  


The simplest way for supernova ejecta to be incorporated into a disk is by direct implantation.  \citet{odh07} noted
that gas-phase supernova ejecta would be deflected by the bow shock that forms as the blast-wave over-runs an obstacle
such as a circumstellar disk.  They suggested that most supernova debris is injected in solid form by grains that move
ballistically through the shock, and implant in the proto-planetary disks {\it after} the proto-Sun formed (the
`aerogel' model).  However, \citet{wg07} modeled the evolution of stars in clusters of various sizes and found the
aerogel model to be inefficient: it requires a very massive supernova progenitor, very close to the nascent Solar
System.  They estimated the probability for any disk to acquire the observed amounts of short-lived species to be less
than a few percent.

The BH accretion scenario proposed here may provide a viable acquisition route for the short-lived radionuclides.
First, in an OB association such as Orion, grains of supernova debris can be implanted into surviving portions of
molecular clouds.  YSOs that form from such material initially would be enriched in supernova ejecta.  Additionally, as
pre-existing young stars pass through these polluted environments, their disks can accrete a veneer of contaminated gas
and dust, adding to an otherwise uncontaminated system.  Large solids (primitive meteorite particles) that formed before
passage through the contaminated zone would be devoid of these products, while those that formed later would be enriched
by them, thus explaining the apparent heterogeneity of the decay products of short lived species \citep[\eg][]{but07}.
This model has the efficiency advantages of using the large cross-section of the entire ISM as a `dam' for such
radionuclides, and using the larger `fishing net' radius of \rbh\ vs. the disk physical radius to capture material.
Thus, the difficulty in accreting such nuclides posed by \citet{wg07} may be addressed.

\subsection{Damping of terrestrial planet eccentricities.}

In order to collisionally grow, planetesimals in the terrestrial planet zone require high eccentricities for orbital
crossing.  However, the planets today have much lower eccentricities ($<$2\% for Earth and Venus) which cannot easily be
explained without a damping mechanism.  It has been suggested that the eccentricities could be damped by a tenuous
remnant gas disk of lifetime $10^6 - 10^7\ \rm{yr}$ \citep{aw02,ki02b}.  For instance, the latter paper studied a gas
disk of surface density $\sim 10^{-3} - 10^{-4}$ that of the MMSN, which exponentially decayed on several Myr
timescales.  The initial gas density was sufficient to maintain circular orbits.  As the disk dispersed, orbital
interactions entered a chaotic regime which allowed for planetary mergers and growth.  The remaining gas slowly damped
the eccentricities of these widely spaced larger bodies until a small number of planets were left on circular orbits.
The position and size of the planets depended on the amount of gas and its decay profile.  

Tail-end accretion fits into this scenario naturally, and can allow for orbital damping on timescales other than the
disk's initial dissipation time.  The lower limit for the planet formation timescale in our solar system corresponds
roughly to the 10~Myr upper limit for cluster lifetimes.  The simulations by \citet{ruf99} showed that BHA easily
produced disks with $10^{-3}$ MMSN density, more than sufficient for orbital damping.  In a cluster, where the accretion
disks are transient and episodic, this may provide multiple stages and/or speeds of damping, as a star passes through
the cluster core multiple times.  As terrestrial planets begin to be detected around other systems, their eccentricities may
give insight into their birth environment as well.

\section{Conclusions}

`Tail-end' Bondi-Hoyle accretion of molecular gas onto young star-disk systems moving through the gravitational
potential of three realistic model star clusters and their molecular clouds are investigated using N-body simulations.
We make the following findings:


\begin{itemize}
\item{Bondi-Hoyle accretion can occur as stars orbit through dense cloud cores located near their birth environments.
Gas is accreted toward the star, but is intercepted by the disk before it hits the star.  Accretion can occur 
for the several Myr between formation of the first YSOs until dispersal of the molecular cloud.}

\item{For solar-mass stars in Taurus-sized clusters (N=30), the accretion rate is $\sim10^{-8}\msolyr$
(one MMSN per Myr); the rate scales with stellar mass as $\dmdtbh \propto M^{2.1\pm0.1}$.}

\item{Accretion onto stars up to 10~\msol\ is robust against stellar winds, outflows, radiation pressure, and turbulence
under most conditions.}

\item{For a fixed cluster diameter, the accretion rate is weakly dependent on the cluster density.  Increasing N by a
factor of $10^2$ decreases accretion rate by only a factor of 3. The lower gas density in small clusters is offset by
the ease of accretion at their lower stellar velocities.}

\item{Accretion is an episodic process.  The accretion rates during brief plunges through the cluster core are up to 10
times higher than the lowest accretion rates for the same star.  Intrinsic variations in gas density and velocity
contribute further to the episodic nature of accretion.}

\item{The scatter in time-averaged \textit{mean} accretion rates for stars of the same mass in the same cluster is a
factor of $\sim 7$; the scatter in \textit{instantaneous} accretion rates for the same stars is a factor of $\sim 50$.
The scatter is due to the spread in positions and velocities of individual stars and by the variations in the density of
ambient medium.}

\item{The mean accretion rate is a few times higher than the median rate, because most accretion occurs where stars
are moving quickly and least likely to be observed.}

\item{Bondi-Hoyle accretion onto disks cannot be solely responsible for the $\mdot \propto M^2$ accretion observed onto
young stars, but these two accretion processes may be linked by mass transport through the disk.  The BH accretion rate
has similar slope and scatter to stellar accretion rates, but is $\sim5$ times lower than observed accretion
rates.}

\item{Accretion onto the disk is a low-energy process and difficult to detect directly, but may be visible in mid-IR
rotational and/or vibrational transitions of H$_2$, CO, CS, and OH.}



\item{Variations in chemical abundances in the gas cloud can be caused by pollution from massive stars.  Accretion of
this gas can result in a disk with substantially different metallicity than its star.  Isotopic and bulk chemical
heterogeneities in meteorites and terrestrial material may be caused by tail-end BH accretion of recent nucleosynthetic
products produced by massive stars and/or supernovae.  Accretion may also explain observed large metallicity variations
in the atmospheres of young stars.}


\item{While we estimate the \textit{amount} of mass deposited during tail-end accretion, the detailed \textit{dynamics}
of accretion onto an existing disk are not well understood, and additional modeling is needed to understand the
deposition of mass and angular momentum onto the disk.}

\end{itemize}

\section{Acknowledgments} %

We are grateful to Sverre Aarseth for providing his NBODY6 code and guiding us on its use.  We thank C. Agnor, P.
Armitage, C. Clarke, D. Hamilton, H. Levison, M. Hedman, T. von Hippel, N. Moeckel, W. Ward, and H. Zinnecker for their
insights and useful discussions.  We also thank the anonymous referee for an extremely careful and detailed review.  HT
and JB acknowledge support from NASA Origins grants NNG06GH33G and NNG05GI43G; HT acknowledges support from NASA
Exobiology grant NNG05GN70G, and JB acknowledges support from the University of Colorado Center for Astrobiology which
is supported by the NASA Astrobiology Institute.

\bibliography{papers}

\clearpage

\onecolumn


\begin{figure}
\centerline{ \scalebox{0.5}{\includegraphics{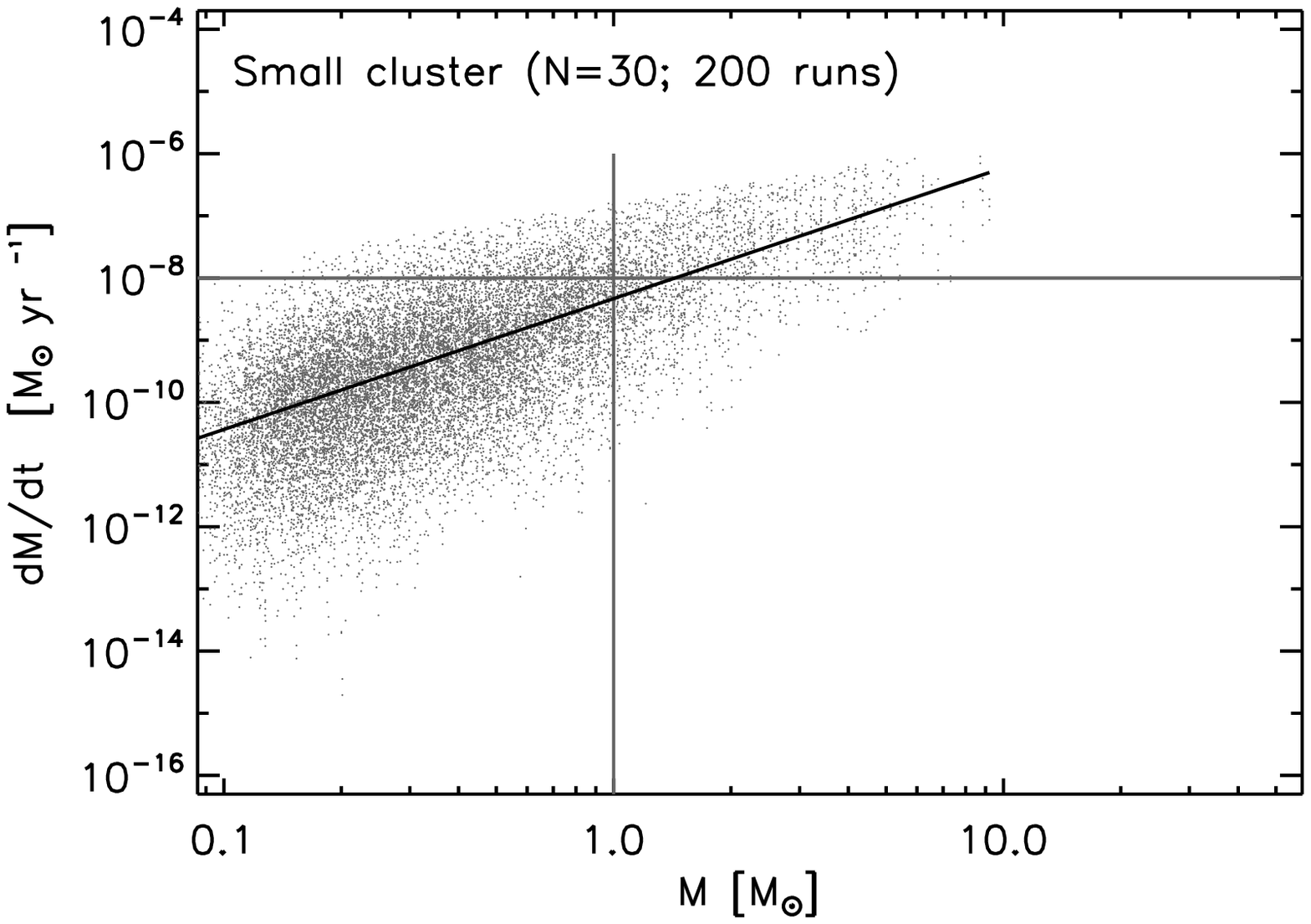}}}
\centerline{ \scalebox{0.5}{\includegraphics{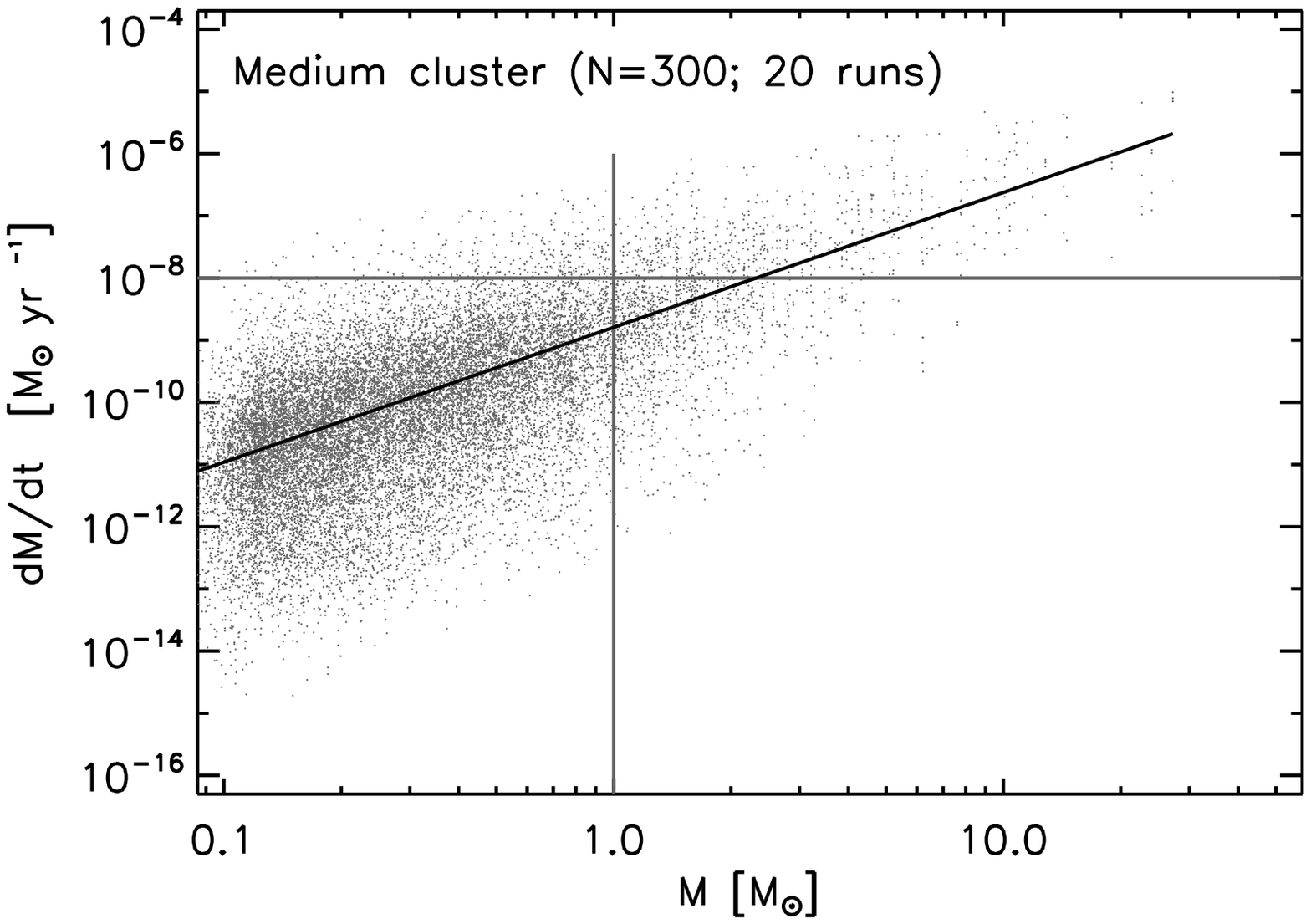}}}
\centerline{ \scalebox{0.5}{\includegraphics{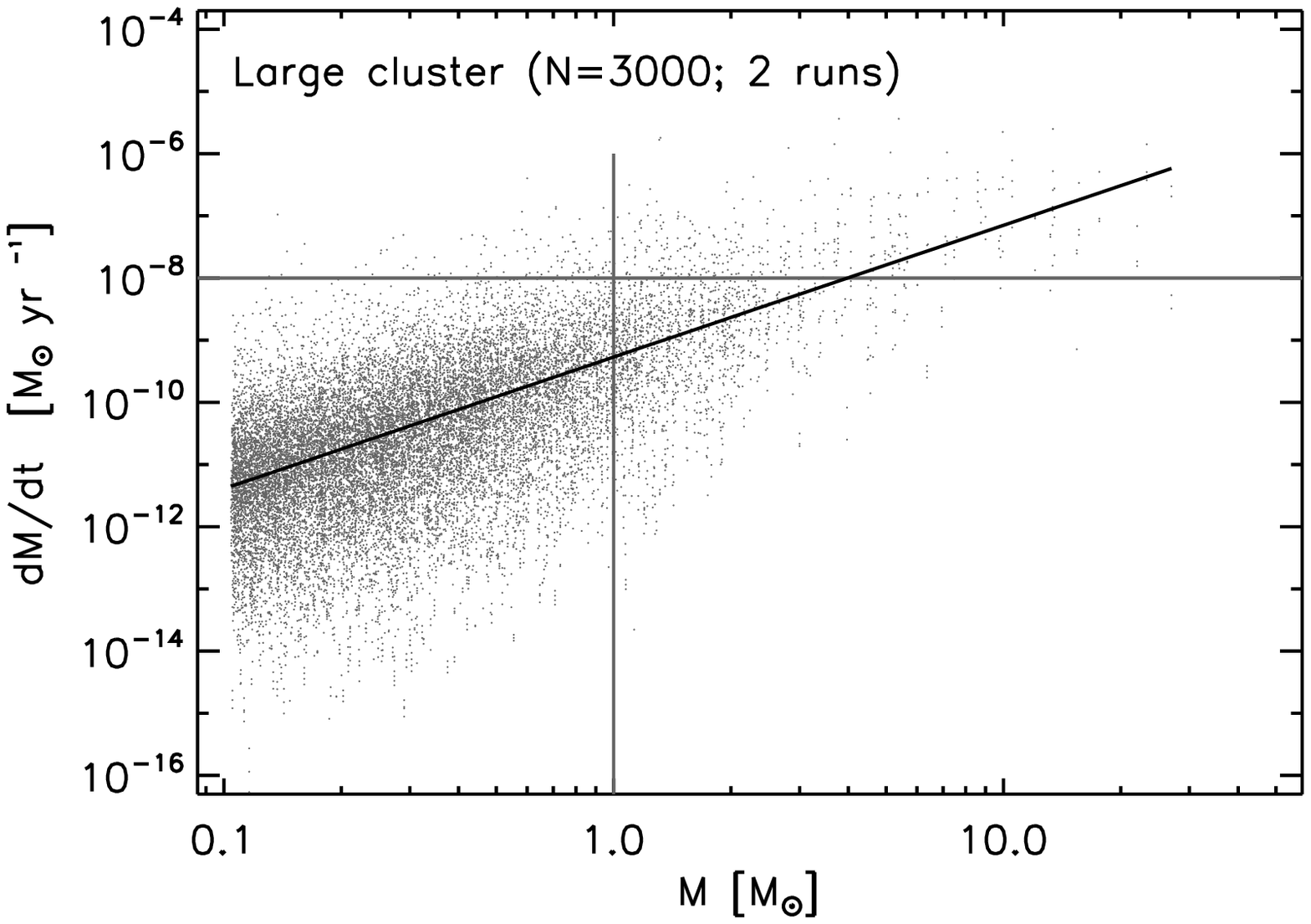}}}
\caption{Instantaneous (`median') values of \dmdtbh.  Each point represents a single star, at a single timestep, and the
vertical lines correspond to the varying accretion rates of individual stars through their orbits.  Solid lines are
numerical fits, which find $\mdot \sim M^{2.1\pm0.1}$.}
\label{fig:histogram_dmdt}
\end{figure}

\begin{small}
\begin{table}[h]
\centerline{
\begin{tabular}{|l|l|l|l|}\hline
{\bf Parameter}           	& {\bf \sc Small}    & {\bf \sc Medium}	 & {\bf \sc Large}		\\ \hline \hline
N				& 30             & 300            & 3000			\\ \hline
Number of runs			& 200		 & 20		  & 2			\\ \hline
$M_{\rm{gas}}$			& 30~\msol       & 300~\msol      & 3000~\msol		\\ \hline
T				& 5~K            & 15~K           & 40~K		\\ \hline
Peak stellar density (N pc$^{-3}$)& $10^4$	 & $10^5$         & $10^6$ 		\\ \hline
Peak gas density (n cm$^{-3}$)  &   $10^4$       & $10^5$         & $10^6$		\\ \hline
Stellar mass range              & 0.1  - 1.2 \msol & 0.1 - 18 \msol & 0.1 - 27 \msol 	\\ \hline
Plummer radius $r_0$		& \multicolumn{3}{c|}{0.22 pc}  \\ \hline
Star formation efficiency (SFE)	& \multicolumn{3}{c|}{33\%}	\\ \hline
Mean stellar mass		& \multicolumn{3}{c|}{0.5 \msol}	\\ \hline
Gas dispersal delay time	& \multicolumn{3}{c|}{2 Myr}		\\ \hline
End of simulation		& \multicolumn{3}{c|}{4 Myr}		\\ \hline 
\end{tabular} }
\caption{Initial conditions for the three cluster models.}
\label{table:params}
\end{table}
\end{small}


\begin{small}
\begin{table}[h]
\centerline{
\begin{tabular}{|l|l|l|l|}\hline
{\bf Result, main runs}        	& {\bf \sc Small}    & {\bf \sc Medium}	 & {\bf \sc Large} 	\\ \hline \hline
\dmbh ($\msol$)			& $ 0.038\, {\left({M \over \msol}\right)}^{2.0} $ 
				& $ 0.018\, {\left({M \over \msol}\right)}^{2.0} $
				& $ 0.008\, {\left({M \over \msol}\right)}^{2.0} $ \\ \hline
Mean \dmdtbh\ ($\msol\ \rm{yr^{-1}}$) 		
				& $9.3 \times 10^{-9}\, {\left({M \over \msol}\right)}^{2.0} $ 
				& $4.3 \times 10^{-9}\, {\left({M \over \msol}\right)}^{2.0} $
				& $2.1 \times 10^{-9}\, {\left({M \over \msol}\right)}^{2.0} $ \\ \hline
Median \dmdtbh\ ($\msol\ \rm{yr^{-1}}$)	
				& $  4.7\times10^{-9}\, {\left({M \over \msol}\right)}^{2.1} $  
                                & $  1.6\times10^{-9}\, {\left({M \over \msol}\right)}^{2.0}  $ 
                                & $ 0.56\times10^{-9}\, {\left({M \over \msol}\right)}^{2.1} $ \\ \hline \hline 
{\bf Result, T=25~K test case} 	          	& {\bf \sc Small} (25 K)    & {\bf \sc Medium } (25 K)	 & {\bf \sc Large} (25 K)\\ \hline 
\dmbh ($\msol$)			& $ 0.027\, {\left({M \over \msol}\right)}^{2.2} $ 
				& $ 0.017\, {\left({M \over \msol}\right)}^{2.1} $
				& $ 0.008\, {\left({M \over \msol}\right)}^{2.0} $ \\ \hline
Mean \dmdtbh\ ($\msol\ \rm{yr^{-1}}$) 		
				& $6.5 \times 10^{-9}\, {\left({M \over \msol}\right)}^{2.2} $ 
				& $3.9 \times 10^{-9}\, {\left({M \over \msol}\right)}^{2.1} $
				& $2.1 \times 10^{-9}\, {\left({M \over \msol}\right)}^{2.0} $ \\ \hline
Median \dmdtbh\ ($\msol\ \rm{yr^{-1}}$)	
				& $  3.6\times10^{-9}\, {\left({M \over \msol}\right)}^{2.2} $  
                                & $  1.5\times10^{-9}\, {\left({M \over \msol}\right)}^{2.0}  $ 
                                & $ 0.56\times10^{-9}\, {\left({M \over \msol}\right)}^{2.1} $ \\ \hline
\end{tabular} 
}
\caption{Results for the three cluster models.  The top table contains our final results; the lower table
investigates the effect of cluster temperature on accretion.}
\label{table:results}
\end{table}
\end{small}

\end{document}